\documentclass[a4paper,11pt]{article}
\usepackage{heppub}
\pdfoutput=1
\usepackage{xspace}

\newcommand{\WidthTwoSubfigs}{0.5\textwidth}
\newcommand{\WidthThreeSubfigs}{0.33\textwidth}

\newcommand{\abs}[1]{\lvert#1\rvert}
\newcommand{\ord}[1]{\mathcal{O}(#1)}
\newcommand{\ORd}[1]{\mathcal{O}\Bigl(#1\Bigr)}

\newcommand{\df}{\mathrm{d}}

\newcommand{\Li}{\textrm{Li}}

\newcommand{\sdt}{\!\cdot\!}

\newcommand{\al}{\alpha}
\newcommand{\eps}{\epsilon}

\newcommand{\cI}{{\mathcal I}}
\newcommand{\cL}{{\mathcal L}}
\newcommand{\Tau}{{\mathcal T}}

\newcommand{\bn}{\bar{n}}

\newcommand{\GeV}{\,\mathrm{GeV}}
\newcommand{\TeV}{\,\mathrm{TeV}}
\newcommand{\ab}{\,\mathrm{ab}}
\newcommand{\pb}{\,\mathrm{pb}}

\newcommand{\nn}{\nonumber}

\newcommand{\ptcut}{p_T^{\rm cut}}
\newcommand{\ptcuttwo}{\tilde{p}_T^{\rm cut}}
\newcommand{\Taucut}{\Tau_{\rm cut}}
\newcommand{\etacut}{\eta_{\rm cut}}
\newcommand{\zetacut}{\zeta_{\rm cut}}
\newcommand{\zetacuttwo}{\tilde{\zeta}_{\rm cut}}
\newcommand{\ptjet}{p_T^{\rm jet}}
\newcommand{\etajet}{\eta_{\rm jet}}

\newcommand{\lqcd}{\Lambda_\mathrm{QCD}}
\newcommand{\as}{\alpha_s}
\newcommand{\Ecm}{E_\mathrm{cm}}
\newcommand{\MSbar}{$\overline{\text{MS}}$\xspace}
\newcommand{\res}{\mathrm{res}}
\newcommand{\mufo}{\mu_\mathrm{FO}}

\allowdisplaybreaks[1]

\title{Jet Veto Resummation with Jet Rapidity Cuts}

\author{Johannes K.\,L.~Michel,}
\author{Piotr Pietrulewicz,}
\author{and Frank J.~Tackmann}

\affiliation{Theory Group, Deutsches Elektronen-Synchrotron (DESY),\\D-22607 Hamburg, Germany}

\emailAdd{johannes.michel@desy.de}
\emailAdd{piotr.pietrulewicz@desy.de}
\emailAdd{frank.tackmann@desy.de}

\abstract{%
Jet vetoes are widely used in experimental analyses at the LHC to distinguish
different hard-interaction processes. Experimental jet selections require a cut
on the (pseudo)rapidity of reconstructed jets, $|\eta_{\rm jet}| \leq \eta_{\rm cut}$. We
extend the standard jet-$p_T$ (jet-veto) resummation, which implicitly works in
the limit $\eta_{\rm cut}\to\infty$, by incorporating a finite jet rapidity cut. We
also consider the case of a step in the required $p_T^{\rm cut}$ at an intermediate
value of $|\eta| \simeq 2.5$, which is of experimental relevance to avoid
the increased pile-up contamination beyond the reach of the tracking detectors.
We identify all relevant parametric regimes, discuss their factorization and
resummation as well as the relations between them, and show that the
phenomenologically relevant regimes are free of large nonglobal logarithms. The
$\eta_{\rm cut}$ dependence of all resummation ingredients is computed to the same
order to which they are currently known for $\eta_{\rm cut}\to\infty$. Our results pave
the way for carrying out the jet-veto resummation including a sharp cut or a
step at $\eta_{\rm cut}$ to the same order as is currently available in the
$\eta_{\rm cut}\to\infty$ limit. The numerical impact of the jet rapidity cut is
illustrated for benchmark $q\bar q$ and $gg$ initiated color-singlet processes
at NLL$'+$NLO. We find that a rapidity cut at high $\eta_{\rm cut} = 4.5$ is safe to
use and has little effect on the cross section. A sharp cut at $\eta_{\rm cut} = 2.5$
can in some cases lead to a substantial increase in the perturbative
uncertainties, which can be mitigated by instead using a step in the veto.
}
\date{October 30, 2018}

\preprint{DESY 18-189}


\arxivnumber{1810.12911}

\journalref[https://doi.org/10.1007/JHEP04(2019)142]{JHEP 04 (2019) 142}

\begin{document}

\maketitle

%%%%%%%%%%%%%%%%%%%%%%%%%%%%%%%%%%%%%%%%%%%%%%%%%%%%%%%%%%%%%%%%%%%%%%%%%%%%%%%%
\vspace{-2ex}
\section{Introduction}
\label{sec:intro}
%%%%%%%%%%%%%%%%%%%%%%%%%%%%%%%%%%%%%%%%%%%%%%%%%%%%%%%%%%%%%%%%%%%%%%%%%%%%%%%%

Measurements that involve a veto on additional jets, or more generally that divide events
into exclusive jet bins, play an important role at the LHC, e.g.\ in Higgs and diboson
measurements or in searches for physics beyond the Standard Model.
The jet binning differentiates between hard processes that differ in the number of hard
signal jets, and hence allows one to separate signal and background processes.
The separation into $0$-jet and $\geq 1$-jet bins also provides a
model-independent way to discriminate between $q\bar q$ and $gg$ initiated
processes~\cite{Ebert:2016idf}.

A veto on jets with transverse momentum $p_T > \ptcut$ gives rise to double
logarithms $\ln^2(\ptcut/Q)$ at each order in $\alpha_s$,
where $Q$ is the characteristic momentum transfer of the hard interaction.
These logarithms dominate the perturbative series when $\ptcut \ll Q$,
and represent an important source of theory uncertainty~\cite{Berger:2010xi, Stewart:2011cf}.
They can be systematically resummed to improve the perturbative predictions and
assess the associated uncertainties, which has been well-developed in Drell-Yan
and Higgs production~\cite{Stewart:2009yx, Stewart:2010pd, Berger:2010xi,
Banfi:2012yh, Becher:2012qa, Tackmann:2012bt, Banfi:2012jm, Liu:2012sz,
Liu:2013hba, Becher:2013xia, Stewart:2013faa, Banfi:2013eda, Boughezal:2013oha,
Gangal:2014qda, Banfi:2015pju}, and has also been applied to several other
color-singlet processes~\cite{Shao:2013uba, Li:2014ria, Moult:2014pja,
Jaiswal:2014yba, Becher:2014aya, Wang:2015mvz, Tackmann:2016jyb, Ebert:2016idf,
Fuks:2017vtl}.

\begin{figure*}
\centering
\includegraphics[width=\WidthThreeSubfigs]{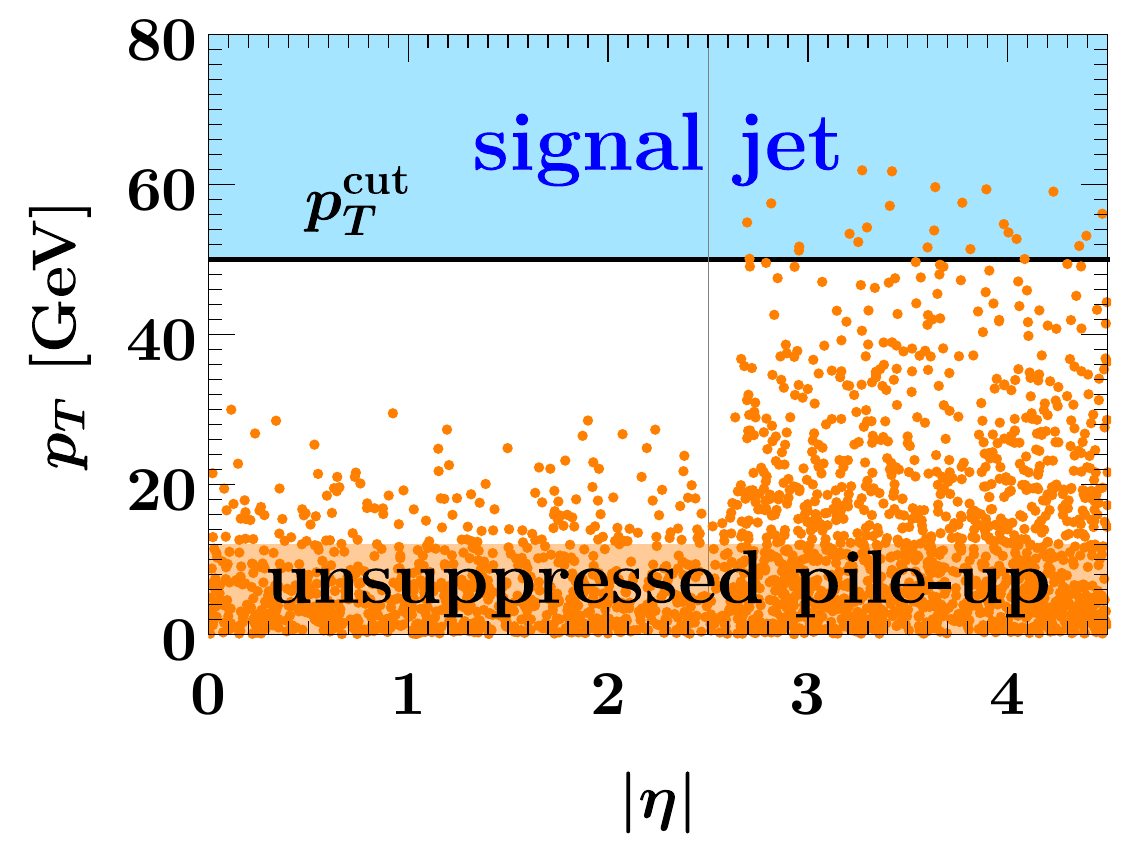}%
\hfill%
\includegraphics[width=\WidthThreeSubfigs]{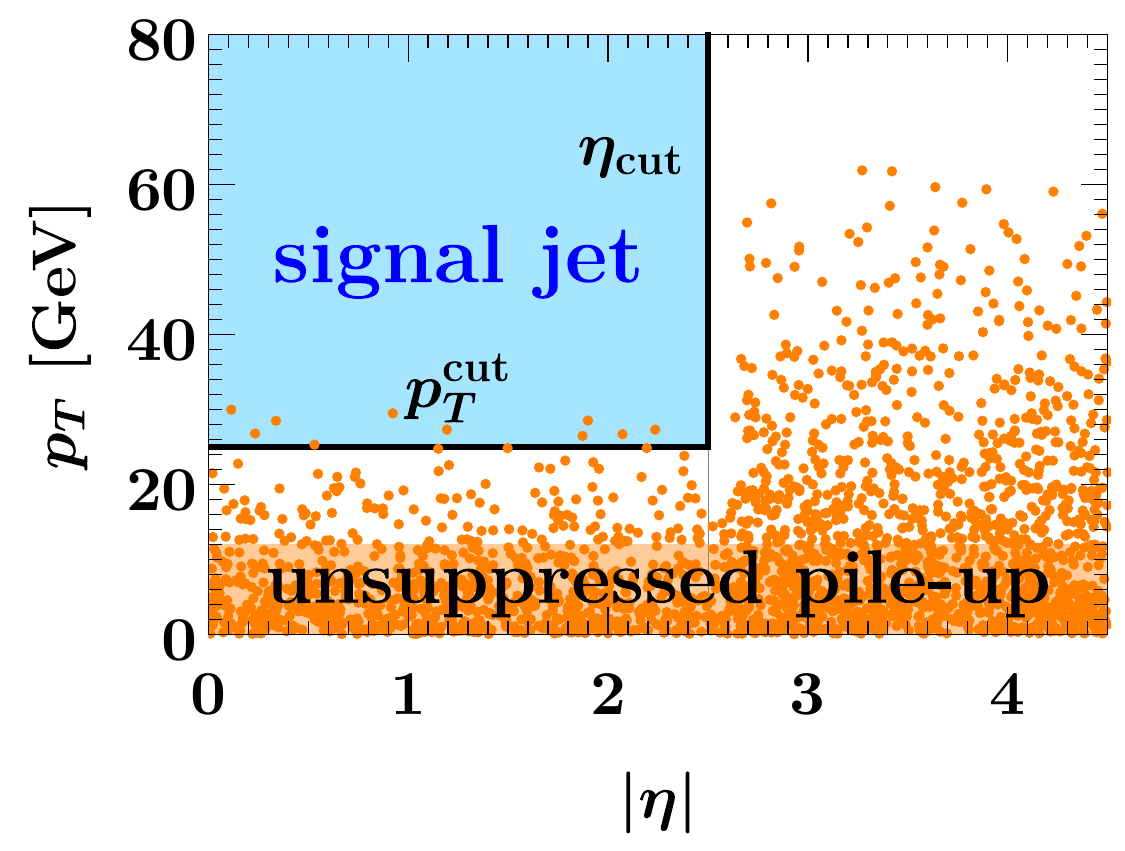}%
\hfill%
\includegraphics[width=\WidthThreeSubfigs]{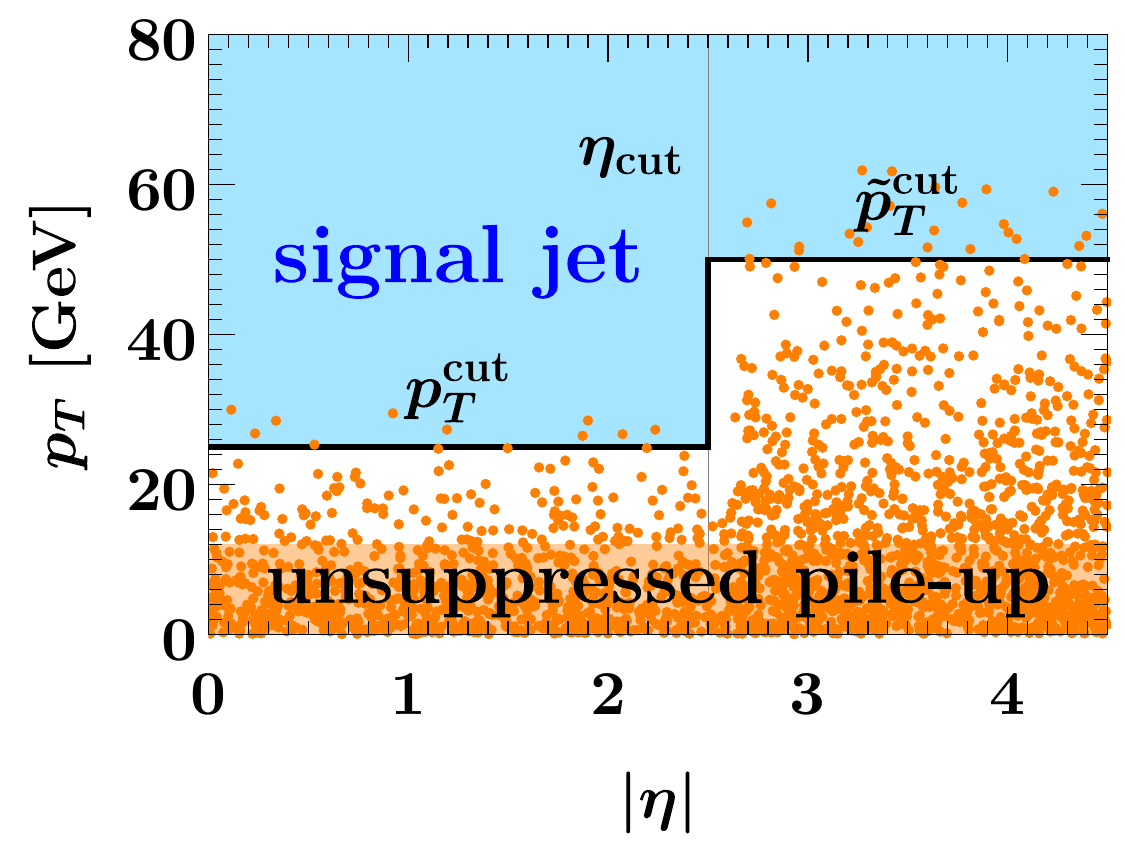}%
\caption{Cartoon of possible strategies to avoid contamination from unsuppressed
pile up in jet-binned analyses. The pile-up suppression is much better in the pseudorapidity range
$\abs{\eta} \lesssim 2.5$, where it can use information from the tracking detectors.
To avoid the higher pile-up contamination in the forward region, one can raise the
jet threshold (left panel), only consider central jets (middle panel), or combine
both approaches by using a step-like jet selection (right panel).}
\label{fig:illustration_pileup}
\end{figure*}

Experiments can only reconstruct jets up to some maximal pseudorapidity
$\abs{\eta} \leq \etacut$ due to the range of the detector, e.g.\ for ATLAS and
CMS $\etacut \sim 4.5$. In principle, the utility of the jet binning to
discriminate between different hard processes increases for a tighter jet veto
(smaller $\ptcut$). However, jets with small transverse momenta are difficult to
reconstruct experimentally, especially for pseudorapidity $\abs{\eta} \gtrsim
2.5$ beyond the reach of the tracking detectors, which are important to suppress
the large contamination from pile up (e.g.\ in the jet vertex tagging algorithm
used by ATLAS~\cite{Aad:2015ina}). This is illustrated in
\fig{illustration_pileup}. As the LHC luminosity increases and pile-up
conditions become harsher, the contamination from unsuppressed pile-up jets
grows worse and must be avoided. One option is to increase the overall $\ptcut$.
For example, in the context of Higgs measurements, the increased pile up in Run
2 has forced raising the jet threshold from $25\GeV$ to $30\GeV$. This however
weakens the jet veto and thus reduces its utility. Alternatively, to avoid
raising the jet threshold, one can consider jets only in a restricted
pseudorapidity range of $\abs{\eta} \lesssim 2.5$. However, this looses the
discrimination power from forward jets, which are a distinguishing feature of
some processes (most notably\ weak-boson fusion topologies in Higgs and diboson
production). The best possible option combines both approaches and performs a
step-like jet selection, with a lower $\ptcut$ threshold for central jets and a
somewhat higher $\ptcuttwo$ threshold for forward jets. For example, recent
ATLAS Higgs measurements~\cite{Aaboud:2018xdt} reconstruct jets using $\ptcut
=25 \GeV$ for $\abs{\eta}<2.4$ and $\ptcuttwo = 30 \GeV$ for $\abs{\eta}>2.4$
(and no jets beyond rapidity $\abs{y}=4.4$).

A discontinuous step in the jet threshold can also pose challenges on its own,
as it makes the experimental measurements more complex. Theoretically, we will see that it can
complicate the resummation of logarithms in some extreme cases. An alternative
to a step is to use jet vetoes that smoothly depend on the jet
rapidity~\cite{Gangal:2014qda, Gangal:2016kuo}, providing a tighter veto at
central rapidities and a looser one at forward rapidities. These rapidity-dependent
vetoes can also be supplemented with an additional sharp jet rapidity cut, which we briefly
discuss in \app{TauCut}.

The usual jet-$p_T$ resummations~\cite{Banfi:2012yh, Becher:2012qa,
Tackmann:2012bt, Banfi:2012jm, Becher:2013xia, Stewart:2013faa} do not account
for any jet rapidity dependence, i.e., the resummation is performed for
$\etacut\to\infty$. Using parton-shower Monte Carlos, one finds that a jet
rapidity cut at $\etacut = 4.5$ has a very small numerical effect, while
$\etacut = 2.5$ has a sizable effect on the jet-$p_T$ spectrum in Higgs
production (see e.g.\ \refscite{Berger:2010xi, Banfi:2012yh}), so it is
important to properly include it in the resummation. This was already pointed
out in \refcite{Tackmann:2012bt}, where it was also speculated that a jet
rapidity cut might change the resummation structure.

Our analysis in this paper fully addresses these questions by systematically
incorporating the jet rapidity cut into the jet-$p_T$ resummation, including in
particular the case of a step-like veto. For this purpose, we extend the
formalism of \refscite{Tackmann:2012bt, Stewart:2013faa}, which uses the
framework of Soft-Collinear Effective Theory (SCET)~\cite{Bauer:2000ew,
Bauer:2000yr, Bauer:2001ct, Bauer:2001yt}. To be concrete, our discussion
focuses on color-singlet production, including the important cases of Higgs and
Drell-Yan production. Our results for how to incorporate the $\etacut$
dependence also carry over to processes with additional signal jets in the final
state to the same extent to which the usual jet-$p_T$ resummation for
color-singlet production carries over to such cases~\cite{Liu:2012sz,
Liu:2013hba}.

We identify all relevant parametric regimes in the veto parameters $\ptcut$,
$\etacut$, $\ptcuttwo$, and discuss the factorization and resummation structure
for each regime. We also study the relations between the different regimes and
perform numerical studies to check their respective ranges of validity. An
important conclusion of our analysis is that all regions of parameter space that
are of phenomenological interest can be described by parametric regimes that are
free of large nonglobal logarithms.

We analytically compute the $\etacut$ dependence of all ingredients at
$\ord{\alpha_s}$ as well as of the dominant $\ord{\alpha_s^2}$ corrections
(those enhanced by jet-veto or jet clustering logarithms), which matches the order to which they are
currently known in the $\etacut\to\infty$ limit. Our results allow for carrying out the
jet-veto resummation including jet rapidity cuts to the same
order as is currently available without such cuts, which for color-singlet
production is NNLL$'+$NNLO. (Reaching this level also requires the
still unknown nonlogarithmic $\ord{\alpha_s^2}$, which can be extracted
numerically from the full NNLO calculation, as was done for $\etacut\to\infty$
in \refcite{Stewart:2013faa}. Carrying out such an analysis is beyond the scope
of this paper.)

The effect of a rapidity cut for transverse momentum vetoes has also been
considered independently in \refscite{Hornig:2016ahz, Hornig:2017pud} for dijet production, and
more recently for the transverse energy event shape in Drell-Yan in
\refcite{Kang:2018agv}. We compare their results to our results for the case of a sharp cut
at $\etacut$ and no measurement beyond in \sec{no_step_literature}.

The paper is organized as follows: In \sec{no_step}, we discuss the parametric
regimes and corresponding effective field theory (EFT) setups for a sharp cut on
reconstructed jets at $\etacut$ and no measurement beyond, as in the middle
panel of \fig{illustration_pileup}. We give the perturbative ingredients at
$\ord{\alpha_s}$ and the leading small-$R$ clustering terms at
$\ord{\alpha_s^2}$ for all partonic channels. We numerically validate the EFT
setup by comparing to the relevant singular limits of full QCD, and also compare
the regimes to each other and identify their respective ranges of validity. In
\sec{yes_step}, we generalize the results of \sec{no_step} to a step in the jet
veto at $\etacut$, as in the right panel of \fig{illustration_pileup}. In
\sec{analysis}, we illustrate the numerical impact of the rapidity cut at
NLL$'+$NLO for Drell-Yan at $Q = m_Z$ and $Q = 1\TeV$ and for $gg\to H$ at $m_H
= 125\GeV$ and $gg\to X$ at $m_X = 1\TeV$ for different values of $\etacut$. We
conclude in \sec{conclusion}. Details of our calculations can be found in
\app{ingredients}. In \app{TauCut}, we briefly discuss how an additional sharp
rapidity cut affects the description of the rapidity-dependent jet vetoes
introduced in~\refcite{Gangal:2014qda}.

%%%%%%%%%%%%%%%%%%%%%%%%%%%%%%%%%%%%%%%%%%%%%%%%%%%%%%%%%%%%%%%%%%%%%%%%%%%%%%%%
\section{\boldmath Factorization with no constraint beyond \texorpdfstring{$\etacut$}{etacut} (\texorpdfstring{$\ptcuttwo = \infty$}{tildepTcut = infinity})}
\label{sec:no_step}
%%%%%%%%%%%%%%%%%%%%%%%%%%%%%%%%%%%%%%%%%%%%%%%%%%%%%%%%%%%%%%%%%%%%%%%%%%%%%%%%

%===============================================================================
\subsection{Overview of parametric regimes}
%===============================================================================

\begin{figure*}
\centering
\includegraphics[width=\WidthTwoSubfigs]{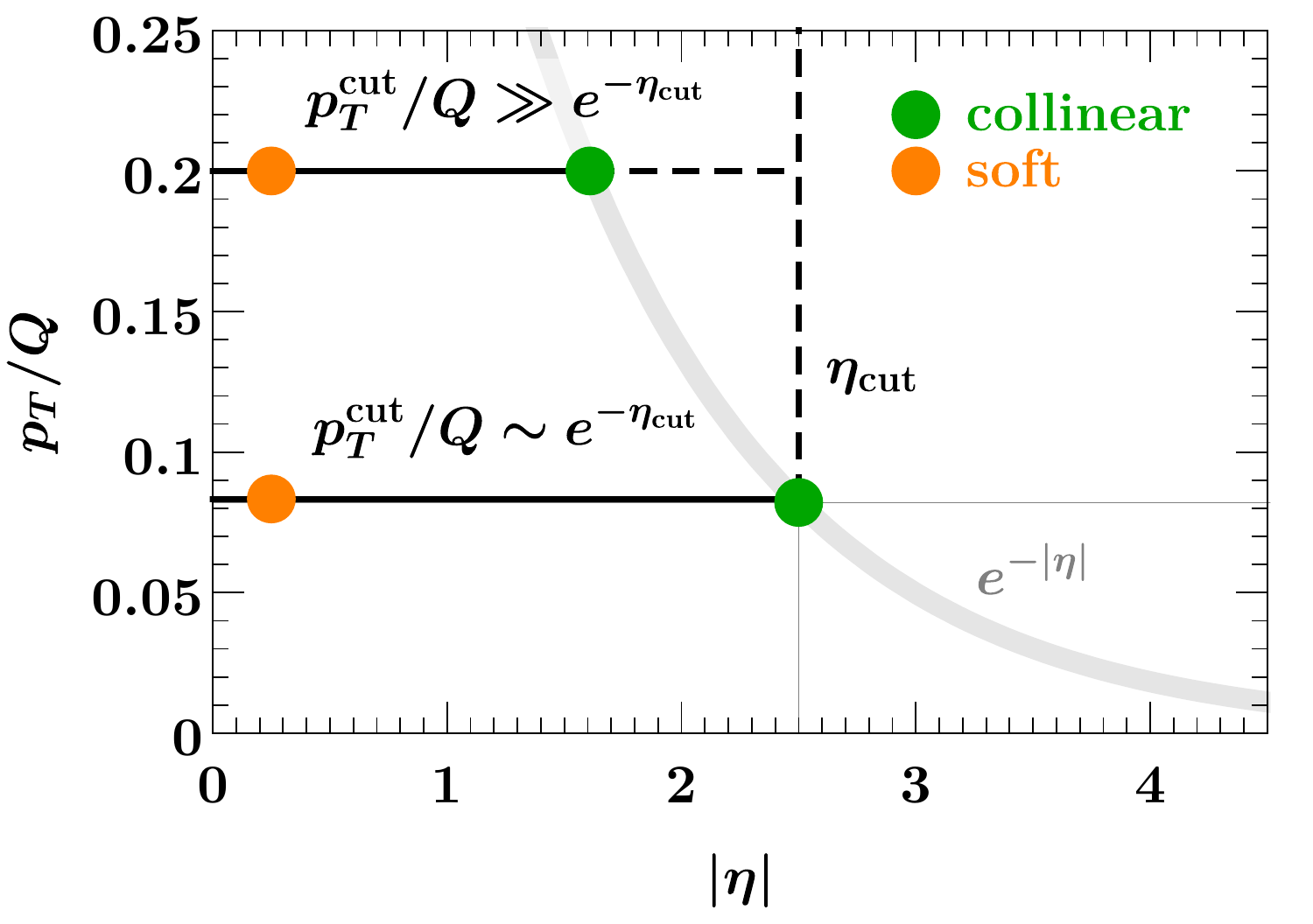}%
\hfill%
\includegraphics[width=\WidthTwoSubfigs]{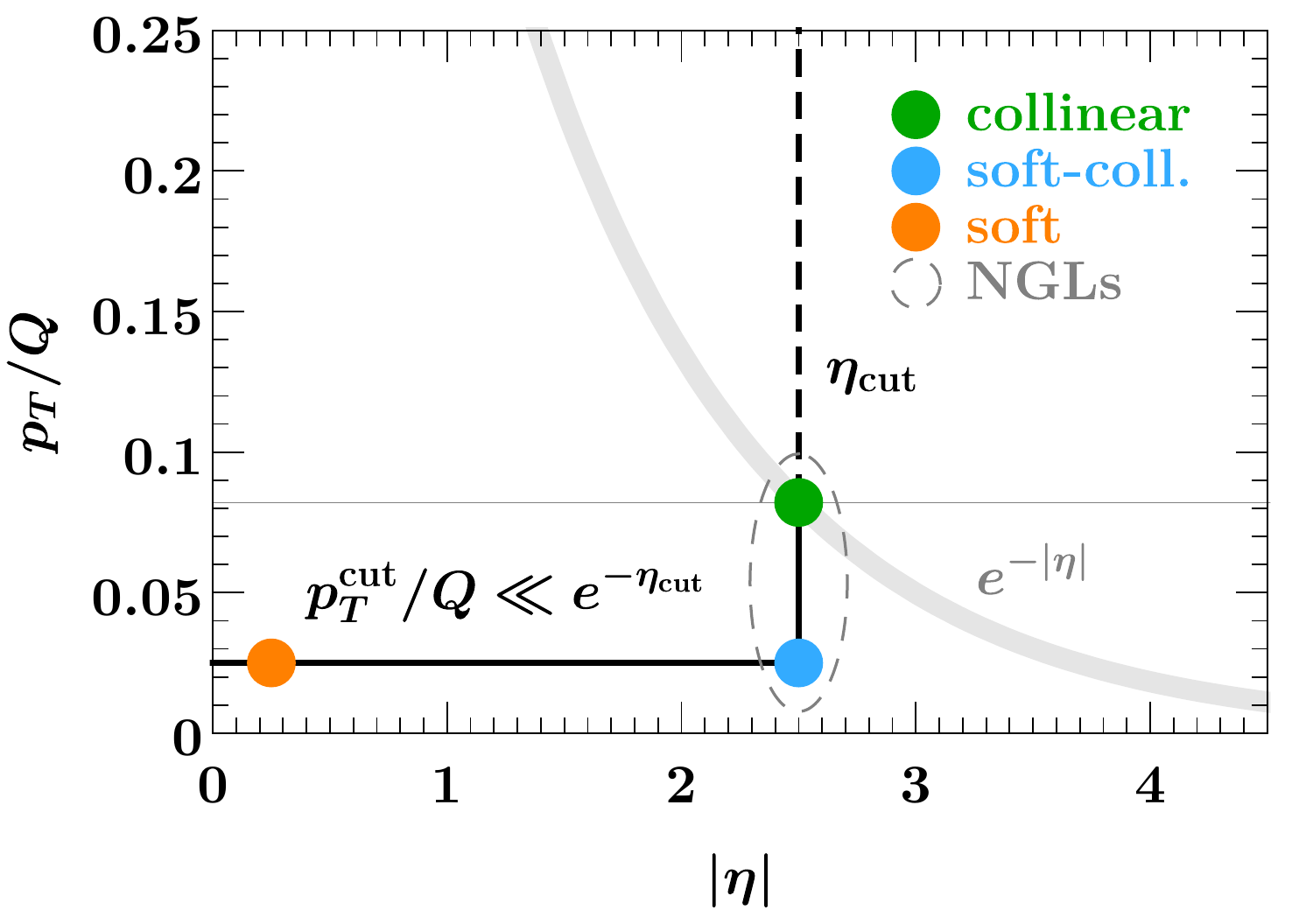}%
\caption{
Illustration of the parametric regimes for a jet veto with a jet rapidity cut.
Emissions above the black solid lines are vetoed as $p_T > \ptcut$ up to
$\abs{\eta} < \etacut = 2.5$. The thick gray line corresponds to $p_T/Q =
e^{-\abs{\eta}}$, and emissions above and to the right of it are power
suppressed. The colored circles indicate the relevant modes in the effective
theory for a given hierarchy between $\ptcut/Q$ and $e^{-\etacut}$. For $\ptcut
= 25 \GeV$, the given examples for $\ptcut/Q$ correspond to $Q = 125 \GeV$ (left
panel, upper case), $Q = 300 \GeV$ (left panel, lower case), $Q = 1\TeV$ (right
panel).
}
\label{fig:sketch_regimes_no_step}
\end{figure*}

We consider exclusive 0-jet cross sections, where the veto is applied by
identifying jets with radius $R$ (the details of the jet-clustering algorithm are
not relevant at the order we are working) and cutting on the transverse momentum
$\ptjet$ of the leading jet within $\abs{\etajet} < \etacut$,
%%%
\begin{equation} \label{eq:def_veto_no_step}
\max_{k \in \text{jets}:\,\abs{\eta_k} < \etacut} \abs{\vec{p}_{T,k}}  < \ptcut
\,.\end{equation}
%%%
The resulting constraints on the rapidities and transverse momenta of
initial-state radiation (ISR) are displayed as black lines in
\fig{sketch_regimes_no_step}. We can identify two distinct power-counting
parameters that govern the typical angular size of energetic collinear ISR with
energy $E \sim Q$, where $Q$ is the momentum transferred in the hard
interaction: First, the $p_T$ of the emissions is constrained by $p_T < \ptcut$
for $\abs{\eta} < \etacut$, corresponding to a maximum opening angle
%%%
\begin{equation}
\frac{p_T}{E} \lesssim \frac{\ptcut}{Q}
\,.\end{equation}
%%%
Second, the $p_T$ of an energetic emission at rapidity $\eta$ is parametrically $p_T \sim Q
e^{-\abs{\eta}}$. The rapidity cut removes the first constraint for $\abs{\eta}
> \etacut$. Hence, if $\etacut$ is central enough, emissions beyond $\etacut$
can reach a characteristic $p_T \lesssim Q e^{-\abs{\etacut}}$, corresponding to
a maximum opening angle
%%%
\begin{equation}
\frac{p_T}{E} \lesssim e^{-\etacut}
\,.\end{equation}
%%%

There are three parametric regimes for $\ptcut/Q$ and $e^{-\etacut}$, which are
illustrated in \fig{sketch_regimes_no_step} for $\etacut = 2.5$. The thick black
lines show the veto for different values of $\ptcut/Q$. The thick gray curve
shows the relation $p_T/Q = e^{-\abs{\eta}}$, while the thin gray lines show the
values of $\etacut$ and $p_T/Q = e^{-\etacut}$.

The first parametric regime is $\ptcut/Q \gg e^{-\etacut}$. As we will
demonstrate in \sec{no_step_reg1}, in this regime effects due to the rapidity
cut are power suppressed by $Q e^{-\etacut}/\ptcut$. Hence, they can be treated
as a fixed-order power correction to the standard jet-veto resummation, which
implicitly works in the limit $\etacut = \infty$. For Higgs measurements with
$\ptcut = 25 \GeV$, $\etacut = 4.5$, $Q \equiv m_H = 125\GeV$, this parametric
assumption is well justified, as $m_H e^{-\etacut}/\ptcut \sim 5 \%$.

For heavier final states and/or more central rapidity cuts the relevant
parametric regime is $\ptcut/Q \sim e^{-\etacut}$. This is the case for example
for $Q= 1 \TeV$ and $\etacut=4.5$ or $Q= 125 \GeV$ and $\etacut=2.5$ at $\ptcut =
25\GeV$. In \sec{no_step_reg2}, we show that in this regime the rapidity cut
effects must be treated as a leading-power correction, and that they can be
seamlessly incorporated into the existing jet-veto resummation without
rapidity cut. We will see that they affect only the boundary terms
in the resummed cross section, but not the anomalous dimensions
and evolution factors. Hence, they start contributing at NLL$'$ or NNLL.

Finally in \sec{no_step_reg3}, we discuss the parametric regime $\ptcut/Q \ll
e^{-\etacut}$. This case is conceptually interesting, since logarithms of the
ratio of scales $Q e^{- \etacut}$ and $\ptcut$ appear, changing the logarithmic
structure already at leading-logarithmic (LL) order. In addition, formally large
nonglobal logarithms of the same ratio appear. This regime is of
very limited phenomenological relevance for typical jet-binned analyses at the
LHC. For example, for $\etacut = 2.3$ corresponding to $e^{-\etacut} = 0.1$, it
would require an extremely tight jet veto $\ptcut \ll 0.1\, Q$, which is
unrealistic as it would leave almost no signal in the $0$-jet cross
section. For the purpose of explicitly probing this regime experimentally, one
could lower $\etacut \simeq 1.0-1.5$, such that the jet veto only acts
on radiation in the very central region.

%===============================================================================
\subsection[Regime 1: \texorpdfstring{$\ptcut/Q \gg e^{-\etacut}$}{pTcut/Q >> exp(-etacut)} (standard jet veto resummation)]
{\boldmath  Regime 1: $\ptcut/Q \gg e^{-\etacut}$ (standard jet veto resummation)}
\label{sec:no_step_reg1}
%===============================================================================

As usual, the scaling of the modes in the EFT follows from the nontrivial constraints imposed on emissions by the measurement. Soft emissions at central rapidities are always restricted by the jet veto. Collinear emissions with energy $\sim Q$ and rapidity $\eta$ have a transverse momentum $\sim Q e^{-\abs{\eta}}$ and are constrained by the measurement if $Q e^{-\abs{\eta}} \sim \ptcut$, which determines their scaling.
Since $Q e^{-\etacut} \ll \ptcut$, these collinear modes are parametrically not forward enough to
be sensitive to the rapidity cut, such that the description of their dynamics is simply governed by the power counting in $\ptcut/Q$. The relevant EFT modes in this regime are thus the same as for a jet veto without any rapidity cut,
%%%
\begin{align}
\text{soft:} 
&\quad
p^\mu\sim \Bigl(\ptcut,\ptcut,\ptcut\Bigr)
\,, \nn \\
n_a \text{-collinear:} 
&\quad
p^\mu \sim \Bigl(\tfrac{(\ptcut)^2}{Q},Q,\ptcut\Bigr)
\,, \nn \\
n_b \text{-collinear:} 
&\quad
p^\mu\sim \Bigl(Q,\tfrac{(\ptcut)^2}{Q},\ptcut\Bigr)
\,.\end{align}
%%%
Here and below, we give the scaling of momenta in terms of light-cone components defined as
(with $n \equiv n_a$, $\bn \equiv n_b$),
%%%
\begin{equation}
p^\mu
= \bn\sdt p\, \frac{n^\mu}{2} + n \sdt p\, \frac{\bn^\mu}{2} + p^\mu_{\perp}
\equiv (n \sdt p,\bar{n} \sdt p,\vec p_{\perp}) \equiv (p^+,p^-,\vec p_\perp)
\,.\end{equation}
%%%
In addition, there are the usual inclusive collinear modes that describe the initial protons
at the scale $\lqcd$, and which are not specific to our discussion here.

In principle, we can consider collinear emissions that are forward enough to resolve rapidities
$\abs{\eta} \sim \etacut$,
%%%
\begin{align}
n_a \text{-collinear ($\etacut$):}
&\quad
p^\mu \sim \Bigl(Qe^{-2\etacut},Q,Qe^{-\etacut}\Bigr)
\,, \nn \\
n_b \text{-collinear ($\etacut$):}
&\quad
p^\mu\sim \Bigl(Q,Qe^{-2\etacut},Qe^{-\etacut}\Bigr)
\,.\end{align}
%%%
However, since $Qe^{-\etacut} \ll \ptcut$, these emissions have too little transverse momentum to be affected
by the jet veto, and are therefore unconstrained and integrated over without
requiring additional modes in the EFT.
To explicitly see that the $\etacut$ dependence is power suppressed, note that the
full jet-veto measurement for the collinear modes contains a $\theta$ function
%%%
\begin{equation}
\theta(\etacut-\abs{\eta}) = \theta(1 - e^{\abs{\eta}-\etacut}) = 1 + \ord{Q e^{-\etacut}/\ptcut}
\,,\end{equation}
%%%
which thus only induces power corrections in $Q e^{-\etacut}/\ptcut$.

Therefore, at leading order in the power expansion,%
\footnote{As discussed in \refscite{Tackmann:2012bt, Stewart:2013faa}, one formally needs to count
$R \ll 1$ to avoid soft-collinear mixing terms of $\ord{R^2}$.
A detailed discussion of possible approaches to include them at $\ord{\alpha_s^2}$
can be found in \refcite{Gangal:2016kuo}.}
we recover the factorization for the 0-jet cross section with $\etacut = \infty$~\cite{Tackmann:2012bt, Becher:2012qa, Stewart:2013faa},
%%%
\begin{align} \label{eq:fact_pTjet}
\sigma_0 (\ptcut,\etacut,R,\Phi)
&=
H_\kappa(\Phi, \mu) \,
B_{a}(\ptcut, R, \omega_a, \mu, \nu ) \,
B_{b}(\ptcut, R, \omega_b, \mu, \nu) \,
S_{\kappa} (\ptcut, R, \mu, \nu) \nn \\
& \quad \times 
\biggl[1+\mathcal{O}\Bigl(\frac{\ptcut}{Q},\frac{Q e^{-\etacut}}{\ptcut},R^2\Bigr)\biggr]
\, .\end{align}
%%%
The hard function $H_\kappa$ contains the short-distance matrix element for producing a color-singlet final state
and depends on the hard kinematic phase space $\Phi$, which encodes e.g.\ the total rapidity $Y$ and invariant mass $Q$
of the color-singlet final state. The soft function $S_\kappa$ encodes soft radiation restricted by $\ptcut$.
The partonic channel is denoted by $\kappa$ and is implicitly summed over (if necessary).
The beam functions $B_{a,b}$ are forward proton matrix elements of collinear SCET fields and
encode the perturbative collinear ISR constrained by $\ptcut$ as well as the
unconstrained ISR below that scale down to the nonperturbative scale of the PDFs~\cite{Stewart:2009yx}.
In \eq{fact_pTjet}, they are evaluated at $\omega_{a,b} =Q e^{\pm Y}$.
They are given by a convolution of perturbative matching coefficients $\mathcal{I}_{ij}$, which encode
the $\ptcut$ constraint, and the standard inclusive quark and gluon PDFs $f_{j}$,
%%%
\begin{align} \label{eq:B}
B_i(\ptcut, R, \omega, \mu, \nu)
= \sum_{j} \int^1_x \frac{\df z}{z}\,
\mathcal{I}_{ij}(\ptcut, R, \omega, z, \mu, \nu) \,
f_j \Bigl(\frac{\omega}{z \Ecm}, \mu\Bigr) \biggl[1+\mathcal{O}\Bigl(\frac{\lqcd}{\ptcut}\Bigr)\biggr]
\,.\end{align}
%%%

As discussed in detail in \refcite{Stewart:2013faa}, all logarithms of the ratio
$\ptcut/Q$ in \eq{fact_pTjet} are resummed by evaluating each of the hard,
beam, and soft functions at their characteristic virtuality and rapidity scales,
%%%
\begin{align} \label{eq:scales1}
\mu_H \sim Q = \sqrt{\omega_a \omega_b}
\, , \quad  \mu_B \sim \mu_S \sim \ptcut
\, , \quad \nu_B \sim Q
\, , \quad \nu_S \sim \ptcut
\,,\end{align}
%%%
and evolving them to common scales $\mu$, $\nu$ using renormalization group (RG)
evolution. The power corrections in \eq{fact_pTjet} can be included at fixed
order in $\alpha_s$ by matching the resummed result to the corresponding
fixed-order result in full QCD. The $\ord{Q e^{-\etacut}/\ptcut}$ corrections
stop being suppressed for large $Q$, small $\ptcut$, or central
$\etacut$. In the next section, we show that they can be incorporated into the
beam functions in \eq{B}.

%===============================================================================
\subsection[Regime 2: \texorpdfstring{$\ptcut/Q \sim e^{-\etacut}$}{pTcut/Q sim exp(-etacut) sim pTcut} (\texorpdfstring{$\etacut$}{etacut} dependent beam functions)]
{\boldmath Regime 2: $\ptcut/Q \sim e^{-\etacut}$ ($\etacut$ dependent beam functions)}
\label{sec:no_step_reg2}
%===============================================================================

In this regime, the scaling of soft and collinear modes is unchanged from the
previous case. However, the characteristic rapidity of the collinear modes now
coincides parametrically with $\etacut$, i.e., %%%
\begin{align} \label{eq:modes_no_step_reg2}
\text{soft:}
&\quad
p^\mu\sim \Bigl(\ptcut,\ptcut,\ptcut\Bigr)
\,, \nn \\
n_a \text{-collinear:}
&\quad
p^\mu \sim \Bigl(\tfrac{(\ptcut)^2}{Q},Q,\ptcut\Bigr) \sim \Bigl(Qe^{-2\etacut},Q,Qe^{-\etacut}\Bigr)
\,, \nn \\
n_b \text{-collinear:} 
&\quad
p^\mu \sim \Bigl(Q,\tfrac{(\ptcut)^2}{Q},\ptcut\Bigr) \sim \Bigl(Q,Qe^{-2\etacut},Qe^{-\etacut}\Bigr)
\,.\end{align}
%%%
Thus, collinear emissions resolve the rapidity cut, and are constrained by the
jet veto for $|\eta|<\etacut$, while for $|\eta|>\etacut$ they are
unconstrained. As a result, the cross section factorizes at leading power as
%%%
\begin{align} \label{eq:fact_pTjet2}
\sigma_0 (\ptcut, \etacut, R, \Phi)
&=
H_\kappa(\Phi, \mu) \,
B_{a}(\ptcut, \etacut, R, \omega_a, \mu, \nu) \,
B_{b}(\ptcut, \etacut, R, \omega_b, \mu, \nu) \nn \\
& \quad \times  S_{\kappa} (\ptcut, \mu, \nu) \,
\biggl[1+\mathcal{O}\Bigl(\frac{\ptcut}{Q}, e^{-\etacut},R^2\Bigr)\biggr]
\,.\end{align}
%%%
The beam functions now explicitly depend on both $\ptcut$ and $\etacut$, while
the hard and soft functions are unchanged (with their characteristic scales still
given by \eq{scales1}).
The RG consistency of the cross section fixes the anomalous dimensions
of the beam function in terms of those for the soft and hard functions. Thus,
the $\etacut$ dependence cannot change the renormalization of the beam function, i.e.,
%%%
\begin{align} \label{eq:RGE_B}
\mu \frac{\df}{\df \mu}  \ln B_i(\ptcut, \etacut, R, \omega, x, \mu, \nu)
&= \gamma^{i}_B (\omega, \mu, \nu)
\,, \nn \\
\nu \frac{\df}{\df \nu}  \ln B_i(\ptcut, \etacut, R, \omega, x, \mu, \nu)
&= \gamma^{i}_{\nu,B} (\ptcut,R,\mu )
\,,\end{align}
%%%
where the anomalous dimensions are the same as in the $\etacut\to\infty$
limit~\cite{Tackmann:2012bt, Stewart:2013faa},
%%%
\begin{align}\label{eq:anom_dim_B}
\gamma^{i}_{B} (\omega, \mu, \nu)
&= 2 \Gamma^{i}_{\rm cusp}[\as(\mu)] \, \ln \frac{\nu}{\omega} + \gamma^{i}_{B}[\as(\mu)]
\,,\nn\\
\gamma^{i}_{\nu,B}(\ptcut, R, \mu)
&= 2\eta_\Gamma^{i}(\ptcut,\mu) + \gamma^{i}_{\nu,B}[\al_s(\ptcut),R]
\,,\end{align}
%%%
and $\eta_\Gamma^{i}$ in the resummed rapidity anomalous dimension is given by
%%%
\begin{equation}
\eta_\Gamma^{i}(\mu_0,\mu)
= \int_{\mu_0}^\mu \! \frac{\df \mu'}{\mu'} \, \Gamma^{i}_{\rm cusp}[\as(\mu')]
\,.\end{equation}
%%%
Hence, the $\etacut$ effects do not affect the RG evolution itself, but only change the
beam function boundary conditions, and therefore first appear at NLL$'$.
The RG evolution between $\mu_B \sim \ptcut \sim Q e^{-\etacut}$ and $\mu_H \sim Q$
now resums all large logarithms of $\mu_B / \mu_H \sim \ptcut/Q \sim $ $e^{-\etacut}$,
while the beam function boundary condition now explicitly
depends on the ratio $Q e^{-\etacut} / \ptcut \sim \ord{1}$, which in contrast to
regime 1 is not power suppressed anymore.

In analogy to \eq{B} the beam functions can be factorized into collinear matching
coefficients, which now also depend on $\etacut$, and the PDFs.
We write the matching coefficients as the sum of the usual $\etacut$-independent
matching coefficients plus a correction term that encodes the $\etacut$ dependence,
%%%
\begin{align}\label{eq:I_tot}
\mathcal{I}_{ij}(\ptcut, \etacut, R, \omega, z,\mu, \nu)
&=\mathcal{I}_{ij}(\ptcut, R, \omega, z, \mu, \nu )
+ \Delta \mathcal{I}_{ij}( \ptcut, \etacut, R, \omega, z, \mu, \nu)
\,.\end{align}
%%%
The $\etacut$-independent $\mathcal{I}_{ij}$ are given in \app{standard_beam_function},
and in the following we focus on the $\Delta \mathcal{I}_{ij}$.

Consistency between the cross sections in \eqs{fact_pTjet}{fact_pTjet2} implies
that $\Delta \mathcal{I}_{ij}$ vanishes as $\etacut \to \infty$.
Specifically, defining
%%%
\begin{equation}
\zeta_{\rm cut} \equiv \omega e^{-\etacut}/\ptcut
\,,\end{equation}
%%%
the $\Delta \cI_{ij}$ scale like
%%%
\begin{align}\label{eq:DeltaI_limit}
\Delta \mathcal{I}_{ij}\bigl(\ptcut, \etacut, R, \omega, z, \mu, \nu \bigr) \sim \ord{\zetacut}
 \qquad \rm{for} \,\, \zetacut \to 0
\,,\end{align}
%%%
which is simply the statement from the previous subsection that the $\etacut$ effects
are power suppressed in $\zeta_{\rm cut}$ for $\zeta_{\rm cut} \ll 1$.

In fact, $\Delta \mathcal{I}_{ij}$ vanishes altogether for $z > \zeta_{\rm
cut}/(1+ \zeta_{\rm cut})$, which can be seen from purely kinematic
considerations as follows: For the $n$-collinear sector the term $\Delta
\mathcal{I}_{ij}$ accounts for the case where at least one jet with $p^{\rm
jet}_{T} \geq \ptcut$ and $\eta_{\rm jet} \geq \etacut$ is reconstructed (and no
jet with $\eta_{\rm jet} < \etacut$). For $R \ll 1$ all radiation in this jet
has $\eta \geq \etacut$, as well. Thus, contributions to $\Delta
\mathcal{I}_{ij}$ can only appear if %%%
\begin{align}
\ptcut \leq |\vec{p}^{\,\text{jet}}_{T}| \leq \sum_{k \in  \text{jets}} |\vec{p}_{T,k}| = \sum_{k \in \text{jets}} p_k^- e^{- \eta_k} \, ,
\end{align}
%%%
where the second equality follows from the jets being massless for $R \ll 1$.
Rewriting this in terms of momentum fractions $p_k^- = z_k \,P_n^- = z_k\,
\omega/z$ yields, with $\sum_k z_k +z =1$ and $P_n^-$ the momentum of the
initial state proton,
%%%
\begin{align}
\ptcut \leq \sum_{k\in \text{jets}} \frac{z_k}{z} \,\omega e^{-\eta_k} \leq \frac{1-z}{z} \, \omega e^{-\etacut} \, .
\end{align}
%%%
The second inequality follows from all reconstructed $n$-collinear jets having
$\eta_k > \etacut$. This implies that \eq{DeltaI_limit} is trivially satisfied
since the domain of integration in $z$ scales as $x \leq z \lesssim \zetacut$.
Hence $\Delta \mathcal{I}_{ij}$ is parametrically important for $\zeta_{\rm cut}
\sim z \sim 1$, but vanishes in the threshold limit $z \to 1$. This leads to an
additional numerical suppression due to the falloff of the PDFs towards larger
partonic momentum fractions.

The RGE of $\Delta \cI_{ij}$ follows from the beam-function RGE \eq{RGE_B} and
the analogue of the matching onto the PDFs in \eq{B}. It is given by (with the
remaining arguments of $\Delta \cI_{ij}$ understood) %%%
\begin{align} \label{eq:rge delta I}
\mu \frac{\df}{\df \mu} \Delta \cI_{ij}(z, \mu, \nu)
&= \gamma^{i}_B(\omega, \mu, \nu) \, \Delta \cI_{ij}(z, \mu, \nu)
- \sum_k \Delta \cI_{ik}(z, \mu, \nu) \otimes_z 2P_{kj}[\as(\mu), z]
\,, \nn \\
\nu \frac{\df}{\df \nu} \Delta \cI_{ij}(z, \mu, \nu)
&= \gamma^{i}_{\nu,B}(\ptcut, R, \mu)\, \Delta \cI_{ij}(z, \mu, \nu)
\,.\end{align}
%%%
The Mellin convolution $\otimes_z$ is defined as
%%%
\begin{equation}
g(z) \otimes_z h(z) = \int_z^1 \! \frac{\df \xi}{\xi} \, g(\xi) \, h\Bigl( \frac{z}{\xi} \Bigr)
\,,\end{equation}
%%%
and $2P_{ij}(\as, z)$ is the standard PDF anomalous dimension with respect to $\mu$,
%%%
\begin{equation} \label{eq:DGLAP}
\mu \frac{\df}{\df \mu} f_i(x, \mu)
= \sum_j \int_x^1 \! \frac{\df z}{z} \, 2P_{ij}[\as(\mu), z]\, f_j\Bigl( \frac{x}{z}, \mu \Bigr)
\,.\end{equation}
%%%
Note that the RGE in \eq{rge delta I} does not mix $\Delta \cI_{ij}$ with $\cI_{ij}$
and therefore does not change the $\zeta_{\rm cut}$ scaling in \eq{DeltaI_limit}.
Solving \eq{rge delta I} order by order in perturbation theory, we find the following structure through two loops:
%%%
\begin{align} \label{eq:delta_I_master_formula}
\Delta \cI_{ij}(z)
&= \frac{\as(\mu)}{4\pi} \, \Delta \cI_{ij}^{(1)}(z) + \frac{\as^2(\mu)}{(4\pi)^2} \, \Delta \cI_{ij}^{(2)}(z) + \ord{\as^3}
\,,\nn \\
\Delta \cI_{ij}^{(1)}(z)
&= \Delta I_{ij}^{(1)} \Bigl( \frac{\omega e^{-\etacut}}{\ptcut}, z \Bigr)
\,,\nn \\
\Delta \cI_{ij}^{(2)}(z)
&= \ln \frac{\mu}{\ptcut} \Bigl[
   2\Gamma^{i}_0 \ln \frac{\nu}{\omega} + 2 \beta_0 + \gamma^{i}_{B\,0}
\Bigr] \,\Delta I_{ij}^{(1)} \Bigl( \frac{\omega e^{-\etacut}}{\ptcut}, z \Bigr)
\nn \\
&\quad -2 \ln \frac{\mu}{\ptcut}
   \sum_k \Delta I_{ik}^{(1)} \Bigl( \frac{\omega e^{-\etacut}}{\ptcut}, z \Bigr)
   \otimes_z P^{(0)}_{kj}(z)
+ \Delta I_{ij}^{(2)} \Bigl( \frac{\omega e^{-\etacut}}{\ptcut}, R, z \Bigr)
\,,\end{align}
%%%
where $\Delta I_{ij}^{(n)}$ is the boundary condition of the RGE at $\mu =
\ptcut$, $\nu = \omega$, and the required anomalous dimension coefficients are
collected in \app{anom_dims}. By dimensional analysis and boost invariance,
$\Delta I_{ij}^{(n)}$ can only depend on $\zetacut = \omega e^{-\etacut}/\ptcut$
in addition to $R$ and $z$.

In \app{etacut_beam_functions_one_loop} we determine the one-loop contribution
$\Delta I_{ij}^{(1)}$, which has the simple form
%%%
\begin{align}\label{eq:DeltaI}
\Delta I^{(1)}_{ij} \bigl(\zeta_{\rm cut}, z\bigr)= \theta\Bigl(\frac{\zeta_{\rm cut}}{1+ \zeta_{\rm cut}} -z\Bigr)  \,2P^{(0)}_{ij}(z) \,\ln \frac{\zeta_{\rm cut}(1-z)}{z}
\,,\end{align}
%%%
with the one-loop splitting functions $P^{(0)}_{ij}(z)$ as given in
\eq{DGLAP_coeff_lo}. The correction vanishes at the kinematic threshold encoded
in the overall $\theta$-function, which also cuts off the singular distributions
in $P^{(0)}_{ij}(z)$ at $z = 1$. The Mellin convolutions of $\Delta
I_{ik}^{(1)}\otimes_z P^{(0)}_{kj}$ appearing in the coefficient of $\ln
(\mu/\ptcut)$ in $\Delta \cI_{ij}^{(2)}(z)$ are given in
\app{two_loop_convolutions}.

While the computation of the full two-loop contribution $\Delta I_{ij}^{(2)}$
is beyond the scope of this paper,
we analytically compute its leading contribution in the small-$R$ limit, which contains
a clustering logarithm of $R$. We write the full two-loop result as
%%%
\begin{align}\label{eq:DeltaI2}
\Delta I^{(2)}_{ij} (\zetacut, R, z)
&= \ln R \, \Delta I^{(2, \ln R)}_{ij}(\zetacut,z) + \Delta I^{(2,c)}_{ij}(\zetacut,z) + \ord{R^2}
\,.\end{align}
%%%
In the limit $R \ll 1$, we exploit that for the emission of two close-by
collinear partons with relative rapidity $\Delta \eta \sim R$, the collinear
matrix element factorizes into two sequential collinear splittings at the scale
$\mu \sim \ptcut$ and $\mu \sim \ptcut R$, respectively. This allows us to
evaluate the coefficient of $\ln R$ in a generic two-loop beam function as a
convolution of a primary on-shell emission and (the anomalous dimension of) the
semi-inclusive jet function of \refcite{Kang:2016mcy}. Specifically, for the
case of $\Delta I^{(2)}_{ij}$ we find
%%%
\begin{align}\label{eq:DeltaI_lnR}
\Delta I^{(2, \ln R)}_{ij}(\zetacut,z)
&= \theta\Bigl(\frac{\zetacut}{1+ \zetacut} -z\Bigr) \, 2 P^{(0)}_{ij} (z)  \Bigl[\theta\Bigl(z-\frac{\zeta_{\rm cut}}{2+\zeta_{\rm cut}}\Bigr) c^{R,\rm cut}_{ij}\Bigl(\frac{z}{\zeta_{\rm cut}(1-z)}\Bigr) - c^{R}_{ij}\Bigr]
\,,\end{align}
%%%
where the coefficient functions $c^{R,\rm cut}_{ij}$ are given by
%%%
\begin{align}
c^{R,\rm cut}_{gg}(x) = c^{R,\rm cut}_{qq}(x)
&= -2 \int_{1/2}^{x} \! \frac{\df z}{z} \, \int_{1/2}^{z} \! \df z_J \Bigl[ P^{(0)}_{gg}(z_J) + 2n_f P^{(0)}_{qg}(z_J) \Bigr]
\,, \nn \\
c^{R,\rm cut}_{gq}(x) = c^{R,\rm cut}_{qg}(x)
&= -2 \int_{1/2}^{x} \! \frac{\df z}{z} \, \int_{1/2}^{z} \! \df z_J \Bigl[ P^{(0)}_{qq}(z_J) + P^{(0)}_{gq}(z_J) \Bigr]
\,,\end{align}
%%%
depending on whether the primary emission we split is a gluon (first line) or a quark (second line).
Their explicit expressions read
%%%
\begin{align} \label{eq:clustering_coefficients_etacut}
c^{R,\rm cut}_{gg}(x) = c^{R,\rm cut}_{qq}(x)
&= 2 C_A\Bigl[\frac{5}{8} +\frac{\pi^2}{3} - 3 x + \frac{9}{2} x^2 - 2 x^3- 2 \ln^2 x-4 \,\Li_2(x)  \Bigr] \nn \\
& \quad +2\beta_0\Bigl[-\frac{29}{24} -\ln 2 + 3 x - \frac{3}{2} x^2 + \frac{2}{3} x^3 -  \ln x\Bigr]
\,, \nn \\
c^{R,\rm cut}_{gq}(x) = c^{R,\rm cut}_{qg}(x) & = 2C_F\Bigl[-3+\frac{\pi^2}{3}-3\ln 2+ 6 x -3 \ln x- 2\ln^2 x -4\,\Li_2(x)\Bigr]
\,.\end{align}
%%%

The coefficients $c^R_{ij}$ in \eq{DeltaI_lnR} are the (in principle known)
coefficients of $\ln R$ in the $\etacut$-independent two-loop beam
function~\cite{Stewart:2013faa, Li:2014ria}, which we also verified.%
\footnote{The coefficient of the $c^R_{gq}$ contribution in eq.~(39) of
\refcite{Stewart:2013faa} has a typo, missing an overall factor of 2.
We also find that the $C_A$ term of the coefficient $c^R_{qq}$ in eq.~(9) of
\refcite{Li:2014ria} misses a factor of $1/2$ compared to \refcite{Stewart:2013faa}
and our result.}
They satisfy
%%%
\begin{equation}
c^R_{ij} = \lim_{x\to 1} c^{R,\rm cut}_{ij}(x)
\,,\end{equation}
%%%
and are given by
%%%
\begin{align} \label{eq:clustering_coefficients_no_etacut}
c^R_{gg} & = c^R_{qq} = \frac{1}{4} \Bigl[ \Bigl(1-\frac{8\pi^2}{3}\Bigr)C_A +\Bigl(\frac{23}{3}-8 \ln 2\Bigr) \beta_0 \Bigr]
\,, \nn \\
c^R_{qg} & = c^R_{gq} = 2 C_F \Bigl(3-\frac{\pi^2}{3}-3 \ln 2\Bigr)
\,.\end{align}
%%%
Our general setup for computing the small-$R$ clustering contributions implies that
the coefficient of the $\ln R$ terms of the two-loop rapidity anomalous dimension must be equal to $c^R_{gg} = c^R_{qq}$,
in agreement with the corresponding result given in \refscite{Tackmann:2012bt, Stewart:2013faa}.
In addition, it also applies to the leading $\ln^2 R$ and $\ln R$ terms in the beam functions for
rapidity dependent jet vetoes in \refcite{Gangal:2016kuo}, with which we agree as well.

The $R$-independent term $\Delta I^{(2,c)}_{ik}(\zeta_{\rm cut}, z)$ and the
$\ord{R^2}$ terms in \eq{DeltaI2} are currently unknown. Their contribution to
the cross section can in principle be obtained numerically from the singular
limit of the full-theory calculation at $\ord{\as^2}$, as was done for the
corresponding $\etacut$-independent pieces in \refcite{Stewart:2013faa}.

\paragraph{Numerical validation.}

\begin{figure*}
\centering
\includegraphics[width=\WidthTwoSubfigs]{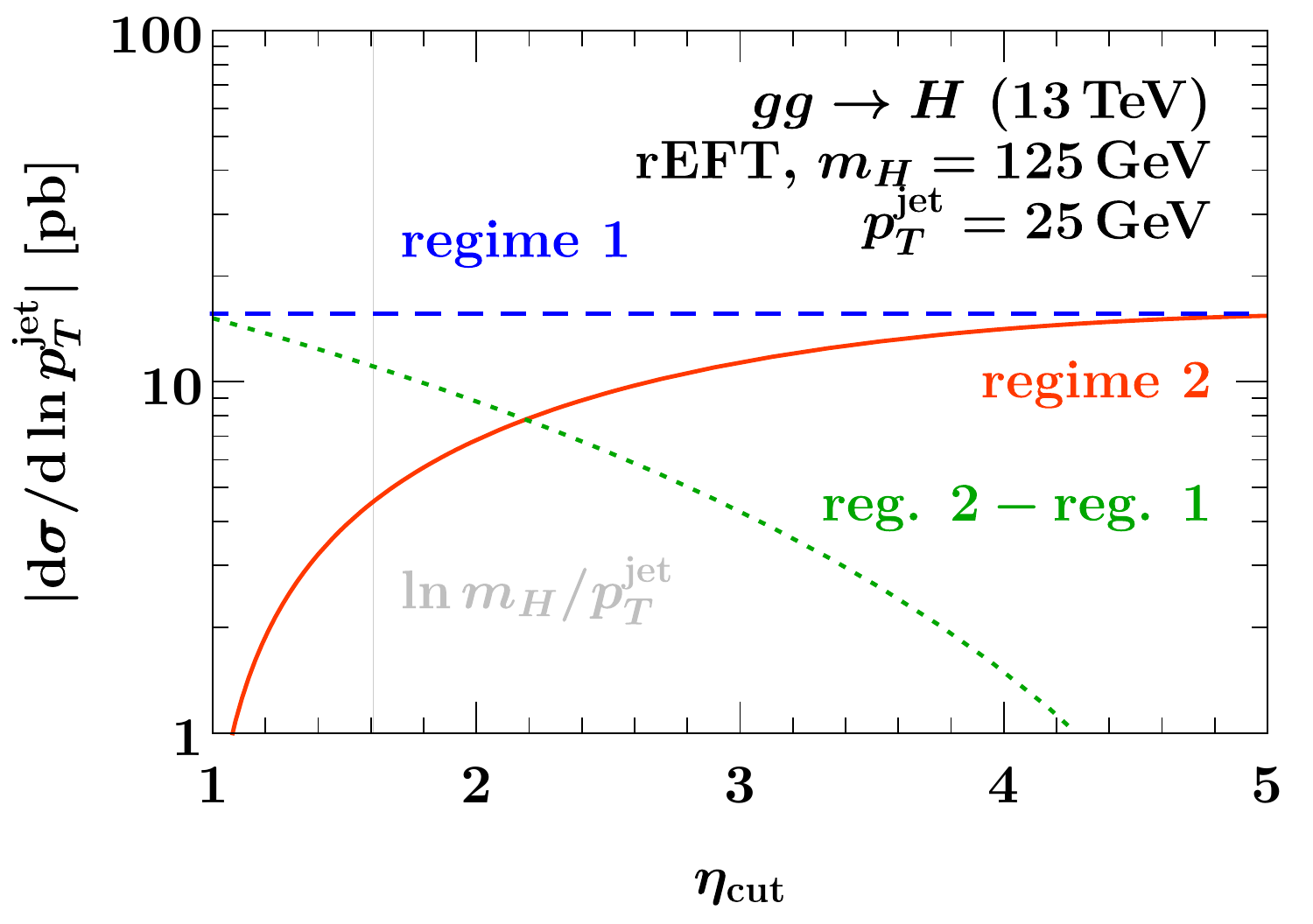}%
\hfill%
\includegraphics[width=\WidthTwoSubfigs]{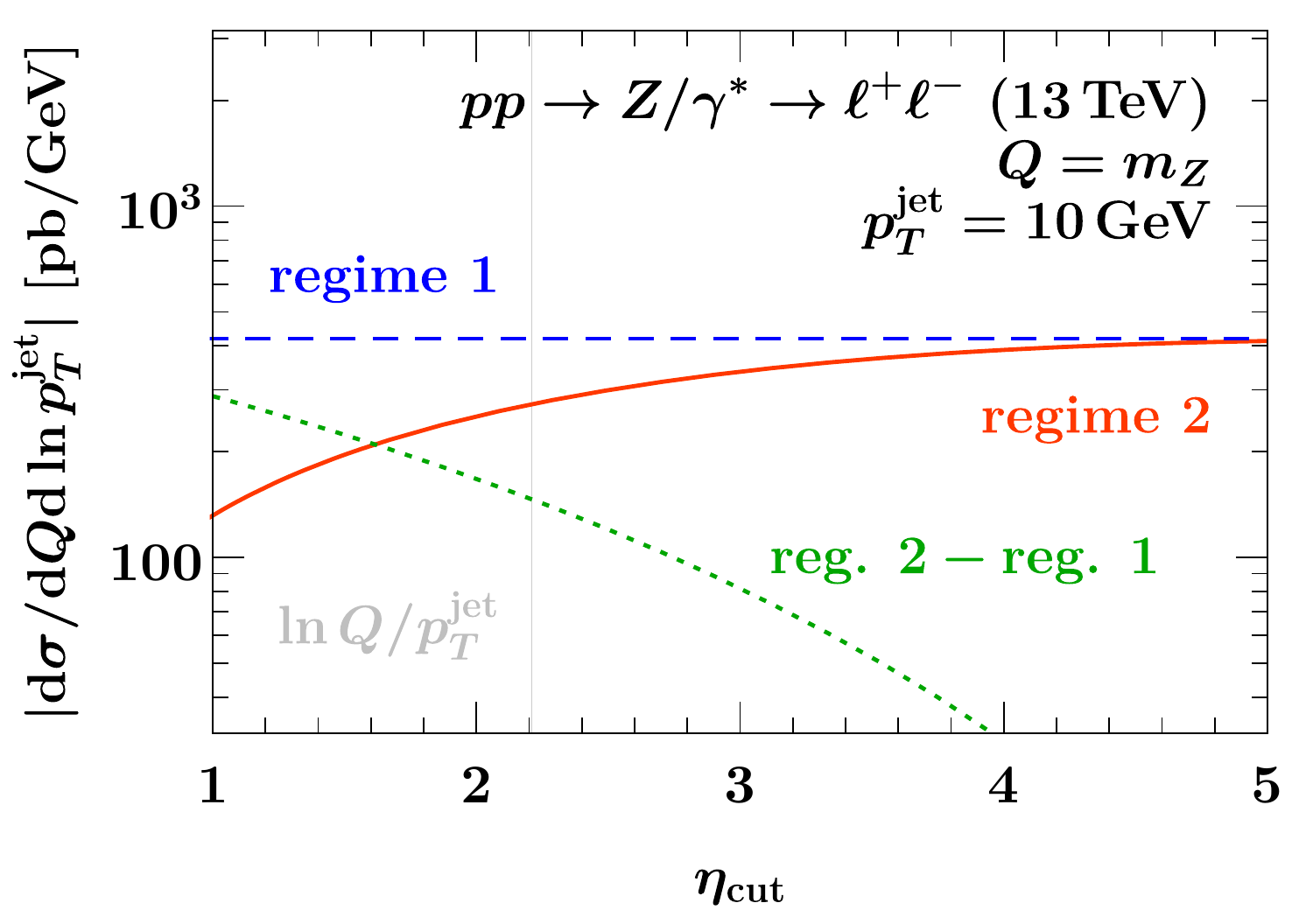}%
\caption{Comparison of the singular contributions to the fixed $\ord{\alpha_s}$
(LO$_1$) $\ptjet$ spectrum for $gg\to H$ (left) and Drell-Yan (right).
The orange solid lines show the singular contributions in regime~2 with $\etacut$ dependent
beam functions. The dashed blue lines show the singular contributions in regime~1
in the limit $\etacut = \infty$, $\ptcut \gg Q e ^{-\etacut}$. Their difference, shown
by the dotted green lines, correctly scales as a power in $Q e^{-\etacut} / \ptjet$.
The vertical lines indicate the point $\ptjet = Q e^{-\etacut}$.}
\label{fig:regimes_1_2}
\end{figure*}

\begin{figure*}
\centering
\includegraphics[width=\WidthTwoSubfigs]{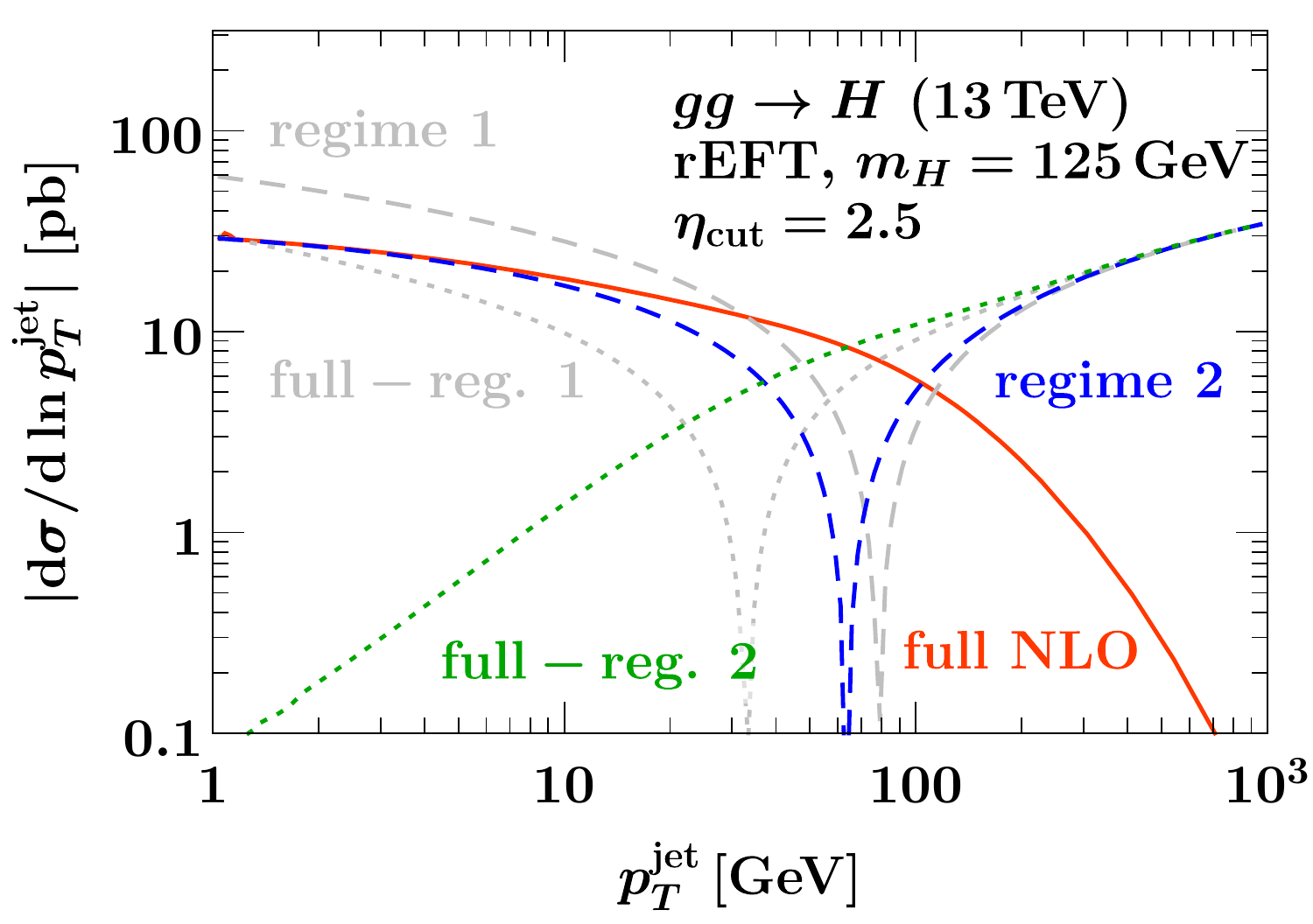}%
\hfill%
\includegraphics[width=\WidthTwoSubfigs]{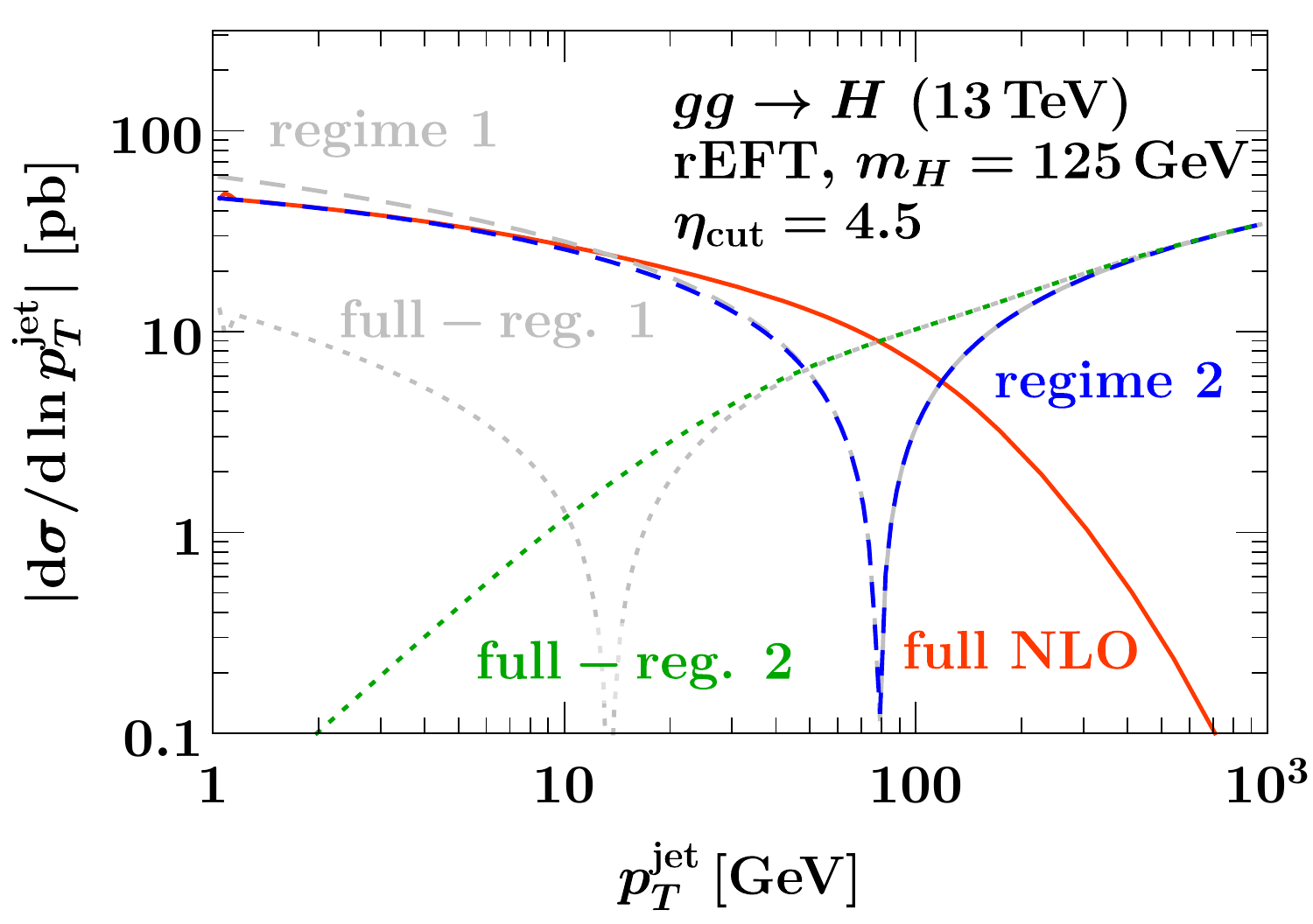}%
\\
\includegraphics[width=\WidthTwoSubfigs]{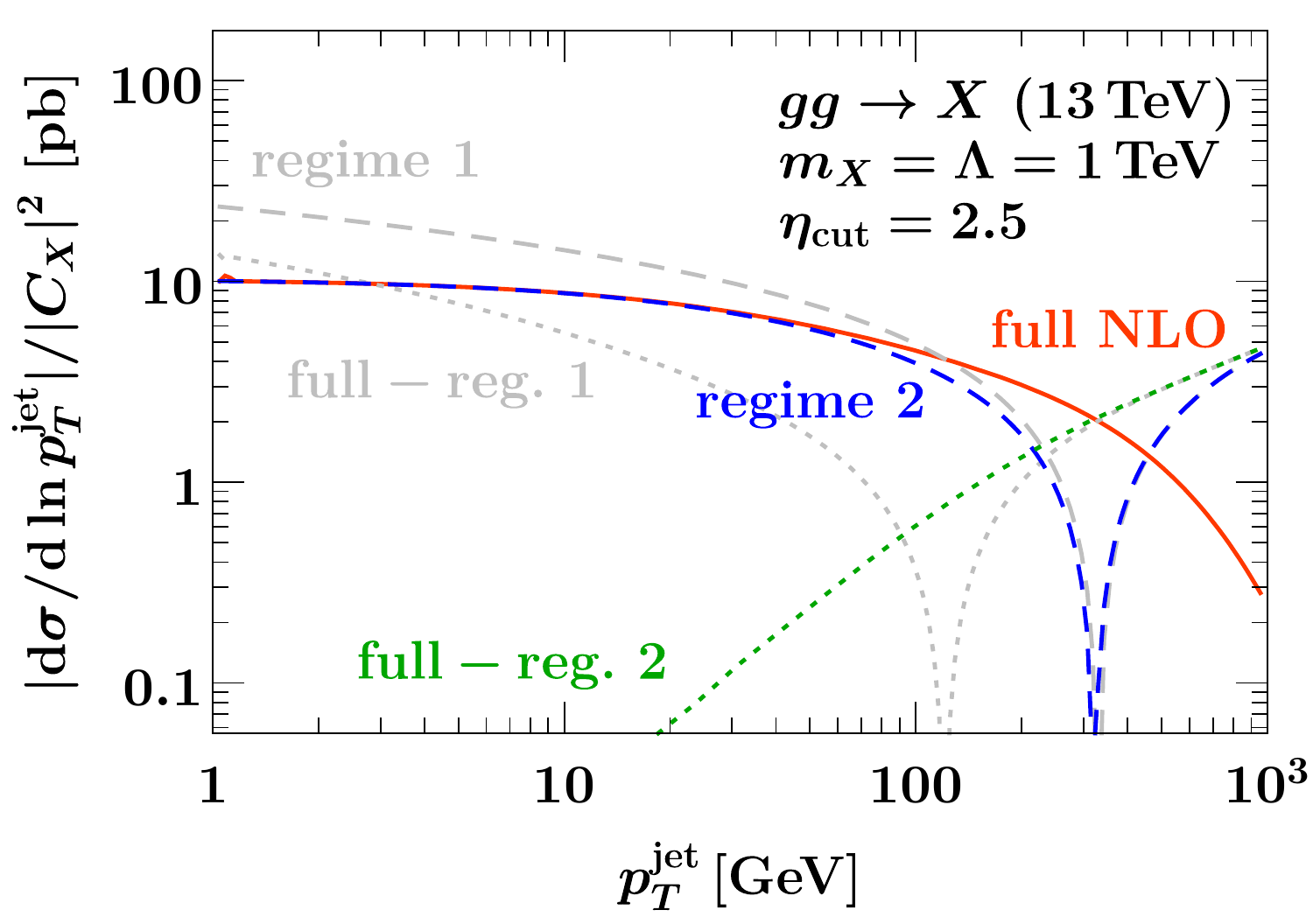}%
\hfill%
\includegraphics[width=\WidthTwoSubfigs]{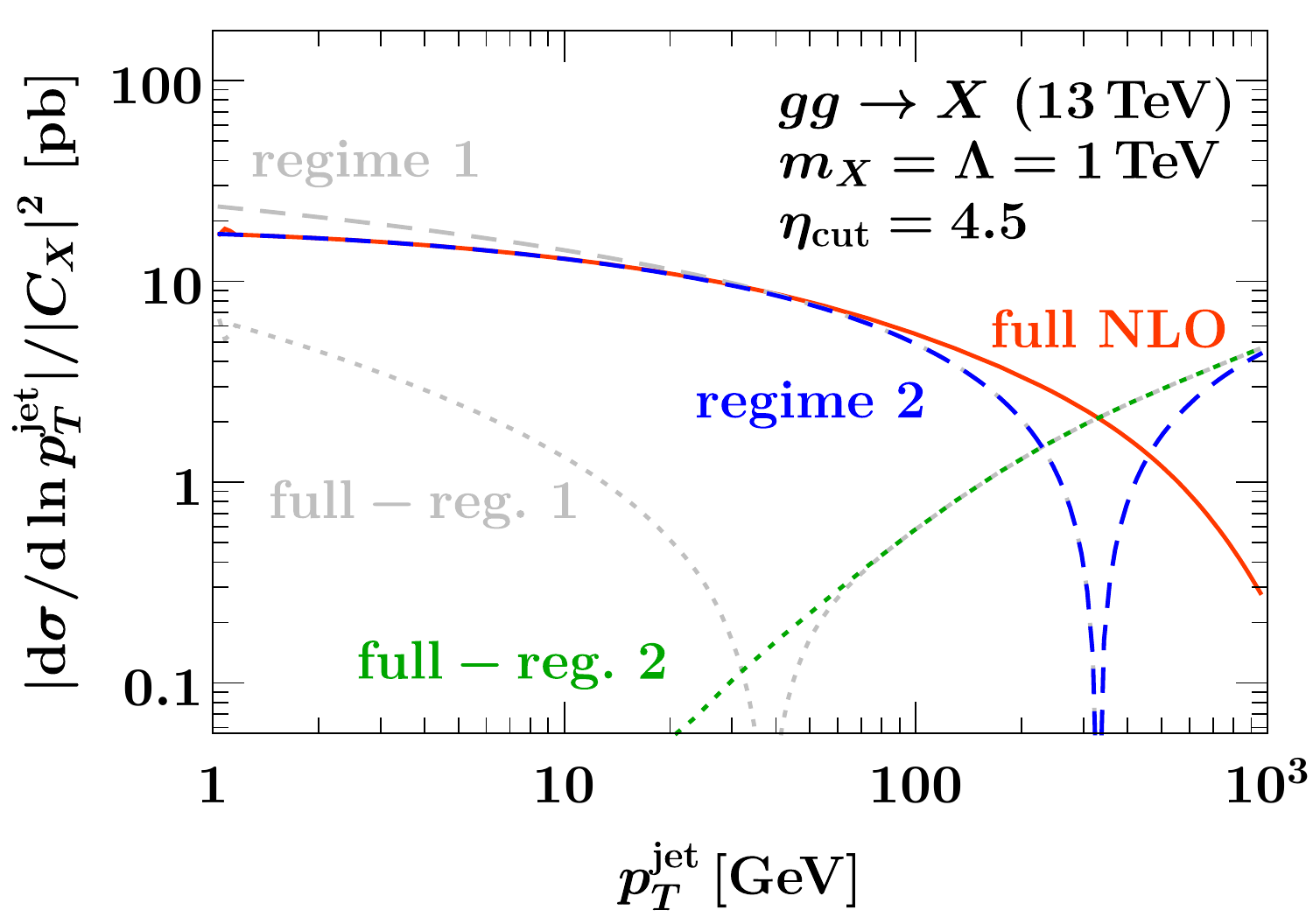}%
\caption{Comparison of singular and nonsingular contributions to the fixed
$\ord{\alpha_s}$ (LO$_1$) $\ptjet$ spectrum with rapidity cut
$\abs{\etajet} < \etacut$ for $gg\to H$ (top row) and $gg\to X$ (bottom row),
$\etacut = 2.5$  (left) and $\etacut = 4.5$ (right). The orange solid lines show the full
results, the dashed blue lines the regime~2 results with $\etacut$ dependent
beam functions, and the dotted green lines their difference. The dashed and
dotted gray lines show the corresponding regime~1 results, which do not describe
the singular behavior of the full cross section for finite $\etacut$.}
\label{fig:sing_nons_ggHX}
\end{figure*}

\begin{figure*}
\centering
\includegraphics[width=\WidthTwoSubfigs]{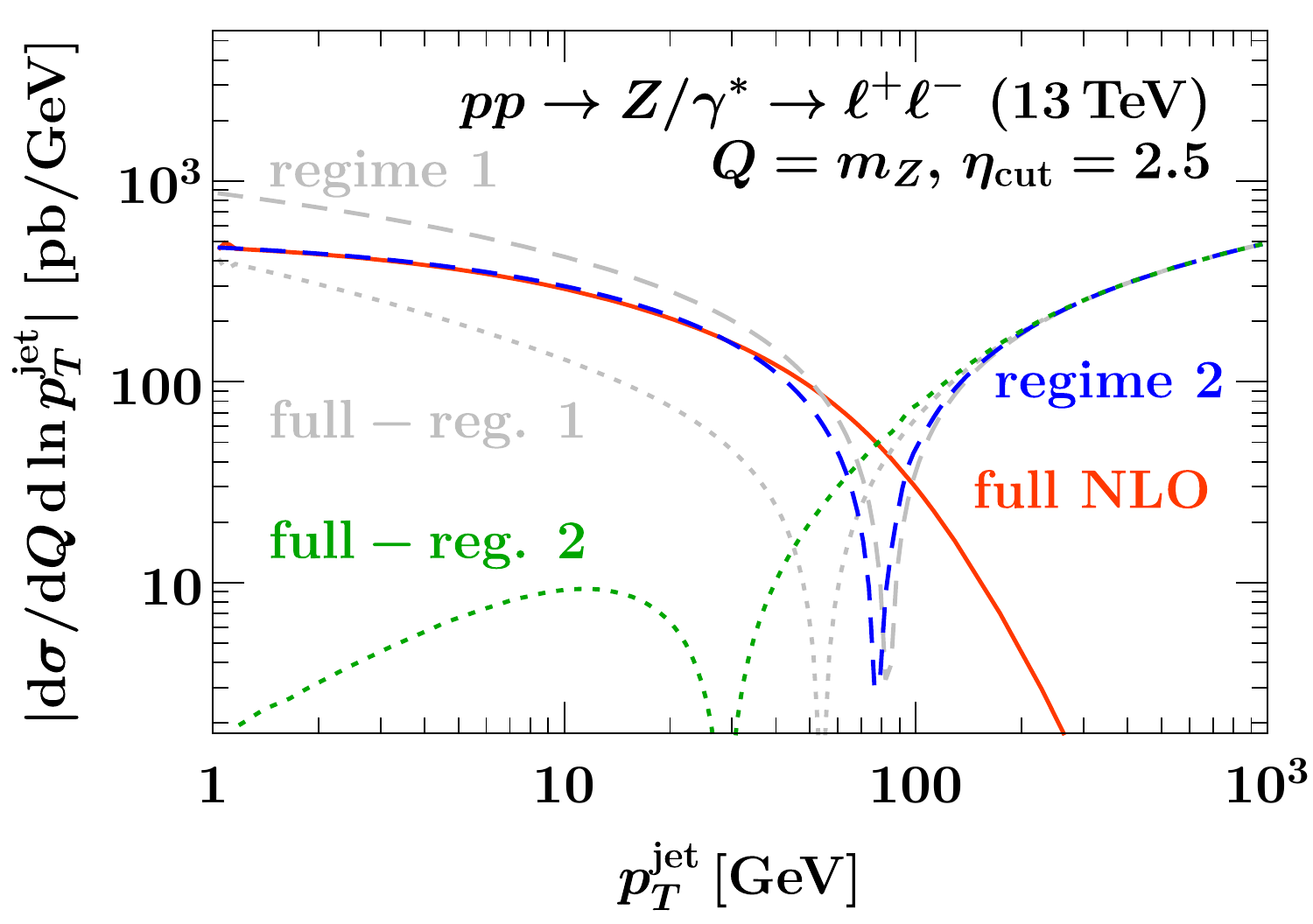}%
\hfill%
\includegraphics[width=\WidthTwoSubfigs]{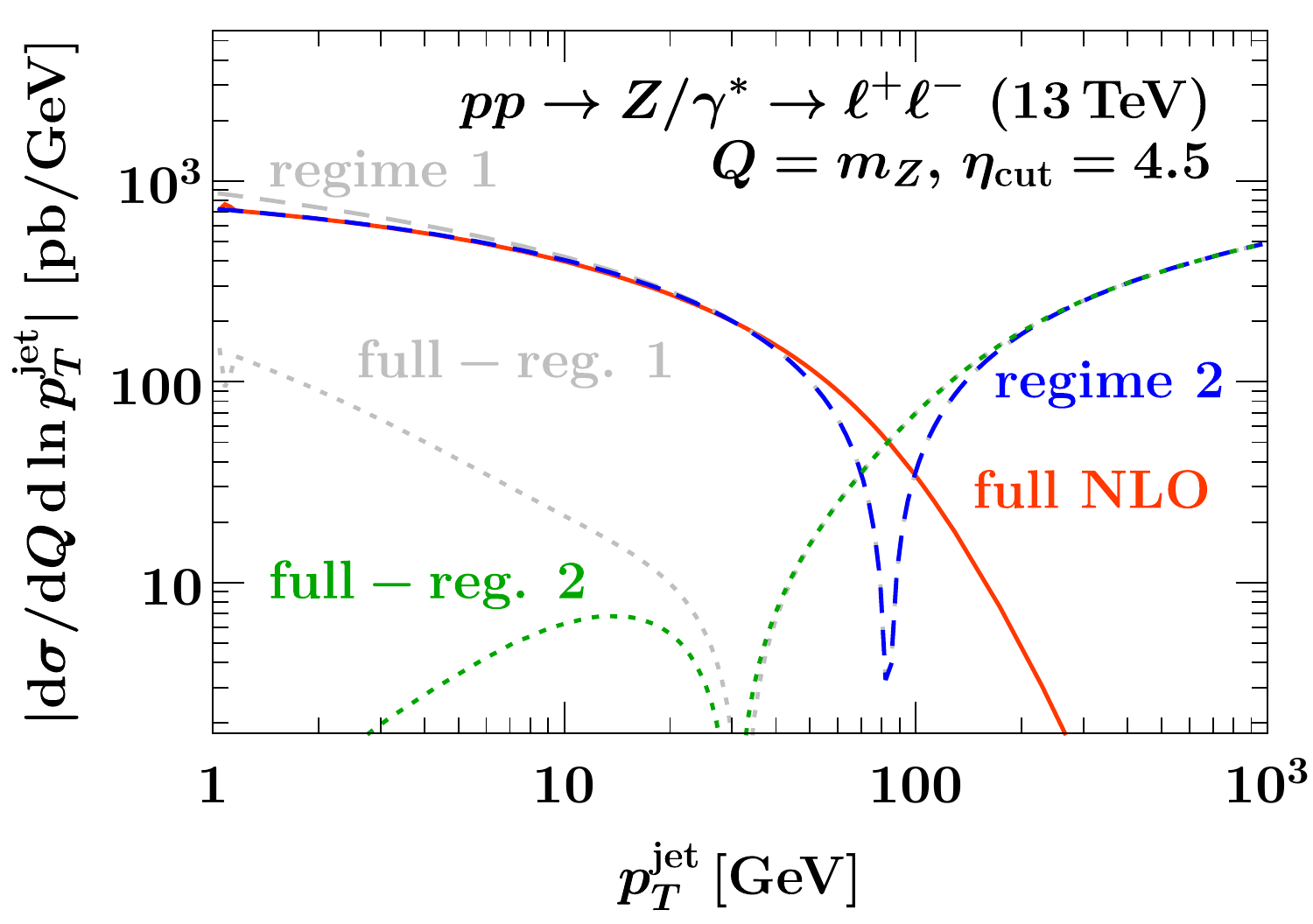}%
\\
\includegraphics[width=\WidthTwoSubfigs]{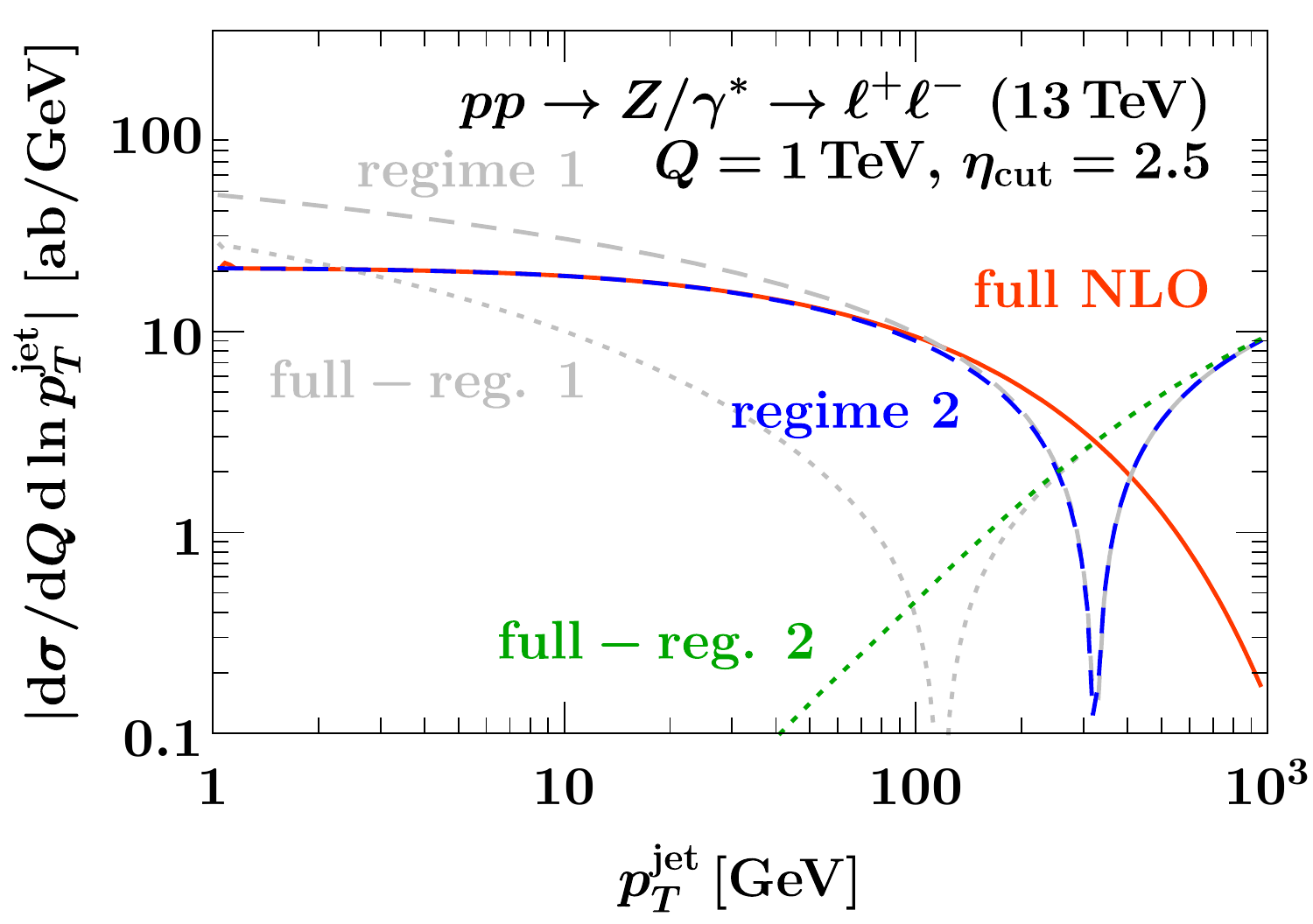}%
\hfill%
\includegraphics[width=\WidthTwoSubfigs]{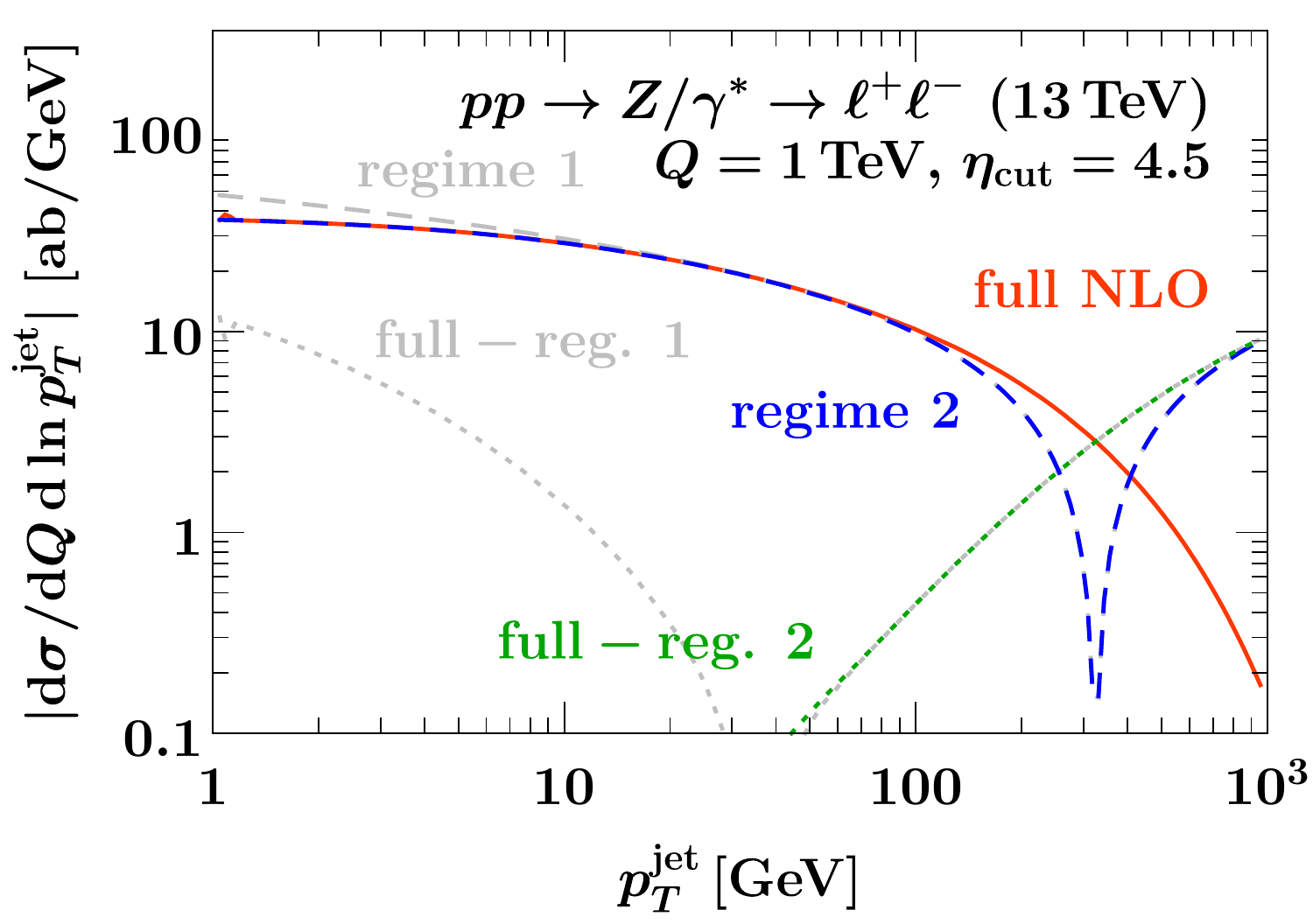}%
\caption{Comparison of singular and nonsingular contributions to the fixed
$\ord{\alpha_s}$ (LO$_1$) $\ptjet$ spectrum with rapidity cut
$\abs{\etajet} < \etacut$ for Drell-Yan at $Q = m_Z$ (top row) and
$Q = 1\TeV$ (bottom row), $\etacut = 2.5$ (left) and $\etacut = 4.5$ (right).
The meaning of the curves are as in \fig{sing_nons_ggHX}.}
\label{fig:sing_nons_DY_narrow_width}
\end{figure*}

To validate our results numerically and highlight the differences in the singular
behavior for regimes 1 and 2, we consider the fixed $\ord{\alpha_s}$ $\ptjet$ spectrum,
$\df\sigma/\df \ptjet$,
where $\ptjet$ is the transverse momentum of the leading jet
within $\abs{\etajet} < \etacut$.
Its relation to the jet veto cross section with a jet rapidity cut is simply
%%%
\begin{equation}
\sigma_0(\ptcut, \etacut, R) = \int_0^{\ptcut} \! \df \ptjet \, \frac{\df \sigma(\etacut, R)}{\df \ptjet}
\,.\end{equation}
%%%
At leading power in $\ptjet/Q$, we obtain it by taking the derivative with respect to $\ptcut$ of either \eq{fact_pTjet2},
retaining the exact dependence on $\etacut$ in the beam functions (regime~2),
or of \eq{fact_pTjet}, incurring power corrections in $Q e^{-\etacut}/\ptjet$ (regime~1).
The numerical results for all singular spectra are obtained with the help
of \texttt{SCETlib}~\cite{scetlib}. The $\ord{\alpha_s}$ spectra in full QCD are obtained
from \texttt{MCFM~8.0}~\cite{Campbell:1999ah,Campbell:2011bn,Campbell:2015qma}.

As representative gluon-induced processes, we consider gluon-fusion Higgs
production $gg\to H$ at $m_H = 125\GeV$ in the infinite top-mass limit,
rescaled with the exact LO top-mass dependence for $m_t = 172.5 \GeV$ (rEFT). In addition, we consider
gluon fusion to a hypothetical heavy color-singlet scalar $X$, $gg\to X$,
mediated by the contact operator
%%%
\begin{equation} \label{eq:L_eff_ggX}
\mathcal{L}_\text{eff} = -\frac{C_X}{\Lambda}\,\as\, G_{\mu\nu}^a G^{a,\mu\nu} X
\,.\end{equation}
%%%
We always choose $m_X = 1\TeV$, $\Lambda = 1 \TeV$, and divide the cross section by
$\abs{C_X}^2$. To the order we are working, this is equivalent to setting
$C_X \equiv 1$, since $C_X$ only starts to run at $\ord{\as^2}$.%
\footnote{%
In \texttt{MCFM 8.0} we mock up this process using a standard-model Higgs with $m_H=1 \TeV$
and manually account for the nonzero one-loop contribution from integrating out
the top quark in the SM, which differs from our choice
of $C_X = 1 + \ord{\as^2}$ for the effective coupling of $X$ to gluons.
We also checked the results against the native $gg \to X$ support
of \texttt{SusHi 1.6.1}~\cite{Harlander:2012pb, Harlander:2016hcx, Harlander:2002wh, Harlander:2005rq}.
}
For quark-induced processes we consider
Drell-Yan $pp \to Z/\gamma^\ast \to \ell^+ \ell^-$ at the $Z$ pole ($Q = m_Z$) and
at $Q = 1\TeV$, where $Q = m_{\ell\ell} $ is the invariant mass of the lepton pair.
Here we set all scales to $\mufo = m_H$, $m_X$, or $Q$, respectively.
We use \texttt{PDF4LHC\_nnlo\_100} \cite{Butterworth:2015oua, Dulat:2015mca,
Harland-Lang:2014zoa, Ball:2014uwa, Gao:2013bia, Carrazza:2015aoa}
NNLO PDFs with $\as(m_Z)= 0.118$ throughout.

In \fig{regimes_1_2}, we compare the regime~2 and regime~1 leading-power (singular) results
for $\df \sigma / \df \ptjet$ at fixed $\ptjet$ as a function of $\etacut$
for $gg\to H$ and Drell-Yan.
The regime~1 result (dashed blue) does not depend on $\etacut$,
while the regime~2 result (solid orange) decreases as $\etacut$ becomes more central.
The difference between the two (dotted green) has the expected behavior, vanishing
as $Q e^{-\etacut}/\ptjet$ for $\etacut \to \infty$. We observe that
regime~1 is applicable beyond $\etacut \gtrsim 4$, where the difference to
regime~2 is suppressed by an order of magnitude.

Another check is provided by comparing the regime~1 and regime~2 singular results
to the full QCD result, which is shown in \figs{sing_nons_ggHX}{sing_nons_DY_narrow_width}
for gluon-fusion and Drell-Yan.
For $\etacut = 2.5$ (left panels), it is clear
that regime~1 (dashed gray) fails to describe the singular limit of full QCD,
with their difference (dotted gray) diverging
for $\ptjet \to 0$ like an inverse power of $\ptjet$ as expected.
While the singular mismatch becomes less pronounced for $\etacut = 4.5$ (right panels), the
uncanceled singular contributions are still clearly visible in the difference.
On the other hand, regime~2 (dashed blue) correctly reproduces the singular limit $\ptjet \to 0$,
with the difference (dotted green) vanishing like a power of $\ptjet$ as it must.
This provides a strong check of the intricate $\ptcut$ dependence
encoded in our $\ord{\as}$ results for $\Delta \mathcal{I}_{ij}$.
(The power corrections in $e^{-\etacut}$, which are present in regime~2,
drop out when taking the derivative of the fixed-order cumulant with respect to $\ptcut$.)

Note that at $m_X = 1\TeV$ or $Q = 1\TeV$, the fixed-order spectrum is completely dominated
by the rapidity-cut dependent singular result up to $\ptjet \lesssim 100 \GeV$.
Hence, the resummation should provide a significant improvement over the fixed-order
result for typical $\ptcut \sim 50 \GeV$, which we will indeed find in \sec{analysis}.

%===============================================================================
\subsection[Regime 3: \texorpdfstring{$\ptcut/Q \ll e^{-\etacut}$}{pTcut/Q << exp(-etacut)} (collinear NGLs)]
{\boldmath Regime 3: $\ptcut/Q \ll e^{-\etacut}$ (collinear NGLs)}
\label{sec:no_step_reg3}
%===============================================================================

The hierarchy $\ptcut \ll Q e^{-\etacut} $ (with $e^{-\etacut} \ll 1$) exhibits
different features than the regimes discussed before. The typical transverse
momentum for emissions with $|\eta|> \etacut$ is parametrically $Q e^{-|\eta|}$,
indicated by the horizontal gray line in \fig{sketch_regimes_no_step}, which is
now much larger than for the strongly constrained emissions at $\abs{\eta} <
\etacut$. While the soft modes at central rapidities are not affected, there are
now two types of collinear modes at forward rapidities with $\abs{\eta} \sim
\etacut$,
%%%
\begin{alignat}{3} \label{eq:modes_no_step_reg3}
n_a\text{-collinear:}
&\quad
p^\mu &&\sim Q \Bigl(e^{-2\etacut},1, e^{-\etacut}\Bigr)
\,, \nn \\
n_a\text{-soft-collinear:}
&\quad
p^\mu&&\sim \Bigl(\ptcut e^{-\etacut},\ptcut e^{\etacut},\ptcut\Bigr)
= \ptcut e^{\etacut} \Bigl(e^{-2\etacut},1,e^{-\etacut}\Bigr)
\,,\end{alignat}
%%%
and analogously for the $n_b$-collinear sector.

The collinear and soft-collinear modes have the same angular resolution and only
differ in their energy. This makes their all-order factorization challenging and
leads to the appearance of nonglobal logarithms $\ln(Q e^{-\etacut}/\ptcut)$
starting at $\mathcal{O}(\alpha_s^2)$. Their factorization and resummation
requires the marginalization over all possible configurations of energetic
collinear emissions, involving soft-collinear matrix elements with a separate
Wilson line along each individual energetic collinear emission, see e.g.\
\refscite{Hatta:2013iba,Caron-Huot:2015bja,Larkoski:2015zka,Becher:2016mmh}.

Since this regime has no immediate phenomenological relevance, we will not carry
out this complete procedure but restrict ourselves to the configuration with
soft-collinear Wilson lines along $n$ and $\bar{n}$, i.e, along the two main
collinear emitters. This is sufficient for the LL resummation, for isolating the
nonglobal effects, and for discussing the relation to the other regimes. Our
discussion here is in close analogy to the regime~3 in the factorization of the
exclusive jet mass spectrum with small jet radius $R$ in
\refcite{Kolodrubetz:2016dzb}, where the rapidity cut $e^{-\etacut}$ here takes
the role of $R$ there.%
\footnote{%
The main difference is that here, emissions for $|\eta|<\etacut$ are constrained
by their $p_T$ relative to the same collinear (beam) direction. In the jet mass
case, emissions outside the jet are not constrained by their $p_T$ relative to
the same collinear (jet) direction (but also relative to the beam direction).}

The factorized cross section takes the form
%%%
\begin{align} \label{eq:fact_pTjet3}
\sigma_0 (\ptcut, \etacut, R, \Phi)
&=
H_\kappa(\Phi, \mu) \,
\mathcal{B}_{a}(\ptcut, \etacut, R, \omega_a, \mu, \nu ) \,
\mathcal{B}_{b}(\ptcut, \etacut, R, \omega_b, \mu, \nu ) \nn \\
& \quad \times  S_{\kappa} (\ptcut, R, \mu,\nu) \,
\biggl[1+\mathcal{O}\Bigl(\frac{\ptcut}{Qe^{-\etacut}}, e^{-\etacut},R^2\Bigr)\biggr]
\,.\end{align}
%%%
The initial-state collinear functions $\mathcal{B}_i$ encode the contributions
of both soft-collinear and energetic collinear modes. They are related to the
$\etacut$ dependent beam functions $B_i$ in \eq{fact_pTjet2} by an expansion in
the limit $\ptcut/(\omega e^{-\etacut}) \ll 1$,
%%%
\begin{align} \label{eq:B_i}
B_i(\ptcut, \etacut, R, \omega, \mu, \nu ) =
\mathcal{B}_{i}(\ptcut, \etacut, R, \omega, \mu, \nu)\biggl[1+\mathcal{O}\Bigl(\frac{\ptcut}{\omega e^{-\etacut}}\Bigr)\biggr]
\,.\end{align}
%%%
Without further factorization, $\mathcal{B}_i$ contains large unresummed Sudakov
double logarithms $\alpha_s^n \ln^{2n}(\ptcut/\omega e^{-\etacut})$. To resum
the leading double logarithms, we can decompose $\mathcal{B}_i$ as
%%%
\begin{align} \label{eq:refactB}
\mathcal{B}_{i}(\ptcut, \etacut, R, \omega, \mu, \nu )
&= B^{(\rm cut)}_{i}(\etacut, \omega, \mu) \,\mathcal{S}^{(\rm cut)}_{i}(\ptcut, \etacut, R, \mu, \nu)
\nn \\
& \quad \times  \biggl[1+ \mathcal{B}^{\rm (NG)}_{i}\Bigl(\frac{\ptcut}{ \omega e^{-\etacut}}, \omega, R\Bigr)\biggr]
\,.\end{align}
%%%
The function $B^{(\rm cut)}_{i}$ mainly describes contributions from the
energetic collinear modes. It was dubbed ``unmeasured" beam function in
\refscite{Hornig:2016ahz, Hornig:2017pud}, in analogy to the unmeasured jet
function~\cite{Ellis:2010rwa}. At one loop its matching coefficients account for
an energetic collinear emission with $|\eta|>\etacut$. They are calculated in
\app{etacut_beam_functions_one_loop} and read
%%%
\begin{align} \label{eq:I_cut}
\mathcal{I}^{(\rm cut)}_{gg}(\etacut, \omega, z, \mu)
&= \delta(1-z) + \frac{\alpha_s(\mu)\, C_A}{4\pi} \biggl[ \delta(1-z) \biggl(4 \ln^2 \frac{\omega e^{-\etacut}}{\mu} - \frac{\pi^2}{6} \biggr)
\nn \\ & \quad
+4P_{gg}(z) \ln \frac{\omega e^{-\etacut}}{\mu\, z}
+ 8 \mathcal{L}_1(1-z)
+ 8\Bigl(\frac{1}{z} - 2 + z - z^2\Bigr)\ln(1-z)\biggr]
\nn \\ & \quad
+ \ord{\as^2}
\,, \nn \\
\mathcal{I}^{(\rm cut)}_{gq}(\etacut, \omega, z, \mu)
&= \frac{\alpha_s(\mu)\, C_F}{4\pi} \biggl[ 4P_{gq}(z) \ln \frac{\omega e^{-\etacut} (1-z)}{\mu\, z} +2z \biggr]
+ \ord{\as^2}
\,, \nn \\
\mathcal{I}^{(\rm cut)}_{qq}(\etacut, \omega, z, \mu)
&= \delta(1-z) + \frac{\alpha_s(\mu)\, C_F}{4\pi} \biggl[ \delta(1-z) \biggl( 4 \ln^2 \frac{\omega e^{-\etacut}}{\mu} -6 \ln \frac{\omega e^{-\etacut}}{\mu} -\frac{\pi^2}{6} \biggr)
\nn\\ &\quad
+4P_{qq}(z) \ln \frac{\omega e^{-\etacut}}{\mu\, z} +8 \mathcal{L}_1(1-z)-4(1+z)\ln(1-z)+2(1-z)\biggr]
\nn \\ & \quad
+ \ord{\as^2}
\,, \nn \\
\mathcal{I}^{(\rm cut)}_{qg}(\etacut, \omega, z, \mu)
&=  \frac{\alpha_s(\mu)\, T_F}{4\pi} \biggl[ 4P_{qg}(z) \ln \frac{\omega e^{-\etacut} (1-z)}{\mu\, z} +4z(1-z) \biggr]
+ \ord{\as^2}
\,,\end{align}
%%%
where $\mathcal{L}_n(1-z) \equiv [\ln^n(1-z)/(1-z)]_+$,
$P_{ij}(z)$ are the color-stripped LO splitting functions given in \eq{p_ij},
and the flavor structure is trivial,
%%%
\begin{equation}
\mathcal{I}^{(\rm cut)}_{\bar{q}_i\bar{q}_j} = \mathcal{I}^{(\rm cut)}_{q_iq_j} = \delta_{ij} \mathcal{I}^{(\rm cut)}_{qq} + \ord{\as^2}
\,, \qquad
\mathcal{I}^{(\rm cut)}_{q_i\bar{q}_j} = \mathcal{I}^{(\rm cut)}_{\bar{q}_iq_j} = \ord{\as^2}
\,.\end{equation}
%%%
As argued in \refcite{Hornig:2016ahz} the results are directly related to the
matching coefficients for fragmenting jet functions in \refcite{Procura:2011aq}.

The function $\mathcal{S}^{(\rm cut)}_{i}$ in \eq{refactB} mainly describes
contributions from soft-collinear modes. At one loop it accounts for a
soft-collinear emission that couples eikonally to the incoming collinear parton
$i$. The emission is constrained to $p_T < \ptcut$ for $|\eta|<\etacut$ by the
jet veto, and is unconstrained for $|\eta| > \etacut$. Using the $\eta$
regulator~\cite{Chiu:2011qc, Chiu:2012ir} it is given by (see \app{soft_coll})
%%%
\begin{align} \label{eq:S_cut}
\mathcal{S}^{(\rm cut)}_{i}(\ptcut, \etacut, R, \mu, \nu)
&=1 + \frac{\alpha_s(\mu)}{4\pi}\,\mathcal{S}^{(\rm cut,1)}_{i} + \frac{\alpha_s^2(\mu)}{(4\pi)^2}\,\mathcal{S}^{(\rm cut,2)} + \ord{\as^3}
\,, \nn \\
\mathcal{S}^{(\rm cut,1)}_{i}(\ptcut, \etacut, R, \mu, \nu)
&= C_i \biggl( 4\ln^2\frac{\ptcut}{\mu} -8 \ln \frac{\ptcut}{\mu} \ln \frac{\nu e^{-\etacut}}{\mu} + \frac{\pi^2}{6} \biggr)
\,,\end{align}
%%%
where $C_i = C_F$ for an incoming quark or antiquark and $C_A$ for an incoming gluon.
We checked explicitly that the above results obey the consistency constraint in \eq{B_i}.
For this purpose, one has to note that \eq{DeltaI} becomes distribution valued in $(1-z)$
when taking the limit $\zeta_{\rm cut} \gg 1$.

At two loops $\mathcal{S}^{(\rm cut)}_i$ contains a $\ln R$ enhanced term.
Focusing on the constant terms not predicted by the RG evolution, we have
%%%
\begin{align} \label{eq:S_cut_ln_R}
\mathcal{S}^{(\rm cut,2)}_{i}(\ptcut, \etacut, R, \mu = \ptcut, \nu = \mu e^{\etacut})
= \ln R \,\mathcal{S}^{({\rm cut},2,\ln R)}_{i}
+ \mathcal{S}^{({\rm cut},2,c)}_{i} + \ord{R^2}
\,,\end{align}
%%%
with $\mathcal{S}^{({\rm cut},2,c)}_{i}$ an unknown two-loop constant.
The coefficient of $\ln R$ is obtained by expanding
the $\ln R$ coefficient in the $\etacut$ dependent beam function [see \eqs{DeltaI_lnR}{I2_lnR}]
to leading power in $1/\zetacut$.
In the limit $\zetacut \gg 1$, the sum $I^{(2, \ln R)}_{ij} + \Delta I^{(2, \ln R)}_{ij}$ becomes proportional to $\delta(1-z)$,
as the arguments of both $\theta$-functions in \eq{DeltaI_lnR} approach $z = 1$.
The coefficient of $\delta(1-z)$ is then given by the $\zetacut \to \infty$ limit
of the integral of $\Delta I^{(2, \ln R)}_{ij}$,
which vanishes for $i \neq j$ and for $i = j$ leaves
%%%
\begin{align} \label{eq:S_cut_ln_R_result}
&\mathcal{S}^{({\rm cut},2,\ln R)}_{i}
=8C_i \int_{1/2}^1 \! \frac{\df x}{x} c^{R,\rm cut}_{ii}(x)
\\
&= C_i \biggl\{
   C_A\Bigl[\frac{1622}{27} - \frac{548}{9} \ln 2 - \frac{88}{3} \ln^2 2- 8 \zeta_3\Bigr]
   + n_f T_F \Bigl[-\frac{652}{27} + \frac{232}{9} \ln 2 + \frac{32}{3} \ln^2 2 \Bigr]
\biggr\}
\,. \nn\end{align}
%%%

The anomalous dimensions of $B^{(\rm cut)}_{i}$ and $\mathcal{S}^{(\rm cut)}_{i}$
have the general structure
%%%
\begin{align}\label{eq:anom_dim_cut}
\gamma^i_{\mathcal{S}^{\rm cut}}(\etacut,\mu,\nu)
&=2 \Gamma^i_{\rm cusp}[\as(\mu)] \, \ln \frac{\nu  e^{-\etacut}}{\mu}
+ \gamma^i_{\mathcal{S}^{\rm cut}}[\as(\mu)]
\,, \nn \\
\gamma^i_{\nu,\mathcal{S}^{\rm cut}}(\ptcut,R,\mu)
&=2 \eta^i_\Gamma(\ptcut, \mu)
+ \gamma^i_{\nu,\mathcal{S}^{\rm cut}}[\as(\ptcut),R]
\,, \nn \\
\gamma^i_{B^{\rm cut}}\bigl(\omega e^{-\etacut},\mu\bigr)
&=2 \Gamma^i_{\rm cusp}[\as(\mu)] \, \ln \frac{\mu}{\omega e^{-\etacut}}
+ \gamma^i_{B^{\rm cut}}[\as(\mu)]
\,,\end{align}
%%%
where the coefficients of the cusp anomalous dimension follow from our explicit one-loop calculation.
Consistency with \eq{anom_dim_B} implies
%%%
\begin{align} \label{eq:consistency_no_step_reg3}
\gamma^i_{\mathcal{S}^{\rm cut}}(\as) + \gamma^i_{B^{\rm cut}}(\as)
&= \gamma^i_B(\as)
\,, \nn \\
\gamma^i_{\nu,\mathcal{S}^{\rm cut}}(\as,R)
&= \gamma^i_{\nu,B}(\as,R)
= - \frac{1}{2} \gamma^i_{\nu}(\as, R)
\,.\end{align}
%%%
All of the above noncusp anomalous dimensions vanish at one loop.
The canonical scales for $B^{(\rm cut)}_{i}$ and $\mathcal{S}^{(\rm cut)}_{i}$ are
%%%
\begin{align}
\mu^{(\rm cut)}_{B}  \sim Q e^{-\etacut}
\,,\quad
\mu^{(\rm cut)}_{\mathcal{S}} \sim \ptcut
\,,\quad
\nu^{(\rm cut)}_{\mathcal{S}} \sim \ptcut e^{\etacut}
\,.\end{align}
%%%
With these choices and the anomalous dimensions in \eq{anom_dim_cut} one may
resum logarithms of $e^{\etacut}$, $\ptcut/Q$ to any logarithmic order, and at
LL also logarithms of $\ptcut/Q e^{-\etacut}$.

Starting at $\mathcal{O}(\alpha_s^2)$, the $\mathcal{B}^{\rm (NG)}_{i}$ term in \eq{refactB}
contains nonglobal logarithms of the form $\alpha_s^n \ln^n(\ptcut/Q  e^{-\etacut})$.
A boost by $\etacut$ translates the measurement
into two hemispheres with one loose ($\eta > \etacut$) and one tight constraint ($\eta < \etacut$) on emissions.
The nonglobal structure in such a scenario is well understood~\cite{Hornig:2011iu}.
Depending on the desired accuracy, the NGLs may be included at fixed order via
$\mathcal{B}^{\rm (NG)}_{i}$ as indicated in \eq{refactB},
or (partially) summed using more steps in a dressed parton expansion~\cite{Larkoski:2015zka}.

Note that beyond one loop there is some freedom in the choice of measurement
that defines the $B^{(\rm cut)}_i$ and $\mathcal{S}^{(\rm cut)}_i$.
In particular, different measurements that reduce to \eqs{I_cut}{S_cut}
for a single emission could give rise to different results for the two-loop
noncusp anomalous dimensions and finite terms
because the difference can be absorbed into $\mathcal{B}^{\rm (NG)}_{i}$.
We stress that the result \eq{S_cut_ln_R_result} for the $\ln R$ coefficient
in the two-loop soft-collinear function is, however, still unique.
This is because a $\ln R$ contribution to $\mathcal{B}^{(\rm NG)}$
requires a collinear parton in the unconstrained region
to emit a soft-collinear gluon into the constrained region,
which then undergoes a further collinear splitting.
This is only possible starting at $\ord{\as^3}$.

\paragraph{Numerical validation.}

\begin{figure*}
\centering
\includegraphics[width=\WidthTwoSubfigs]{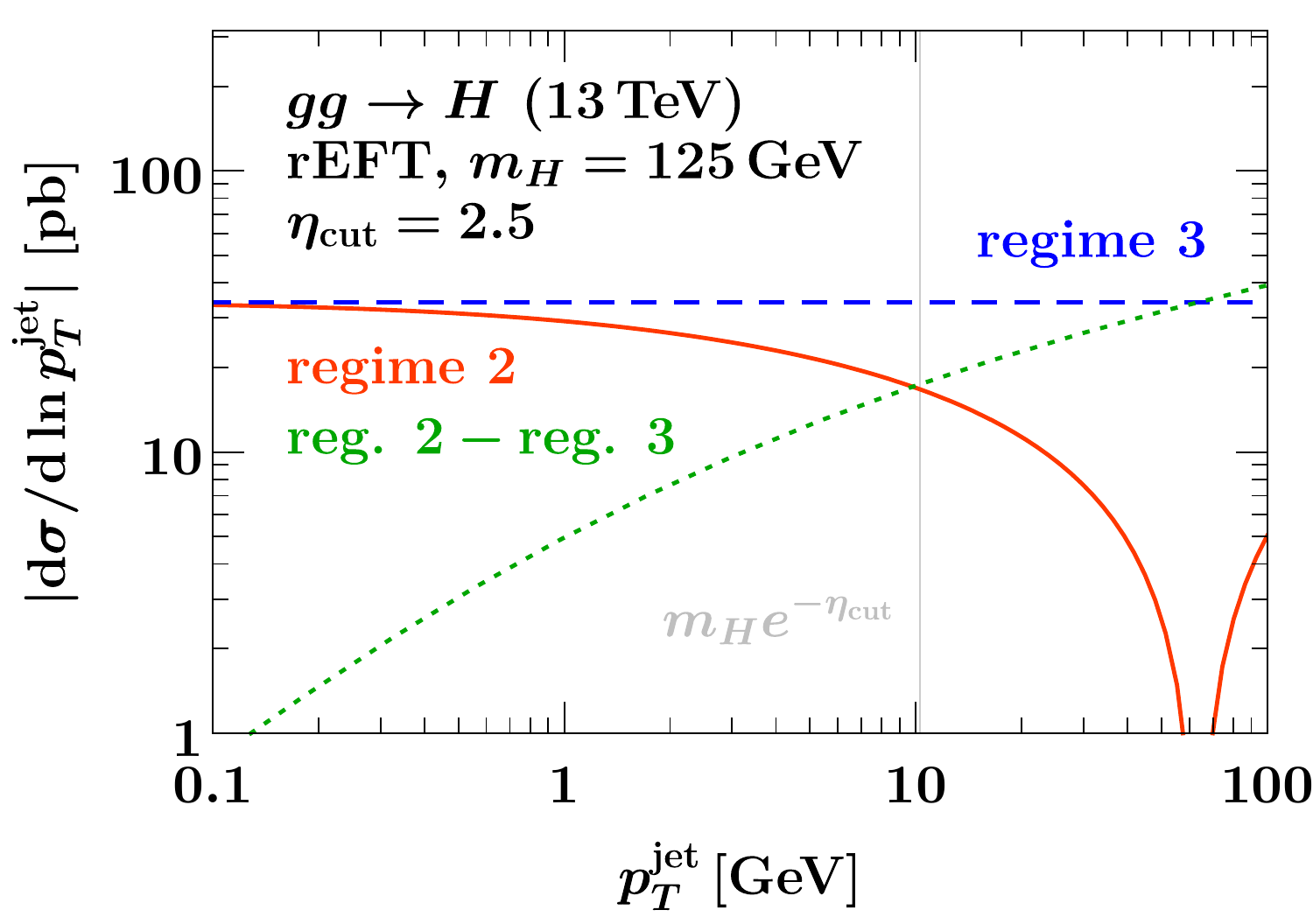}%
\hfill%
\includegraphics[width=\WidthTwoSubfigs]{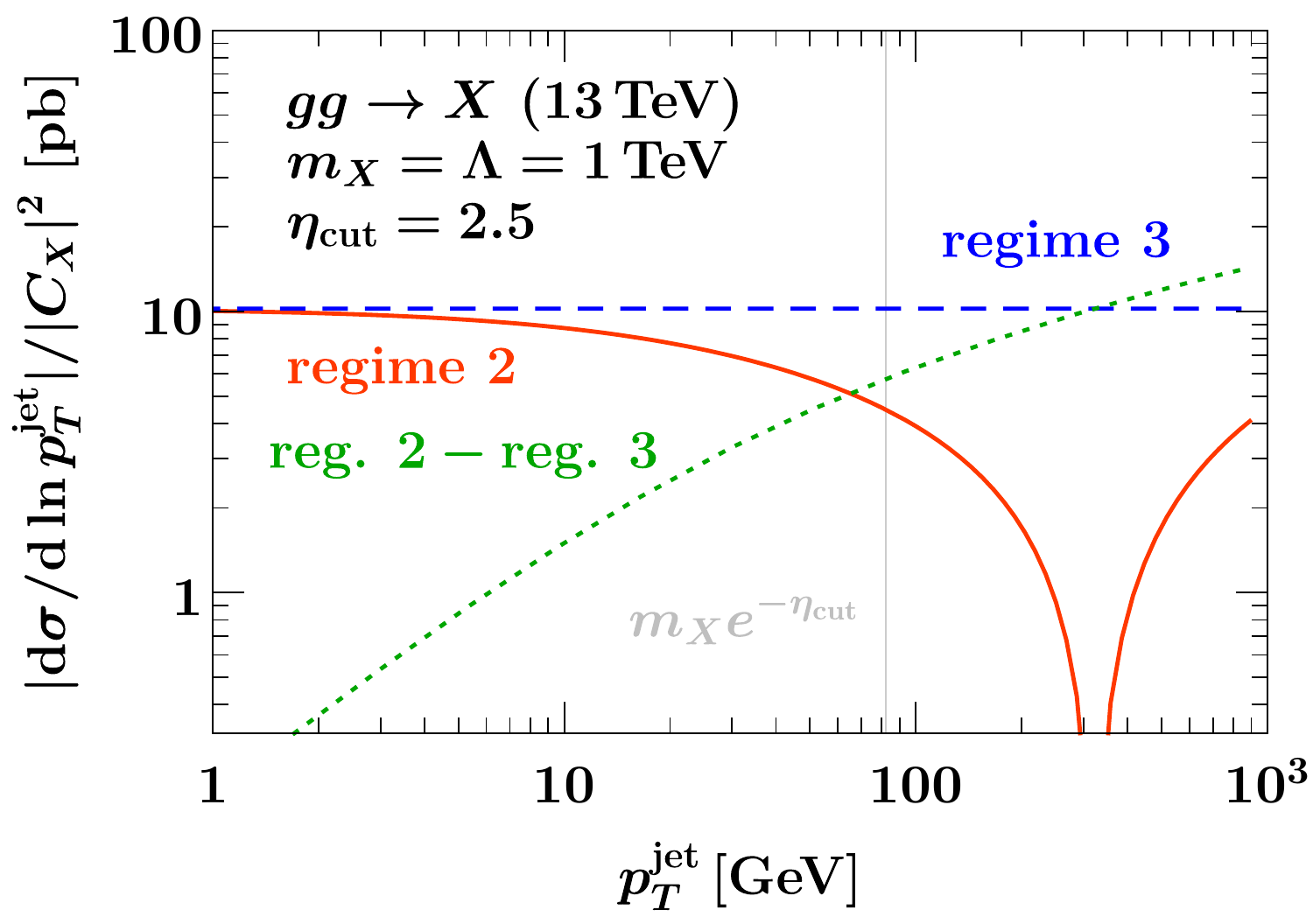}%
\\
\includegraphics[width=\WidthTwoSubfigs]{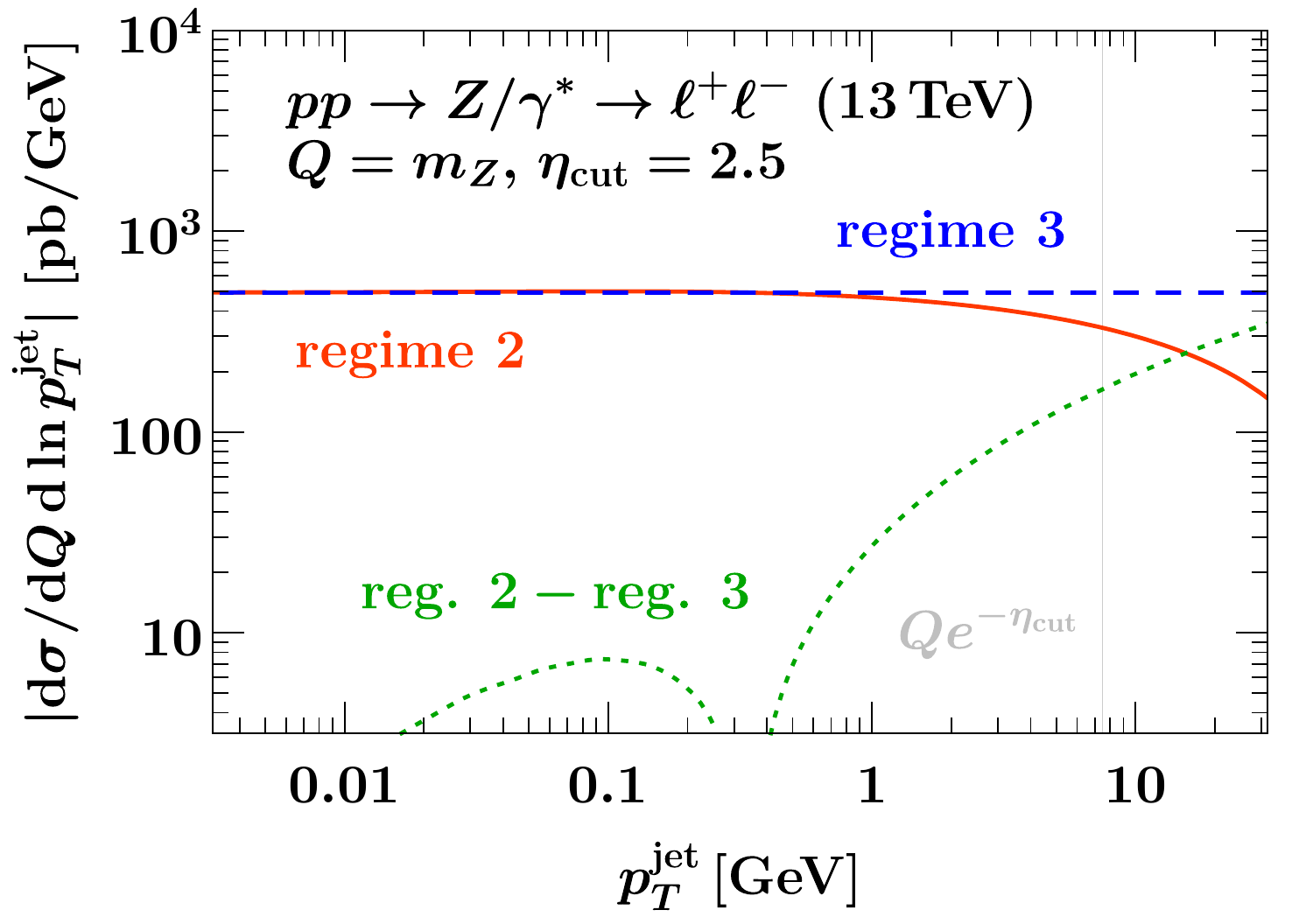}%
\hfill%
\includegraphics[width=\WidthTwoSubfigs]{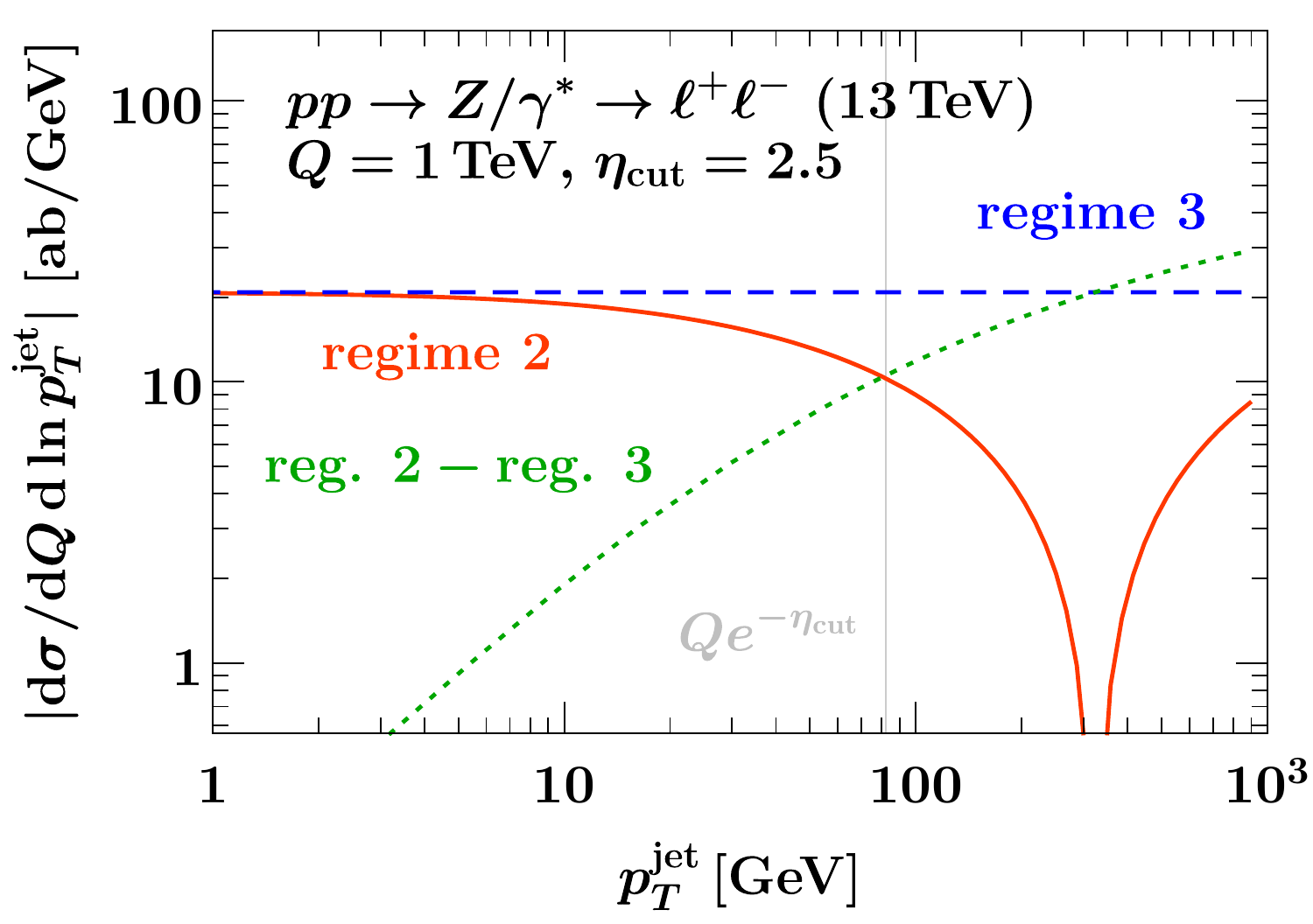}%
\caption{Comparison of the singular contributions to the fixed $\ord{\alpha_s}$
$\ptjet$ spectrum for $gg\to H$ (top left), $gg\to X$ (top right), and Drell-Yan at $Q = m_Z$ (bottom left) and $Q = 1 \TeV$ (bottom right).
The solid orange lines show the full regime~2 singular spectrum, the blue dashed lines the
further factorized regime~3 result. Their difference shown by the dotted green lines vanishes
as a power in $\ptjet / Q e^{-\etacut}$ for small $\ptjet$.
The vertical lines indicate where the relation $\ptjet = Q e^{-\etacut}$ is satisfied.}
\label{fig:regimes_2_3}
\end{figure*}

To illustrate the numerical relevance of regime~3,
we again consider the fixed $\ord{\alpha_s}$ $\ptjet$ spectrum.
In regime~2, it is given to leading power in $\ptjet/Q$ by the derivative of \eq{fact_pTjet2},
while in regime~3, it is given to leading power in $\ptjet/(Q e^{-\etacut})$ by the derivative of \eq{fact_pTjet3}.

In \fig{regimes_2_3} we compare the two results for $\etacut = 2.5$.
In regime~3, the 0-jet cross section at $\ord{\as}$ contains only single logarithms of $\ptcut$,
because the double logarithms cancel between the soft and soft-collinear functions.
For this reason, the dashed-blue regime~3 spectrum with respect to $\ln \ptjet$ is just a constant.
The exact regime~2 result (solid orange)
becomes well approximated by the further factorized regime~3 expression for $\ptjet\to 0$,
with their difference (dotted green) behaving like a power in $\ptjet$.
This provides a strong check of the regime~3 ingredients,
more precisely, of the $\ptcut$ dependence encoded in the soft-collinear function.
(Since the beam function in regime~3 is independent of $\ptcut$,
it drops out when computing the fixed-order spectrum.)

We also observe that for $gg\to H$ and Drell-Yan at $Q = m_Z$, the regime~3 limit
is applicable only at very small $\ptjet \lesssim 1 \GeV$ and
already at $\ptjet \sim 10-20 \GeV$ the power corrections with respect to regime~2
are of the same size as the full regime~2 result. This means that
one would have to turn off the additional regime~3 resummation above this region.
For $gg\to X$ with $m_X = 1\TeV$ and Drell-Yan at $Q = 1\TeV$, the canonical regime~3 resummation
region, i.e., the region where the regime~3 singular corrections clearly dominate, extends
up to $\ptjet \lesssim 10 \GeV$,
while regime~2 power corrections become $\ord{1}$ around $\ptjet \sim 60 \GeV$.

Hence, we find that the additional resummation of logarithms of $\ptjet/(Q
e^{-\etacut})$ in regime~3 is not relevant for jet veto analyses at the LHC,
where the lowest jet cuts are $\ptcut \sim 25 \GeV$, for $\etacut = 2.5$ and
final states in the $Q \sim 100 \GeV$ range. This also holds for final states at
very high invariant mass, e.g.\ in new physics searches, since in this case one
would typically also apply higher jet thresholds to retain enough signal in the
$0$-jet bin. Realistically, one would not go below $\ptcut \sim 0.1 Q$, which
means one never enters the limit where the regime~3 resummation is necessary.
This of course does not exclude the possibility that measurements designed to
probe simultaneously very high $Q$ and very low $\ptjet$ could benefit from the
regime~3 resummation. To explicitly explore this regime experimentally, the best
option is to restrict the jet veto to the very central region with $\etacut \sim
1-1.5$.

%===============================================================================
\subsection{Comparison to the literature}
\label{sec:no_step_literature}
%===============================================================================

Jet vetoes in a restricted rapidity range were already encountered in
\refcite{Hornig:2016ahz} for the case of dijet production. Without spelling it
out explicitly, \refcite{Hornig:2016ahz} used a factorization for the regime~3 hierarchy $\ptcut \ll
Q e^{-\etacut} \ll Q$, but did not distinguish between the soft and
soft-collinear modes necessary in this regime. As a result,
parametrically large rapidity logarithms $\ln e^{\etacut}$ were not captured, which are
relevant starting at NLL. The numerical results in \refcite{Hornig:2016ahz} were obtained for $Q \sim 1
\TeV$, $\etacut = 5$, and $\ptcut= 20$ GeV, which rather corresponds to the opposite regime~1,
$\ptcut \gg Q e^{-\etacut}$. The difference between regimes~1 and 3 already
matters at LL.

In \refcite{Hornig:2017pud}, the soft and soft-collinear modes in regime~3 are
distinguished and the presence of nonglobal logarithms in this regime is
recognized. Their factorization for dijet production is carried out at a level analogous to ours in
the previous subsection. That is, at NLL and beyond it only captures
logarithms of ``global'' origin, but does not capture nonglobal logarithms that are
parametrically of the same size. Our results for the one-loop quark matching
coefficients in \eq{I_cut} and the one-loop soft-collinear function in
\eq{S_cut} agree with \refcite{Hornig:2017pud} [see their eqs.~(3.27), (B.3),
and (B.5)]. Our results for the gluon channels and the two-loop clustering
corrections are new.

Ref.~\cite{Hornig:2017pud} does not consider regime~2 as a separate parametric
regime. Instead, it attempts to extend the validity of the regime~3
factorization into regime~2.
This is done by effectively adding the regime~2 nonsingular corrections
appearing in \eq{B_i} to the unmeasured beam functions.
Since some of the regime~3 modes become redundant in regime~2, this
also requires them to account for a nontrivial soft-collinear zero bin.
At fixed order, the sum of all their contributions must reproduce our result for the
regime~2 beam function; in \app{compare_Hornig_et_al} we check that this is
indeed the case for the quark matrix elements given in \refcite{Hornig:2017pud}.
As we have seen in \fig{regimes_2_3}, outside the canonical regime~3,
there are large cancellations between the terms that are singular in
the regime~3 limit and the remaining regime~2 nonsingular contributions. This means
that the distinction between these contributions becomes arbitrary in regime~2
and that they must not be treated differently, as otherwise one risks inducing
large miscancellations.
(This is completely analogous to the situation when matching to full QCD, in which
case the $\ptcut$ resummation must be turned off when entering the fixed-order
region at large $\ptcut$ to properly recover the full-QCD result.)
In particular, in regime~2 all contributions that belong to the full
$\etacut$-dependent regime~2 beam function must be evaluated at a common scale
$\mu \simeq \ptcut$ and evolved together according to \eq{RGE_B}.
This is not the case in \refcite{Hornig:2017pud},
where individual contributions to the regime~2 beam function are evaluated at different scales
throughout ($\mu_B^\mathrm{cut}$ and $\mu_\mathcal{S}^\mathrm{cut}$ in our notation).

Recently, the setup of \refcite{Hornig:2017pud} was applied in \refcite{Kang:2018agv}
to the case of transverse energy $E_T$ in a restricted rapidity range in Drell-Yan.
In \refcite{Kang:2018agv}, profile scales are used to combine regimes~3 and
1, requiring that asymptotically $\mu^{(\rm cut)}_{B} = \mu^{(\rm cut)}_{\mathcal{S}}$
in the regime~1 limit $E_T \gg Q e^{-\etacut}$. While this can alleviate
the issue raised above, formally this relation must be satisfied already
in regime~2 for $E_T \sim Q e^{-\etacut}$.

As we have seen in \sec{no_step_reg2}, there is no need to distinguish collinear
and soft-collinear modes in regime~2. Since for jet-veto analyses regimes~1 and
2 are the phenomenologically relevant ones, doing so unnecessarily complicates
the description. Recovering the NNLL$'$ structure in regime~2
[see \eq{delta_I_master_formula}] based on regime~3
would be quite challenging due to the intricate nonglobal structure in regime~3.
Our dedicated treatment of regime~2 makes the absence of
nonglobal logarithms manifest, avoiding the associated complications,
and automatically ensures the correct treatment of the regime~2 nonsingular terms.
Furthermore, it shows how regime~2 generalizes the well-understood regime~1,
and as we will see in the next section allows for the generalization to a step
in the jet veto.

Concerning regime~1, \refcite{Kang:2018agv} also gave an argument that regime~1
holds up to power corrections in $Qe^{-\etacut}/E_T$, which was more intricate
due to immediately comparing regime~1 to regime~3.
The power suppression of $\etacut$
effects at sufficiently large $\etacut$ was also pointed out briefly in a
somewhat different context in \refcite{Balsiger:2018ezi}.

%%%%%%%%%%%%%%%%%%%%%%%%%%%%%%%%%%%%%%%%%%%%%%%%%%%%%%%%%%%%%%%%%%%%%%%%%%%%%%%%
\section{\boldmath Generalization to a step in the jet veto at \texorpdfstring{$\etacut$}{etacut}}
\label{sec:yes_step}
%%%%%%%%%%%%%%%%%%%%%%%%%%%%%%%%%%%%%%%%%%%%%%%%%%%%%%%%%%%%%%%%%%%%%%%%%%%%%%%%

%===============================================================================
\subsection{Overview of parametric regimes}
%===============================================================================

We now generalize our results to the experimentally relevant scenario of the step-like jet veto
illustrated in the right panel of \fig{illustration_pileup}.
Here, jets with $\ptjet > \ptcut$ are vetoed if $|\etajet|<\etacut$,
while for $|\etajet|>\etacut$ the veto is loosened to $\ptjet > \ptcuttwo > \ptcut$.
The 0-jet cross section is thus defined by the following measurement:
%%%
\begin{equation} \label{eq:def_veto_step}
\max_{k \in \text{jets}:\,\abs{\eta_k} < \etacut} \abs{\vec{p}_{T,k}}  < \ptcut
\qquad\text{and}\qquad
\max_{k \in \text{jets}:\,\abs{\eta_k} > \etacut} \abs{\vec{p}_{T,k}}  < \ptcuttwo
\,.\end{equation}
%%%

\begin{figure*}
\centering
\includegraphics[width=\WidthTwoSubfigs]{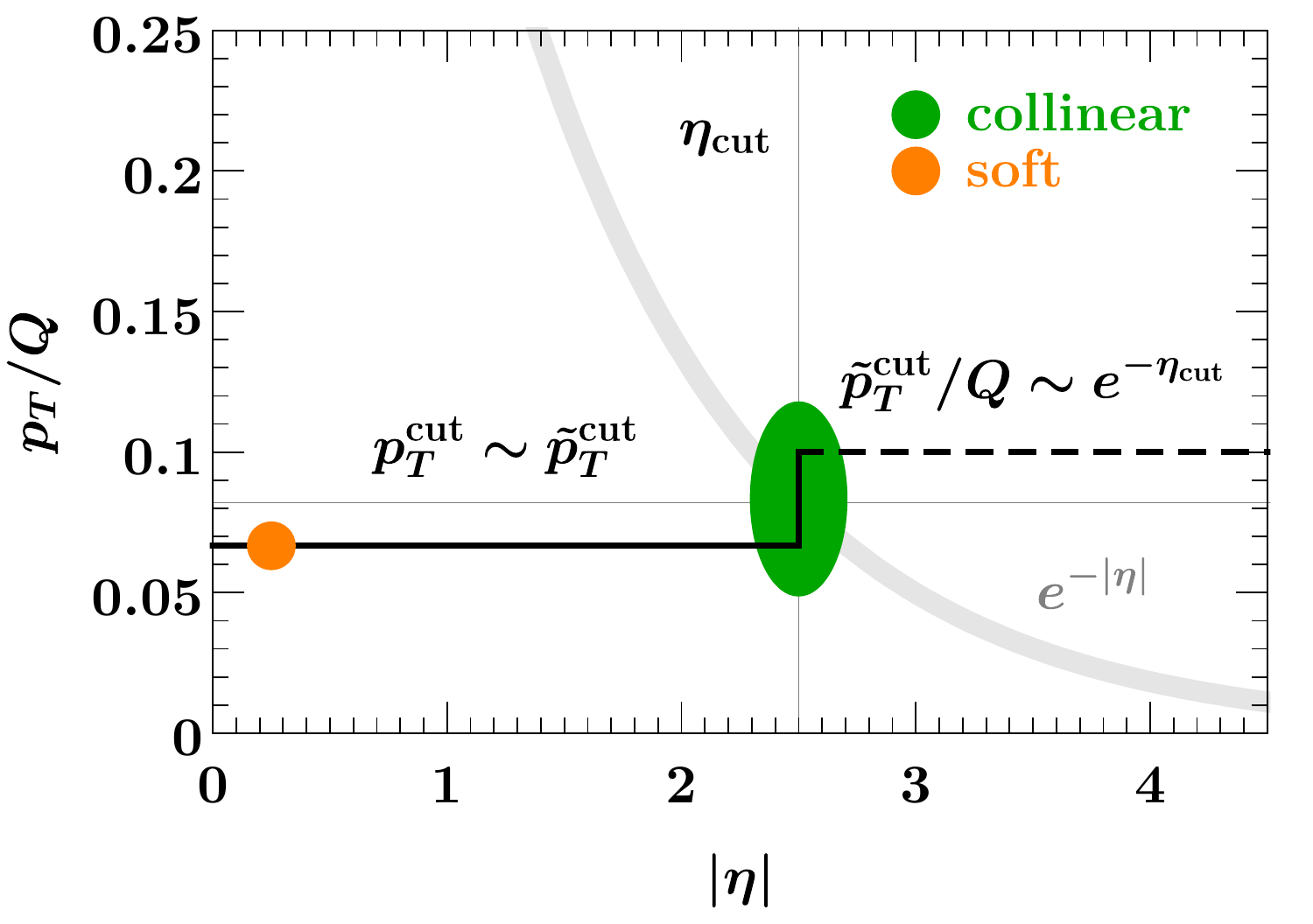}%
\hfill%
\includegraphics[width=\WidthTwoSubfigs]{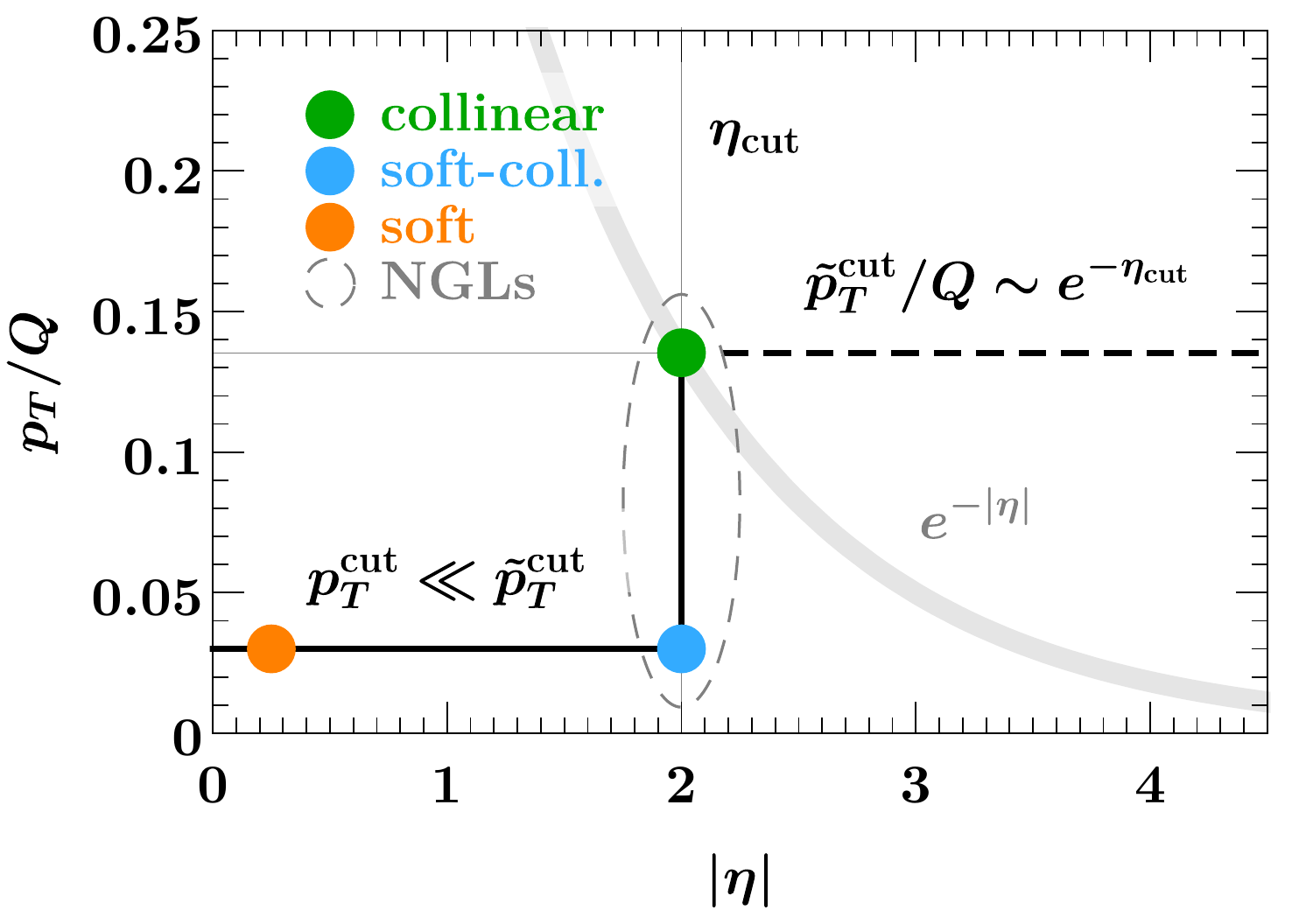}%
\\
\includegraphics[width=\WidthTwoSubfigs]{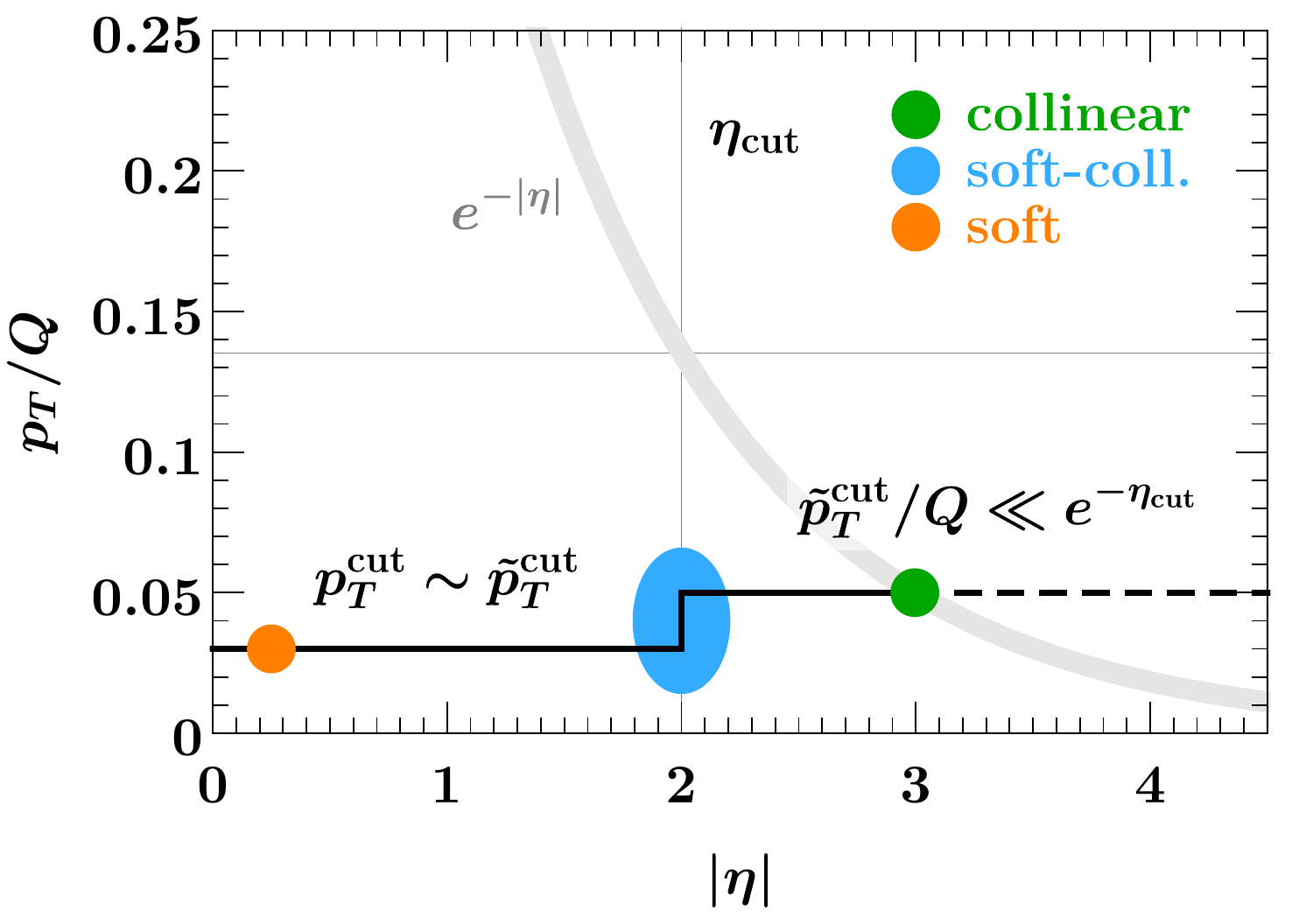}%
\hfill%
\includegraphics[width=\WidthTwoSubfigs]{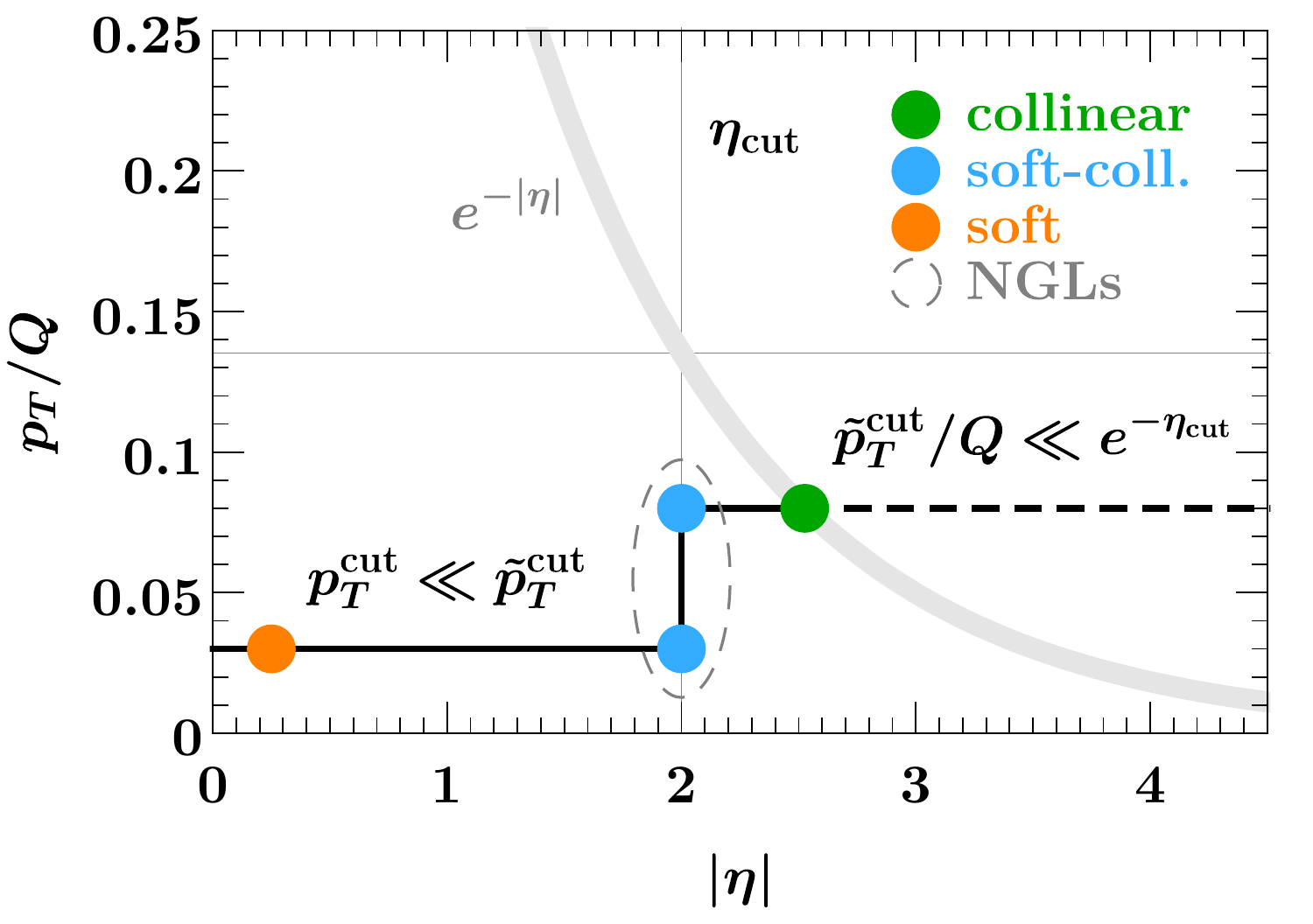}%
\caption{Illustration of the parametric regimes for a jet veto with a step.
Emissions above the black lines are vetoed, and the thick gray line corresponds
to $p_T/Q = e^{-\abs{\eta}}$.
The colored circles indicate the relevant modes in the effective theory.
The regimes in the top row are characterized by $\ptcuttwo \sim e^{-\etacut}$,
while those in the bottom row have $\ptcuttwo \ll e^{-\etacut}$.
The regimes on the left have $\ptcut \sim \ptcuttwo$, while those on
the right have $\ptcut \ll \ptcuttwo$ and involve parametrically large non-global logarithms.}
\label{fig:extend_by_step}
\end{figure*}

There are now three relevant power-counting parameters $\ptcut/Q$, $\ptcuttwo/Q$, and $e^{-\etacut}$
with four distinct parametric regimes (assuming $\ptcut \leq \ptcuttwo$),
illustrated in \fig{extend_by_step}:
%%%
\begin{itemize}
   \item $\ptcut/Q \sim \ptcuttwo/Q \sim e^{-\etacut}$ (collinear step, top left),
   \item $\ptcut/Q \ll \ptcuttwo/Q \sim e^{-\etacut}$ (collinear NGLs, top right),
   \item $\ptcut/Q \sim \ptcuttwo/Q \ll e^{-\etacut}$ (soft-collinear step, bottom left),
   \item $\ptcut/Q \ll \ptcuttwo/Q \ll e^{-\etacut}$ (soft-collinear NGLs, bottom right).
\end{itemize}
%%%
We discuss each of them in turn in the following subsections.
For $\ptcut/Q \sim e^{-\etacut}$ (top left) the only relevant case is $\ptcuttwo \sim \ptcut$,
leading to a modified measurement on the collinear modes, a collinear step,
compared to the case without a step ($\ptcuttwo = \ptcut$).

For $\ptcut/Q \ll e^{-\etacut}$, we have to distinguish three
cases depending on $\ptcuttwo$. Keeping $\ptcuttwo \sim e^{-\etacut}$ implies
the hierarchy $\ptcut/Q \ll \ptcuttwo/Q \sim e^{-\etacut}$ (top right). Here, the mode setup is the same
as for regime~3 without step (corresponding to $\ptcuttwo = \infty$). As in that
case, the large difference in the constraints on collinear radiation above and
below $\etacut$ gives rise to collinear NGLs.

For $\ptcuttwo/Q \ll e^{-\etacut}$, we can then have either
$\ptcut/Q \sim \ptcuttwo/Q \ll e^{-\etacut}$ (bottom left)
or $\ptcut/Q \ll \ptcuttwo/Q \ll e^{-\etacut}$ (bottom right).
For the former, the standard jet veto factorization
is recovered except that there are additional soft-collinear modes that resolve the
shallow step at $\etacut$. For the latter, the steep step $\ptcut \ll \ptcuttwo$ at $\etacut$
gives rise to two distinct sets of soft-collinear modes with parametrically large
soft-collinear NGLs between them.

%===============================================================================
\subsection[\texorpdfstring{$\ptcut/Q \sim \ptcuttwo/Q \sim e^{-\etacut}$}{pTcut/Q sim tildepTcut/Q sim exp(-etacut)} (collinear step)]
{\boldmath $\ptcut/Q \sim \ptcuttwo/Q \sim e^{-\etacut}$ (collinear step)}
\label{sec:collinear_step}
%===============================================================================

We first note that the hierarchy $\ptcut / Q \sim e^{-\etacut} \ll \ptcuttwo/Q$
is effectively equivalent to the case without any jet veto beyond $\etacut$
(regime~2 in \sec{no_step_reg2}).
Since collinear emissions with $\abs{\eta} > \etacut$ cannot resolve the loose veto
at $\ptcuttwo$, its effect is suppressed by $1/\ptcuttwo$ and vanishes for $\ptcuttwo \to\infty$.

The first nontrivial hierarchy is $\ptcut/Q \sim \ptcuttwo/Q \sim e^{-\etacut}$,
illustrated in the top left panel of \fig{extend_by_step}.
In this regime, the required modes are the same as in regime~2 in \sec{no_step_reg2}.
The collinear radiation resolves the step at $\etacut$ while soft emissions are
insensitive to it, leading to a generalization of \eq{fact_pTjet2},
%%%
\begin{align} \label{eq:fact_collinear_step}
\sigma_0 (\ptcut, \ptcuttwo, \etacut, R, \Phi)
&= H_\kappa(\Phi, \mu) \,
\nn \\ & \quad\times
B_{a}(\ptcut, \ptcuttwo, \etacut, R, \omega_a, \mu, \nu) \,
B_{b}(\ptcut, \ptcuttwo, \etacut, R, \omega_b, \mu, \nu) \,
\nn \\ & \quad \times
S_{\kappa} (\ptcut, R, \mu, \nu)
\biggl[1+\mathcal{O}\Bigl(\frac{\ptcut}{Q}, \frac{\ptcuttwo}{Q}, e^{-\etacut},R^2\Bigr)\biggr]
\,,\end{align}
%%%
with the beam functions now additionally depending on $\ptcuttwo$.
In analogy to \eq{I_tot} we write the modified beam function matching coefficients as
%%%
\begin{align} \label{eq:I_tot_ptcuttwo}
\mathcal{I}_{ij}(\ptcut, \ptcuttwo, \etacut, R, \omega, z,\mu, \nu)
&=\mathcal{I}_{ij}(\ptcut, R, \omega, z, \mu, \nu)
+ \Delta \mathcal{I}_{ij}(\ptcut, \ptcuttwo, \etacut, R, \omega, z, \mu, \nu)
\,.\end{align}
%%%
The first term on the right-hand side is again the matching coefficient for a single
veto at $\ptcut$ without any rapidity dependence.
The second term is the correction due to the step in the jet veto at $\abs{\eta} = \etacut$,
which vanishes for $\ptcut = \ptcuttwo$.
The correction is again renormalized according to \eq{rge delta I}, which as before
follows from RG consistency.
In particular, its two-loop structure predicted by the RGE is the same as in
\eq{delta_I_master_formula}, where the finite terms now depend on two dimensionless ratios,
%%%
\begin{equation}
\zetacut = \frac{\omega e^{-\etacut}}{\ptcut}
\,,\qquad
\zetacuttwo = \frac{\omega e^{-\etacut}}{\ptcuttwo}
\,.\end{equation}
%%%
The one-loop and $\ln R$ enhanced two-loop finite terms in $\Delta \mathcal{I}_{ij}$
can be written in terms of the results in \eqs{DeltaI}{DeltaI_lnR} as
%%%
\begin{align} \label{eq:delta_I_step}
\Delta I^{(1)}_{ij}(\zetacut, \zetacuttwo, z)
&= \Delta I^{(1)}_{ij}(\zetacut, z) - \Delta I^{(1)}_{ij}(\zetacuttwo, z)
\,, \nn \\
\Delta I^{(2)}_{ij} (\zetacut, \zetacuttwo, R, z)
&= \ln R \, \Bigl[
\Delta I^{(2, \ln R)}_{ij}(\zetacut, z) - \Delta I^{(2, \ln R)}_{ij}(\zetacuttwo, z) \Bigr]
\,, \nn \\ & \quad
+ \Delta I^{(2,c)}_{ij}(\zetacut, \zetacuttwo, z) + \ord{R^2}
\,,\end{align}
%%%
since for a single (primary) $n_a$-collinear emission at $(\eta, p_T)$ the measurement
function for the step correction can be rewritten as
%%%
\begin{align}
&\theta(\eta - \etacut) \bigl[ \theta(\ptcuttwo - p_T) - \theta(\ptcut - p_T) \bigr]
\nn \\ &\qquad
= \theta(\eta - \etacut) \,\theta(p_T - \ptcut) - \theta(\eta - \etacut)\,\theta(p_T - \ptcuttwo)
\,.\end{align}
%%%
Due to the presence of correlated emissions with rapidities smaller and larger than $\etacut$ at two loops,
this decomposition no longer applies for the full two-loop finite term $\Delta I^{(2,c)}_{ij}$,
which therefore needs to be determined separately.

This regime is free of large nonglobal logarithms and is of direct phenomenological
interest. The parametric assumptions are satisfied e.g.\ for high-mass searches,
$Q \gtrsim 300 \GeV$, a realistic rapidity cut $\etacut = 2.5$, and veto
parameters $\ptcut = 25 \GeV$, $\ptcuttwo = 50 \GeV$, which clearly warrant
resummation of logarithms of $\ptcut/Q \sim \ptcuttwo / Q \sim e^{-\etacut}$.
Evolving the beam function from $\mu_B \sim \ptcut \sim \ptcuttwo \sim Q e^{-\etacut}$ to $\mu_H \sim Q$
achieves this resummation for all of the above large ratios in the cross section,
while the full (logarithmic and nonlogarithmic) dependence on all of the $\ord{1}$ ratios
$\ptcut/\ptcuttwo$, $Q e^{-\etacut}/\ptcut$, and $Q e^{-\etacut}/\ptcuttwo$
is included at fixed order via the beam function boundary condition.

\paragraph{Numerical validation.}

\begin{figure*}
\centering
\includegraphics[width=\WidthTwoSubfigs]{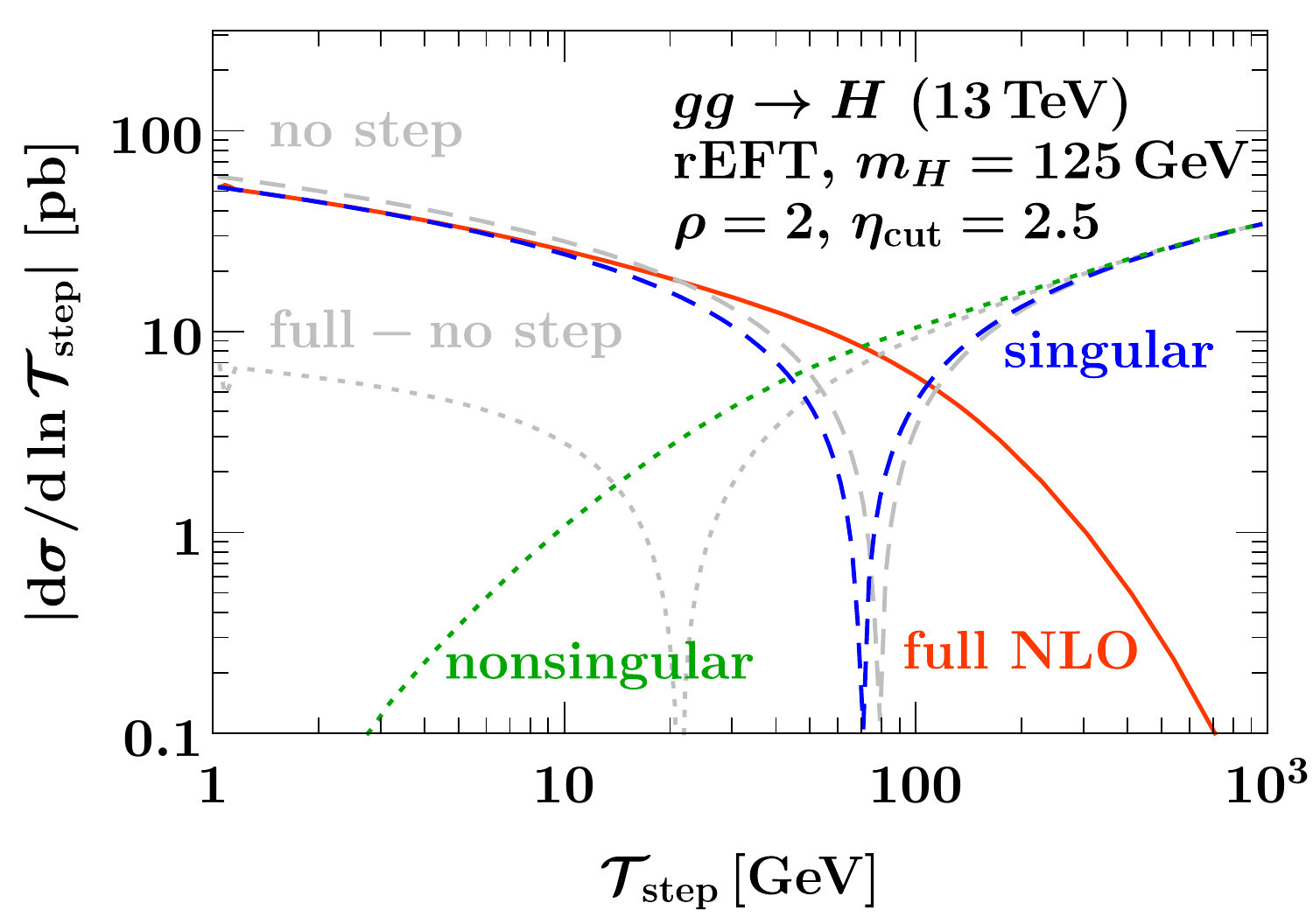}%
\hfill%
\includegraphics[width=\WidthTwoSubfigs]{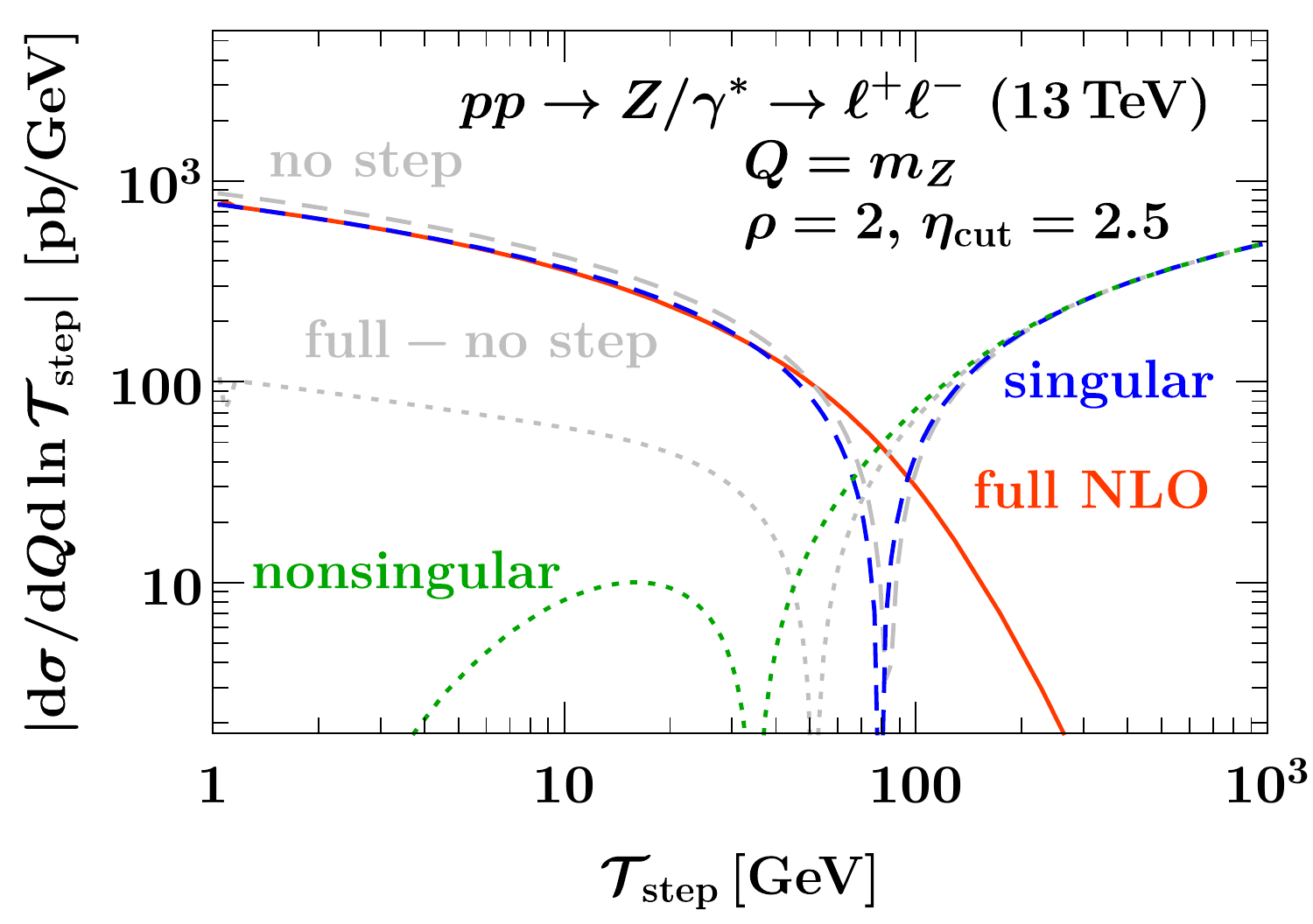}%
\caption{Comparison of singular and nonsingular contributions to the fixed
$\ord{\alpha_s}$ (LO$_1$) $\Tau_\mathrm{step}$ spectrum
with a step at $\etacut = 2.5$ and $\rho = \ptcuttwo/\ptcut = 2$
for $gg\to H$ (left) and Drell-Yan at $Q = m_Z$ (right).
The orange solid lines show the full results, the dashed blue lines the
singular result that accounts for the jet veto step at $\etacut$ in the beam function,
and the dotted green lines their difference.
The dashed and dotted gray lines show the corresponding results without taking into
account the step in the jet veto, which do not describe the singular behavior
of the full cross section.}
\label{fig:sing_nons_step}
\end{figure*}

We now check that the factorized 0-jet cross section in \eq{fact_collinear_step}
reproduces the singular limit of full QCD.
For this purpose, we construct an observable
that simultaneously forces $\ptcut \to 0$ and $\ptcuttwo \to 0$
as it approaches its singular limit. Following the rapidity-dependent jet vetoes
in \refcite{Gangal:2014qda}, we define
%%%
\begin{equation}
\Tau_\mathrm{step}
= \max_{k \in \text{jets}} \abs{\vec{p}_{T,k}} f_\mathrm{step}(\eta_k)
\,,\qquad
f_\mathrm{step}(\eta) = \begin{cases}
   \frac{1}{\rho},\,&\abs{\eta} > \etacut
   \,, \\
   1,\,& \abs{\eta} < \etacut
\,,\end{cases}
\end{equation}
%%%
i.e., we can express the step veto by ordering the jets with respect to their weighted transverse momenta,
where for $\abs{\eta} > \etacut$ the corresponding step weight function $f_\mathrm{step}(\eta)$
is given by the ratio of veto parameters,
%%%
\begin{equation}
\rho \equiv \frac{\ptcuttwo}{\ptcut} > 1
\,.\end{equation}
%%%
The differential spectrum in $\Tau_\mathrm{step}$ is then related to the jet-vetoed cross section with a step by
the relation
%%%
\begin{equation} \label{eq:relate_Tau_step_to_step_veto}
\sigma_0(\ptcut,\, \rho\,\ptcut, \, \etacut, R )
= \int_0^{\ptcut}\! \df\Tau_\mathrm{step}\,
\frac{\df \sigma(\rho, \etacut, R)}{\df \Tau_\mathrm{step}}
\,.\end{equation}
%%%

In \fig{sing_nons_step} we compare $\df\sigma(\rho, \etacut)/\df \Tau_\mathrm{step}$
at fixed $\ord{\alpha_s}$ in full QCD to the singular spectrum
predicted by \eq{fact_collinear_step} as well as the standard factorization \eq{fact_pTjet} without a step
for $gg\to H$ (left panel) and Drell-Yan at the $Z$ pole (right panel).
The singular result using the full $\ptcuttwo$ and $\etacut$ dependent beam functions (dashed blue)
correctly reproduces the singular behavior of full QCD (solid orange) in the limit $\Tau_\mathrm{step}\to 0$,
with the difference to the full QCD spectrum (dotted green) vanishing like a power in
$\Tau_\mathrm{step}$ as it should.
On the other hand, the standard factorization without step (dashed gray) does not reproduce
the correct singular behavior of full QCD,
with the difference (dotted gray) diverging for $\Tau_\mathrm{step} \to 0$.
Note that the mismatch here is reduced compared to the $\ptcuttwo = \infty$ case
shown in \figs{sing_nons_ggHX}{sing_nons_DY_narrow_width},
owing to the larger phase space available to unconstrained radiation at
$\abs{\eta} > \etacut$ for $\ptcuttwo = \infty$.

%===============================================================================
\subsection[\texorpdfstring{$\ptcut/Q \ll \ptcuttwo/Q \sim e^{-\etacut}$}{pTcut/Q << tildepTcut/Q sim exp(-etacut)} (collinear NGLs)]
{\boldmath $\ptcut/Q \ll \ptcuttwo/Q \sim e^{-\etacut}$ (collinear NGLs)}
\label{sec:collinear_ngls}
%===============================================================================

This regime is a direct extension of regime~3 without a step in \sec{no_step_reg3}.
For $e^{-\etacut} \ll \ptcuttwo/Q$, the effect of $\ptcuttwo$ is again
suppressed by $1/\ptcuttwo$ and vanishes for $\ptcuttwo\to\infty$,
yielding the same result as in \sec{no_step_reg3}.
The nontrivial new hierarchy is $\ptcut/Q \ll \ptcuttwo/Q \sim e^{-\etacut}$,
shown in the top right panel of \fig{extend_by_step}.
In this regime, the mode setup is as in \sec{no_step_reg3}.
However, the collinear modes are now additionally constrained for $\abs{\eta}>\etacut$
by the jet veto at $\ptcuttwo$, making them sensitive to both $\ptcuttwo$ and the
kinematic scale $Q e^{-\etacut}$.
This leads to a modification of the overall initial-state collinear functions in \eqs{fact_pTjet3}{B_i} by
%%%
\begin{align}\label{eq:beam_refactorization2}
\mathcal{B}_{i}(\ptcut, \ptcuttwo, \etacut, R, \omega, \mu, \nu)
&= B^{(\rm cut)}_{i}(\ptcuttwo, \etacut, R, \omega, \mu) \,\mathcal{S}^{(\rm cut)}_{i}(\ptcut, \etacut, R, \mu, \nu) \nn \\
&\quad \times  \biggl[1+ \mathcal{B}^{\rm (NG)}_{i}\Bigl(\frac{\ptcut}{ \omega e^{-\etacut}},\frac{\ptcut}{\ptcuttwo}, \omega, R\Bigr)\biggr]
\,.\end{align}
%%%
Here $\mathcal{S}^{(\rm cut)}_{i}$ is the same soft-collinear function as in \eq{refactB}.
By RG consistency the functions $B^{(\rm cut)}_{i}$ have the same renormalization as those in \eq{refactB},
i.e., the additional dependence on $\ptcuttwo$ does not change their renormalization.
The associated matching coefficients at one loop are given by subtracting the correction term
$\Delta I_{ij}^{(1)}$ in \eq{DeltaI}, which accounts for an $n$-collinear emission with $\eta > \etacut$ and $p_T > \ptcuttwo$,
from the coefficient $\mathcal{I}^{({\rm cut})}_{ij}$ in \eq{I_cut},
which accounts for an $n$-collinear emission with $\eta > \etacut$ without constraints from a jet veto, such that
%%%
\begin{align}
\mathcal{I}^{({\rm cut})}_{ij}(\ptcuttwo, \etacut ,R, \omega, z, \mu)
= \mathcal{I}^{({\rm cut})}_{ij}(\etacut, \omega, z, \mu)
- \frac{\as(\mu)}{4\pi} \Delta I^{(1)}_{ij}\Bigl(\frac{\omega e^{-\etacut}}{\ptcuttwo},z,R\Bigr)
+ \ord{\as^2}
\,.\end{align}
%%%
The $\mathcal{B}^{\rm (NG)}_{i}$ term in \eq{beam_refactorization2} contains nonglobal
logarithms of $\ptcut/\ptcuttwo \sim \ptcut/Q e^{-\etacut}$.

%===============================================================================
\subsection[\texorpdfstring{$\ptcut/Q \sim \ptcuttwo/Q \ll e^{-\etacut}$}{pTcut/Q sim tildepTcut/Q << exp(-etacut)} (soft-collinear step)]
{\boldmath $\ptcut/Q \sim \ptcuttwo/Q \ll e^{-\etacut}$ (soft-collinear step)}
\label{sec:soft_collinear_step}
%===============================================================================

In this regime (bottom left panel of \fig{extend_by_step}), the mode setup in
\sec{no_step_reg1} is extended by soft-collinear modes that resolve the step in
the jet veto at $\etacut$,
%%%
\begin{align}
n_a\text{-soft-collinear:}
&\quad
p^\mu\sim  \ptcut (e^{-\etacut},e^{\etacut},1) \sim  \ptcuttwo (e^{-\etacut},e^{\etacut},1)
\,, \nn \\
n_b\text{-soft-collinear:}
&\quad
p^\mu\sim  \ptcut (e^{\etacut},e^{-\etacut},1) \sim  \ptcuttwo (e^{\etacut},e^{-\etacut},1)
\,.\end{align}
%%%
At the same time, the collinear modes only see the jet
veto at $\ptcuttwo$, while the soft modes only see the veto at $\ptcut$.
This yields the factorized cross section
%%%
\begin{align}\label{eq:fact_soft_collinear_step}
\sigma_0 (\ptcut,\ptcuttwo,\etacut,R,\Phi)
&=
H_\kappa(\Phi, \mu) \,
B_{a}(\ptcuttwo, R, \omega, \mu, \nu) \,
B_{b}(\ptcuttwo, R, \omega, \mu, \nu) \,
S_{\kappa} (\ptcut, \mu, \nu)
\nn \\ &\quad
\times \mathcal{S}_{a} (\ptcut, \ptcuttwo, \etacut, R, \mu, \nu) \,\mathcal{S}_{b} (\ptcut, \ptcuttwo, \etacut, R, \mu, \nu)
\nn \\ &\quad
\times
\biggl[1+\mathcal{O}\Bigl(\frac{\ptcut}{Q},\frac{\ptcuttwo}{Q},\frac{\ptcut}{Q e^{-\etacut}},\frac{\ptcuttwo}{Q e^{-\etacut}},R^2\Bigr)\biggr]
\,.\end{align}
%%%
The soft-collinear function $\mathcal{S}_{i}$ encodes the actual step at $\etacut$ and is
defined by the measurement \eq{def_veto_step}.
For $\ptcuttwo=\ptcut$ there is no step in the jet veto and $\mathcal{S}_{i}$ has to vanish.
The RG consistency of the cross section implies that its $\mu$ anomalous dimension vanishes in general, while its resummed $\nu$ anomalous dimension is given by
%%%
\begin{align} \label{eq:gamma_nu_soft_collinear_step}
\gamma^i_{\nu,\mathcal{S}}(\ptcut, \ptcuttwo, R)
= 2\eta_\Gamma^i(\ptcut, \ptcuttwo) + \frac{1}{2}\Bigl\{ \gamma_\nu^i[\alpha_s(\ptcuttwo), R] - \gamma_\nu^i[\alpha_s(\ptcut), R] \Bigr\}
\,.\end{align}
%%%
It does not depend on $\mu$ at all, as required by exact path independence in the $(\mu, \nu)$ plane.
Note that the beam functions in \eq{fact_soft_collinear_step} depend on $\ptcuttwo$ (rather than $\ptcut$)
because collinear radiation is too forward to be constrained by the tighter central veto.
This is reflected in the somewhat curious rapidity anomalous dimension of $\mathcal{S}_{i}$ in \eq{gamma_nu_soft_collinear_step},
which accounts for the mismatch between the logarithms of $\ptcut$ and $\ptcuttwo$
generated by the soft and beam rapidity evolution, respectively.

Solving \eq{gamma_nu_soft_collinear_step} order by order in $\as$
we find the following very simple structure of the soft-collinear function through two loops:
%%%
\begin{align} \label{eq:S_step}
\mathcal{S}_{i} (\ptcut, \ptcuttwo, \etacut, R, \mu, \nu)
&= 1 + \frac{\alpha_s(\mu)}{4\pi} \Bigl[
   2\Gamma^i_0 \ln \rho \, L^\nu_\mathcal{S}
   + \mathcal{S}_{i,1}(\rho)
\Bigr]
\\
& \quad + \frac{\alpha_s^2(\mu)}{(4\pi)^2} \Bigl\{
   2 (\Gamma^i_0)^2 \! \ln^2 \!\! \rho \, (L^\nu_\mathcal{S})^2
   \!+\! 2 \ln \rho \, L^\nu_\mathcal{S} \bigl[ 2 L^\mu_\mathcal{S} \beta_0 \Gamma^i_0 \!+\! \Gamma^i_0 \mathcal{S}_{i,1}(\rho) \!+\! \Gamma^i_1 \bigr]
\nn \\
&\qquad\qquad\qquad
   + 2 \beta_0 L^\mu_\mathcal{S} \,\mathcal{S}_{i,1}(\rho)
   + \mathcal{S}_{i,2}(\rho, R)
\Bigr\} + \ord{\alpha_s^3}
\nn
\,,\end{align}
%%%
where
%%%
\begin{equation}
\rho \equiv \frac{\ptcuttwo}{\ptcut}
\,, \qquad
L^\nu_\mathcal{S} \equiv \ln \frac{\nu}{\sqrt{\ptcut \ptcuttwo} e^{\etacut}}
\,,\qquad
L^\mu_\mathcal{S} \equiv \ln \frac{\mu}{\sqrt{\ptcut \ptcuttwo}}
\,.\end{equation}
%%%
It is straightforward to check that the one-loop finite term vanishes (see \app{soft_coll}),
%%%
\begin{equation}
\mathcal{S}_{i,1} = 0
\,.\end{equation}
%%%
The two-loop finite term is a generic function of the dimensionless ratio $\rho$
and the jet radius parameter $R$, which must satisfy $\mathcal{S}_{i,2}(\rho = 1, R) = 0$.
As usual, we can decompose it according to its $R$ dependence as
%%%
\begin{equation}
\mathcal{S}_{i,2}(\rho, R) = - 8 C_i c_{ii}^R\, \ln \rho \ln R + \mathcal{S}_{i,2}^{(c)}(\rho) + \ord{R^2}
\,,\end{equation}
%%%
where $c_{ii}^R$ is given by \eq{clustering_coefficients_no_etacut} and $C_i = C_F\,(C_A)$ for $i = q\,(g)$.
The coefficient of $\ln R$ at this order is completely determined
by the $R$ dependence of the noncusp rapidity anomalous dimensions in \eq{gamma_nu_soft_collinear_step}.
The full two-loop finite term $\mathcal{S}_{i,2}(\rho, R)$ could readily be obtained
numerically using the methods of \refscite{Bell:2015lsf, Bell:2018jvf},
which would enable the full NNLL$'$ resummation.

This regime is again free of nonglobal logarithms
and hence can easily be applied to phenomenological studies.
It can be used to supplement the EFT setup from \sec{collinear_step},
which enables the resummation of logarithms of the ratio $\ptcut/Q\sim \ptcuttwo/Q$,
with an additional resummation of logarithms of the ratio
$\ptcut / Q e^{-\etacut} \sim \ptcuttwo / Q e^{-\etacut}$ by choosing the canonical scales
%%%
\begin{alignat}{3}
\mu_B &\sim \ptcuttwo \,,\quad
&\mu_\mathcal{S} &\sim \sqrt{\ptcut \ptcuttwo} \,,\quad
&\mu_S &\sim \ptcut
\,, \nn \\
\nu_B &\sim Q \,,\quad
&\nu_\mathcal{S} &\sim \sqrt{\ptcut \ptcuttwo} e^{\etacut} \,,\quad
&\nu_S &\sim \ptcut
\,.\end{alignat}
%%%
Here, the rapidity evolution between $\nu_\mathcal{S}$ and $\nu_S$ is responsible
for resumming the large logarithms of $e^{-\etacut} \sim \nu_S/\nu_\mathcal{S}$.

\paragraph{Numerical Validation.}

\begin{figure*}
\centering
\includegraphics[width=\WidthTwoSubfigs]{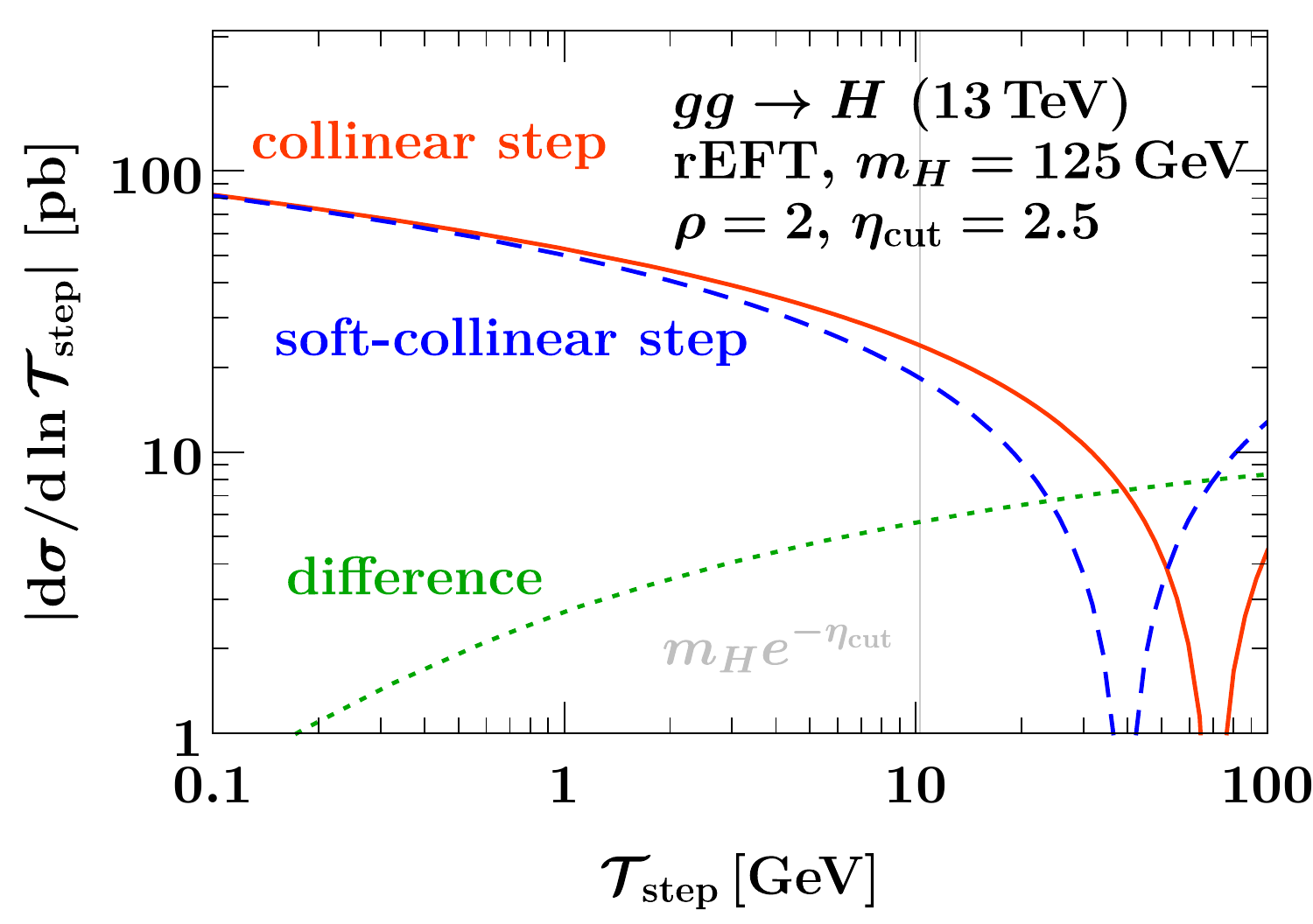}%
\hfill%
\includegraphics[width=\WidthTwoSubfigs]{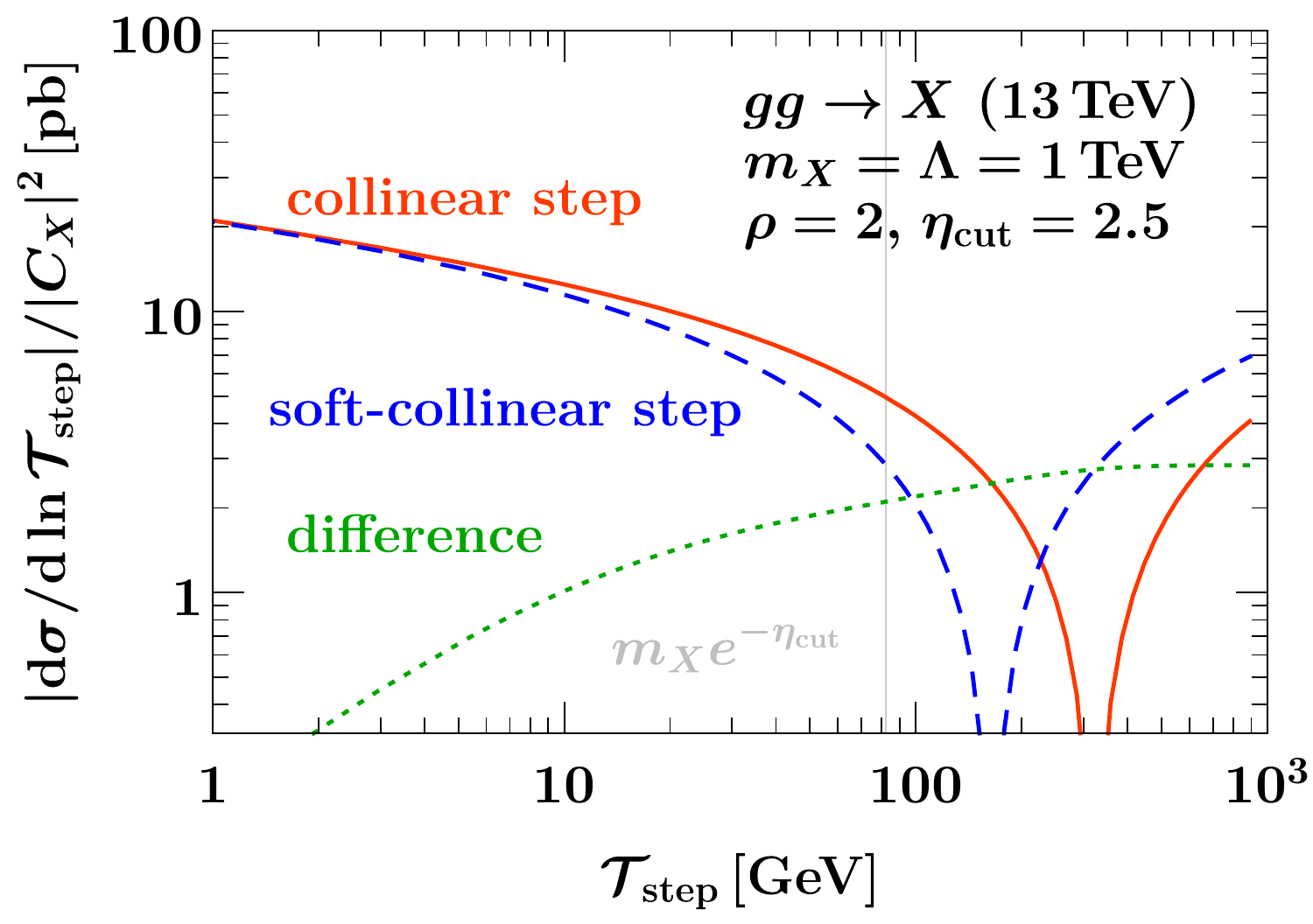}%
\\
\includegraphics[width=\WidthTwoSubfigs]{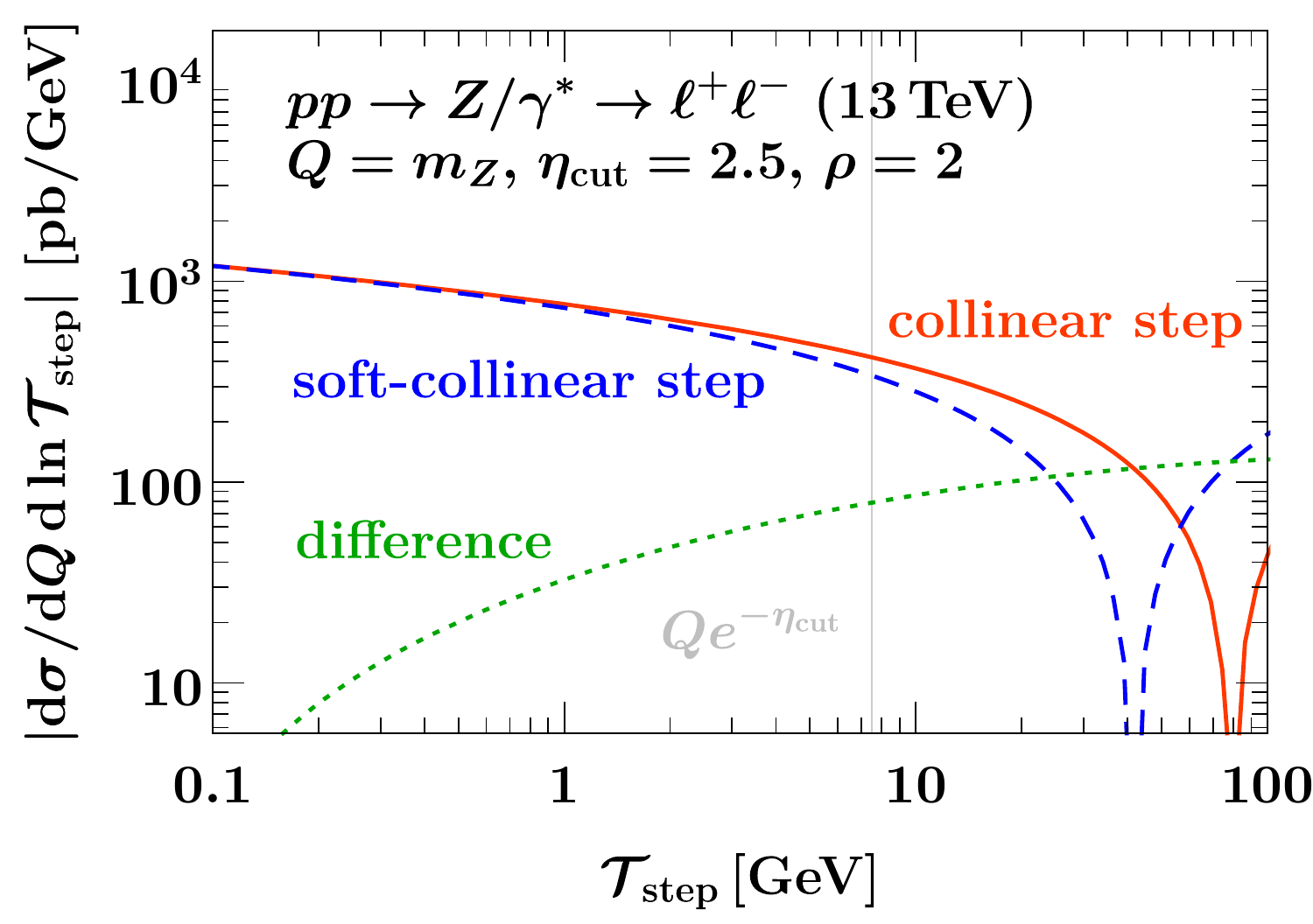}%
\hfill%
\includegraphics[width=\WidthTwoSubfigs]{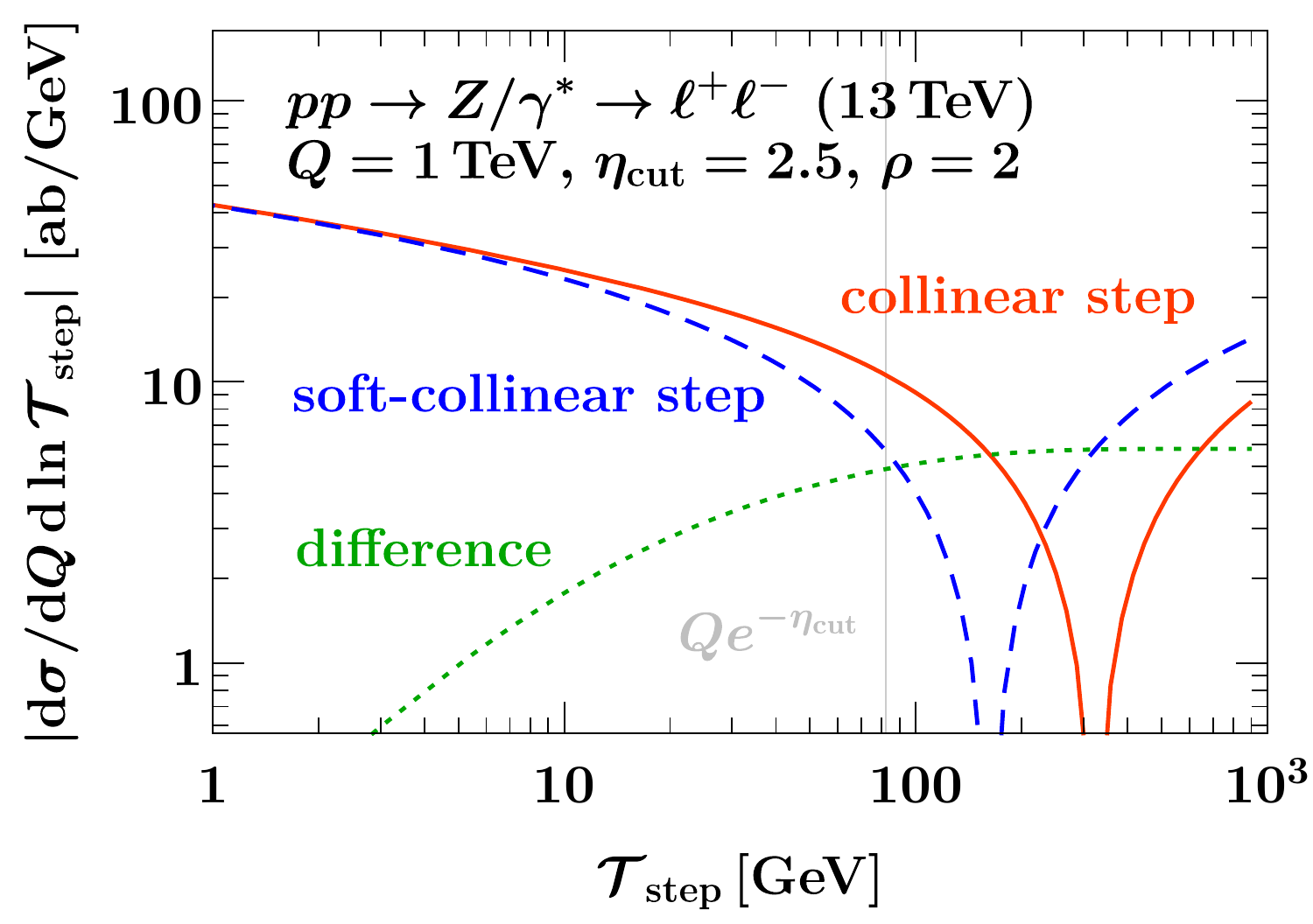}%
\caption{Comparison of the singular contributions to the fixed $\ord{\alpha_s}$ (LO$_1$)
$\Tau_\mathrm{step}$ spectrum for $\etacut = 2.5$ and $\rho = 2$ for $gg\to H$ (top left), $gg\to X$ (top right),
and Drell-Yan at $Q = m_Z$ (bottom left) and $Q = 1 \TeV$ (bottom right).
The solid orange lines show the singular spectrum for the collinear-step regime and the blue dashed lines
the further factorized result in the soft-collinear-step regime. Their difference, shown by
the dotted green lines vanishes as a power of $\Tau_\mathrm{step}$.
The vertical lines indicate where the parametric relation $\Tau_\mathrm{step}/Q = e^{-\etacut}$ is satisfied.}
\label{fig:regimes_step}
\end{figure*}

To validate our setup in this regime,
we exploit that \eq{fact_soft_collinear_step} provides a refactorization of the
collinear step in \eq{fact_collinear_step}, where
%%%
\begin{align}
\mathcal{I}_{ij}(\ptcut, \ptcuttwo, \etacut, R, \omega, z, \mu, \nu)
&= \mathcal{S}_i(\ptcut, \ptcuttwo, \etacut, R, \mu, \nu) \, \mathcal{I}_{ij}(\ptcut, R, \omega, z, \mu, \nu)
\nn \\ &\quad
\times \biggl[1+\mathcal{O}\Bigl(\frac{\ptcut}{\omega e^{-\etacut}},\frac{\ptcuttwo}{\omega e^{-\etacut}},R^2\Bigr)\biggr]
\,.\end{align}
%%%
In particular, \eq{fact_soft_collinear_step}
must reproduce \eq{fact_collinear_step}
up to power corrections in $\ptcut / Q e^{-\etacut}$ and $\ptcuttwo / Q e^{-\etacut}$.
We can test this numerically using the $\Tau_\mathrm{step}$ observable defined in \sec{collinear_step},
which simultaneously probes both classes of power corrections.
In \fig{regimes_step}, we show the fixed $\ord{\alpha_s}$ $\Tau_\mathrm{step}$ spectra
for the collinear step (solid orange) and soft-collinear step (dashed blue).
In all cases their difference (dotted green) vanishes like a power in $\Tau_\mathrm{step}$.

The additional resummation using the soft-collinear step
may be applicable up to values of $\ptcut = 20 \GeV$ ($\ptcut = 80 \GeV$) for $Q \sim 100 \GeV$ ($Q = 1 \TeV$),
for the choice of $\rho = 2, \etacut = 2.5$ displayed in \fig{regimes_step}.
This can be read off from the relative size of leading-power (soft-collinear step)
and subleading power (difference) contributions,
which leave some room where resummation in the leading-power cross section can improve the prediction.
We find a slightly larger potential resummation region than for the analogous refactorization
in the $\ptcuttwo = \infty$ case, where an earlier onset of the power corrections
was observed in \fig{regimes_2_3}.

%===============================================================================
\subsection[\texorpdfstring{$\ptcut/Q \ll \ptcuttwo/Q \ll e^{-\etacut}$}{pTcut/Q << tildepTcut/Q << exp(-etacut)} (soft-collinear NGLs)]
{\boldmath $\ptcut/Q \ll \ptcuttwo/Q \ll e^{-\etacut}$ (soft-collinear NGLs)}
\label{sec:soft_collinear_ngls}
%===============================================================================

For this hierarchy (bottom right panel of \fig{extend_by_step}), two types of soft-collinear modes arise,
%%%
\begin{align}
n_a\text{-soft-collinear ($\ptcut$):}
&\quad
p^\mu\sim  \ptcut (e^{-\etacut},e^{\etacut},1)
\,, \nn \\
n_a\text{-soft-collinear ($\ptcuttwo$):}
&\quad
p^\mu\sim  \ptcuttwo (e^{-\etacut},e^{\etacut},1)
\,,
\end{align}
%%%
and analogously for the $n_b$-soft-collinear sectors,
which are both parametrically distinct from the energetic collinear modes.
Compared to the regime $\ptcut \sim \ptcuttwo \ll Q e^{-\etacut}$ there are now parametrically
large logarithms $\ln(\ptcut/\ptcuttwo)$ in the soft-collinear function $\mathcal{S}_{i}$ in \eq{fact_soft_collinear_step}.
The cross section can be written as in \eq{fact_soft_collinear_step},
where the soft-collinear function is refactorized as
%%%
\begin{align}\label{eq:softcoll_refact}
\mathcal{S}_{i}(\ptcut, \ptcuttwo, \etacut, R, \mu, \nu)
 &= \mathcal{S}^{(\rm cut)}_{i}(\ptcut, \etacut, R, \mu, \nu) \,\Bigl[\mathcal{S}^{(\rm cut)}_{i}(\ptcuttwo, \etacut, R, \mu, \nu)\Bigr]^{-1}  \nn \\
 & \quad \times  \biggl[1+ \mathcal{S}^{\rm (NG)}_{i}\Bigl(\frac{\ptcut}{\ptcuttwo},R\Bigr)\biggr]
         \times  \biggl[1+ \ORd{\frac{\ptcut}{\ptcuttwo}}\biggr]
\,,\end{align}
%%%
with $\mathcal{S}^{(\rm cut)}_{i}$ the same soft-collinear function as in
\eqs{refactB}{beam_refactorization2}. Both the power corrections and the
nonglobal piece $\mathcal{S}_i^{(\rm NG)}$ are absent at one loop and at
$\ord{\alpha_s^2 \ln R}$. Equivalently this regime can be interpreted as a
refactorization of \eq{beam_refactorization2}, where compared to the hierarchy
for $\ptcut \ll \ptcuttwo \sim Q e^{-\etacut}$ there are large (rapidity)
logarithms $\ln(\ptcuttwo e^{\etacut}/Q)$ in the beam function $B^{(\rm
cut)}_{i}$. Evolving the two soft-collinear functions to separate
renormalization scales $\mu_{\mathcal{S},1}=\ptcut$, $\nu_{\mathcal{S},1}=\ptcut
e^{\etacut}$ and $\mu_{\mathcal{S},2}=\ptcuttwo$, $\nu_{\mathcal{S},2}=\ptcuttwo
e^{\etacut}$ resums Sudakov logarithms of $\ptcut/\ptcuttwo$, but does not
account for the nonglobal logarithms of the same ratio in $\mathcal{S}^{\rm
(NG)}_{i}$.

%%%%%%%%%%%%%%%%%%%%%%%%%%%%%%%%%%%%%%%%%%%%%%%%%%%%%%%%%%%%%%%%%%%%%%%%%%%%%%%%
\section{Numerical results}
\label{sec:analysis}
%%%%%%%%%%%%%%%%%%%%%%%%%%%%%%%%%%%%%%%%%%%%%%%%%%%%%%%%%%%%%%%%%%%%%%%%%%%%%%%%

In \sec{no_step} we discussed in detail how to incorporate the jet rapidity cut
into the resummed 0-jet cross section. In particular, in the regime $\ptcut / Q \sim e^{-\etacut}$ (regime~2),
the dependence on $\etacut$ is incorporated into the resummation via the RG evolution of the
$\etacut$ dependent beam functions. In this section, we illustrate these results
by presenting numerical predictions for the resummed cross section at NLL$'+$NLO.

In \sec{matching_and_uncertainties}, we outline how the resummed results are
combined with the full QCD results, as well as
our estimation of perturbative uncertainties.
In \sec{compare_approaches}, we assess the impact of the additional perturbative ingredients
by comparing the different treatments of $\etacut$.
In \sec{resummed_predictions}, we show the predictions for selected
$\etacut$ as a function of $\ptcut$.

In the following, we consider the four cases
of gluon-fusion Higgs production $gg\to H$ at $m_H = 125\GeV$,
gluon fusion to a generic heavy scalar $gg\to X$ with $m_X = 1\TeV$,
and Drell-Yan production at $Q = m_Z$ and $Q = 1\TeV$, with the same setup
and inputs as described in \sec{no_step_reg2}.
The numerical results for the resummed predictions for all processes are obtained from our
implementation in \texttt{SCETlib}~\cite{scetlib}. The NLO results in full QCD
are obtained from \texttt{MCFM~8.0}~\cite{Campbell:1999ah,Campbell:2011bn,Campbell:2015qma}.

%===============================================================================
\subsection{Fixed-order matching and perturbative uncertainties}
\label{sec:matching_and_uncertainties}
%===============================================================================

The resummed cross section obtained from \eq{fact_pTjet2} describes the 0-jet cross section up
to power corrections in $\ptcut/Q$, which become relevant when $\ptcut \sim Q$.
We account for them by the usual additive matching,
%%%
\begin{equation} \label{eq:matching}
\sigma_0(\ptcut, \etacut) = \sigma_0^\mathrm{res}(\ptcut, \etacut) + \bigl[\sigma_0^\mathrm{FO}(\ptcut, \etacut) - \sigma_0^\mathrm{sing}(\ptcut, \etacut)\bigr]
\,.\end{equation}
%%%
Here, $\sigma_0^\mathrm{res}$ is the resummed singular cross section obtained from \eq{fact_pTjet2},
$\sigma_0^\mathrm{sing}$ is its fixed-order expansion, and
$\sigma_0^\mathrm{FO}$ is the fixed-order result in full QCD.
By construction, the difference in square brackets is nonsingular and
vanishes as $\ptcut \to 0, \etacut \to \infty$ and can therefore be included at
fixed order even at small $\ptcut$. The dominant corrections at small $\ptcut$
are resummed in $\sigma_0^{\rm res}$. At large $\ptcut$, fixed-order perturbation theory is
the appropriate description, so \eq{matching} should recover $\sigma_0^{\rm FO}$.
This is achieved by turning off the resummation in $\sigma_0^{\rm res}$ as a function
of $\ptcut$, and by constructing $\sigma_0^{\rm res}$ such that
it precisely reproduces $\sigma^\mathrm{sing}$ when the resummation is fully
turned off.

To smoothly turn off the resummation as we approach $\ptcut \to Q$,
we use profile scales~\cite{Ligeti:2008ac, Abbate:2010xh}, following the setup developed
in \refcite{Stewart:2013faa}.
We stress that the profile scales for regime~2 are in one-to-one correspondence
with the standard treatment in regime~1, since both regimes have the same RG structure.
Similarly, our treatment of perturbative uncertainties is based on profile scale variations
following \refcite{Stewart:2013faa}.
We distinguish an overall yield uncertainty $\Delta_{\mu 0}$,
which is determined by a collective variation of all scales up and down,
and a resummation (jet bin migration) uncertainty $\Delta_\mathrm{resum}$
from varying individual scales in the beam and soft functions.
For the gluon-induced processes, we follow \refcite{Ebert:2017uel} and include an
additional uncertainty $\Delta_\varphi$ from varying the complex phase of the hard scale,
which was not considered in \refcite{Stewart:2013faa}.
The total uncertainty is then obtained by considering the
different uncertainty sources as independent, and hence uncorrelated, and
adding them in quadrature,
%%%
\begin{equation}
\Delta_\mathrm{total}
= \Delta_{\mu 0} \oplus \Delta_\varphi \oplus \Delta_\mathrm{resum}
\equiv \bigl( \Delta_{\mu 0}^2 + \Delta_\varphi^2 + \Delta_\mathrm{resum}^2 \bigr)^{1/2}
\,.\end{equation}
%%%

%===============================================================================
\subsection{Comparing different treatments of the jet rapidity cut}
\label{sec:compare_approaches}
%===============================================================================

\begin{figure*}
\centering
\includegraphics[width=\WidthTwoSubfigs]{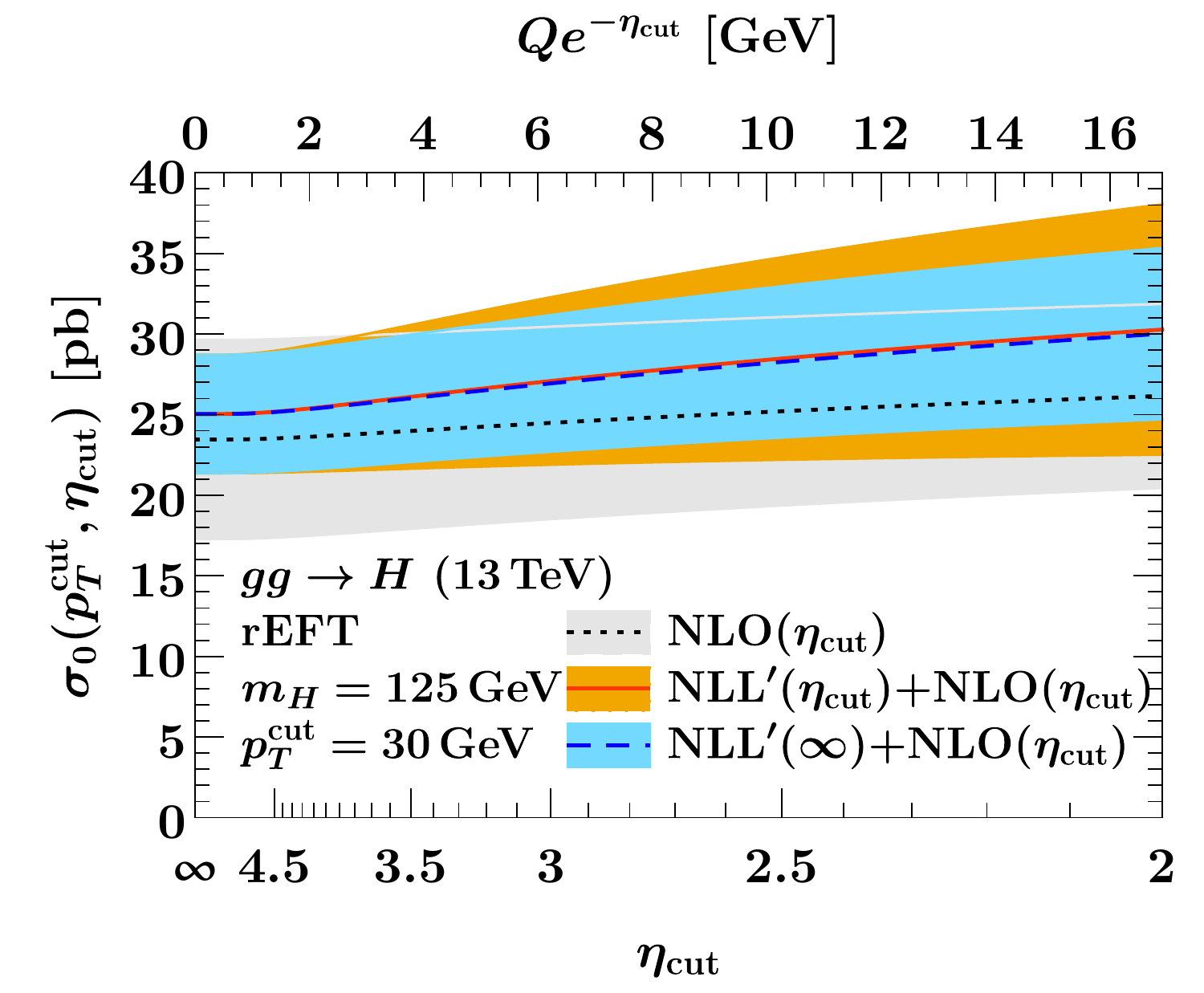}%
\hfill%
\includegraphics[width=\WidthTwoSubfigs]{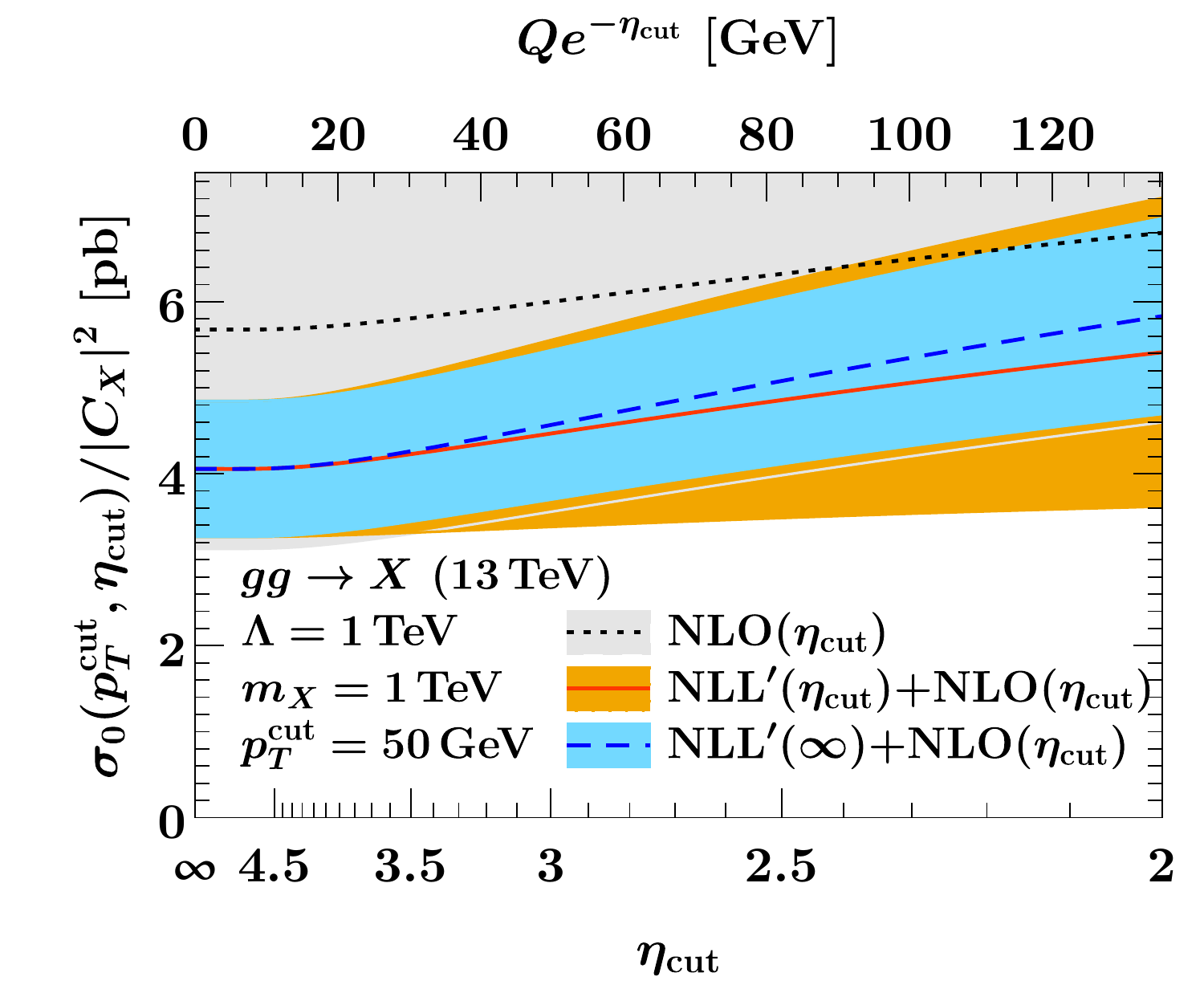}%
\caption{The 0-jet cross section for $gg\to H$ at $m_H = 125\GeV$ for $\ptcut = 30\GeV$ (left) and
$gg\to X$ at $m_X = 1\TeV$ and $\ptcut = 50\GeV$ (right) as a function of $\etacut$.
The same observable ($\sigma_0$) is calculated in three different ways, shown by the different bands, as described in the text.}
\label{fig:compare_approaches_ggfus}
\end{figure*}

It is interesting to consider the impact of the additional perturbative ingredients
in the $\etacut$ dependent beam function on the prediction,
e.g.\ compared to treating the rapidity cut effects purely at fixed order.
In \figs{compare_approaches_ggfus}{compare_approaches_DY},
we plot the results for fixed $\ptcut$ as a function of $\etacut$ starting
at $\etacut=\infty$ on the left and decreasing toward the right.
The corresponding values of the $Q e^{-\etacut}$ scale are shown
at the top.

Our result for the 0-jet cross section using the matching in \eq{matching}
is shown as orange bands.
We refer to this prediction as NLL$'(\etacut)+$NLO$(\etacut)$,
because both the NLL$'$ resummed singular cross section
and the fixed-order matching are exact in $\etacut$.
To highlight the effect of the additional $\etacut$ dependence in the regime~2
beam function, we consider two more alternative treatments of $\etacut$.
For the regime~1 result, shown by the blue bands and denoted by NLL$'(\infty)+$NLO$(\etacut)$,
the $\etacut$ dependence in the resummed cross section is dropped,
%%%
\begin{alignat}{3} \label{eq:modified_matching_nll_infty}
\sigma_0(\ptcut, \etacut)
&= \sigma_0^\mathrm{res}(\ptcut, \infty)
&&+ \bigl[\sigma_0^\mathrm{FO}(\ptcut, \etacut) - \sigma_0^\mathrm{sing}(\ptcut, \infty)\bigr]
\,.\end{alignat}
%%%
The resummation then only acts on the singular cross section for $\etacut = \infty$,
while all $\etacut$ effects are included purely at fixed order via the
matching term in square brackets.
Note that the matching term is now no longer nonsingular,
i.e., it no longer vanishes like a power in $\ptcut$ as $\ptcut \to 0$,
as we saw in \figs{sing_nons_ggHX}{sing_nons_DY_narrow_width}.
The plain fixed-order calculation without any resummation,
%%%
\begin{alignat}{3} \label{eq:modified_matching_plain_fo}
\sigma_0(\ptcut, \etacut)
&= \sigma_0^\mathrm{FO}(\ptcut, \etacut)
\,,\end{alignat}
%%%,
is denoted by NLO$(\etacut)$ and shown by the gray bands. In this case, the uncertainties are evaluated
using the ST procedure~\cite{Stewart:2011cf}.

We first consider gluon-fusion Higgs production shown in the left panel of
\fig{compare_approaches_ggfus}, where we set $\ptcut = 30 \GeV$.
The NLO$(\etacut)$ prediction (gray band) exhibits a slight, physical rise in the
cross section as $\etacut$ decreases towards the right.
This is not surprising as at fixed order, decreasing $\etacut$ simply amounts
to accumulating the squared LO$_1$ matrix element over a larger part of phase space.
The rise is less pronounced than for the resummed results (orange and blue bands),
but still compatible with them within each others' uncertainties.
Comparing NLL$'(\etacut)+$NLO$(\etacut)$ (orange) to NLL$'(\infty)+$NLO$(\etacut)$ (blue)
we find that the additional tower of logarithms predicted by NLL$'(\etacut)$
on top of the fixed NLO $\etacut$ dependence barely affects the central value
of the prediction down to $\etacut = 2$. This is perhaps not surprising since
$Q e^{-\etacut}$ is at most half of $\ptcut$, which means we are not far from
regime~1. However, we do observe a noticeable increase in the perturbative uncertainty estimate.
This is mainly due to the resummation uncertainty, which is reasonable: $\Delta_\mathrm{resum}$
probes the unknown higher-order finite terms (the RGE boundary condition)
and is therefore sensitive to a change of the beam function boundary condition
by the $\etacut$ correction $\Delta I_{ij}^{(1)}$ (see \sec{no_step_reg2}).
On the other hand, $\Delta I_{ij}^{(1)}$ must be large enough to accommodate
--- up to power corrections ---
the fixed-order difference to $\etacut = \infty$
(roughly $2 \pb$ at $\etacut = 2.5$, as can be read off from the gray line),
so we expect an impact on $\Delta_\mathrm{resum}$ of similar size.
Hence, the conclusion is not that the NLL$'(\infty)+$NLO$(\etacut)$ result
is more precise, but rather that its uncertainty is potentially
underestimated because it cannot capture the $\etacut$ dependence.

In the right panel of \fig{compare_approaches_ggfus}, we show the same results
for a hypothetical color-singlet scalar resonance $gg\to X$ at $m_X = 1 \TeV$ using
$\ptcut = 50\GeV$.
[The dimension-five operator mediating the production of $X$ is given in \eq{L_eff_ggX}.]
The NLO$(\etacut)$ result (gray)
is now off by a large amount already at $\etacut = \infty$,
where it is not covered by the resummed predictions.
This is expected because the high production energy of $1 \TeV$
implies we are deep in the resummation region, even for the larger value of
$\ptcut = 50 \GeV$. The central values of the two resummed treatments start to differ below $\etacut = 3$
or above $Qe^{-\etacut} \simeq 50\GeV$, where we are now fully in regime~2.
However, the main difference is
again the larger and likely more reliable uncertainty estimate in the NLL$'(\etacut)$ prediction.

\begin{figure*}
\centering
\includegraphics[width=\WidthTwoSubfigs]{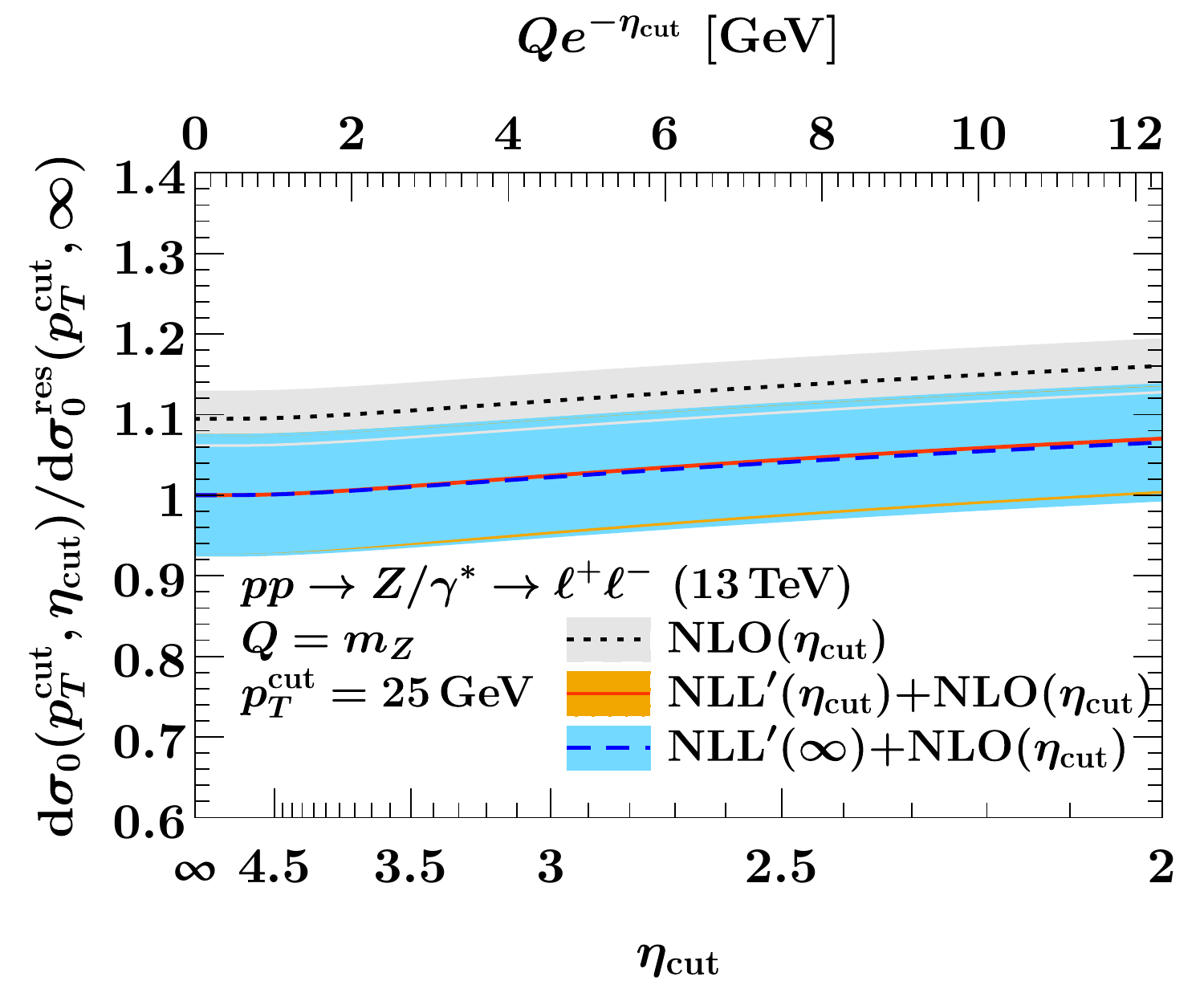}%
\hfill%
\includegraphics[width=\WidthTwoSubfigs]{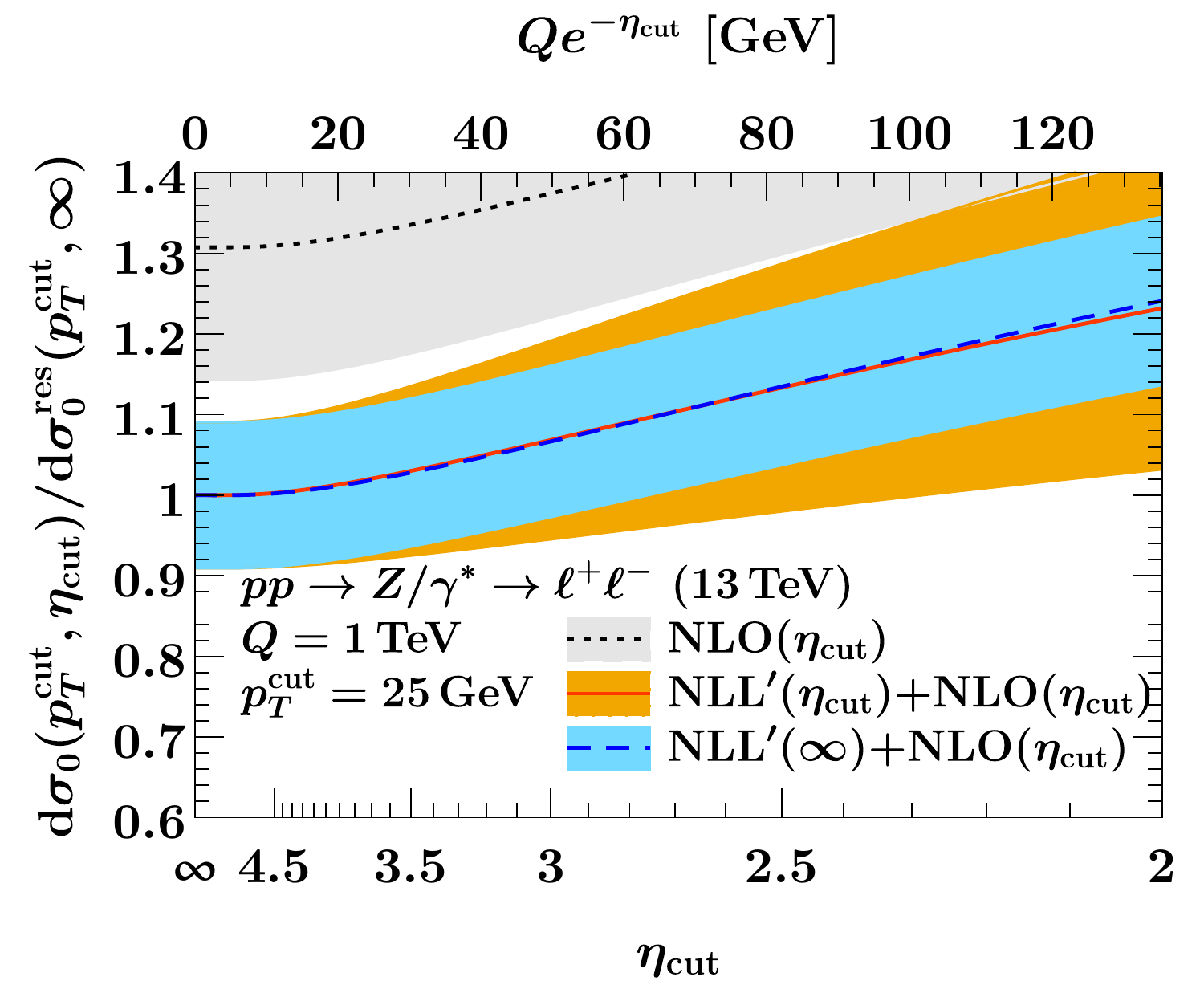}%
\caption{The 0-jet cross section for Drell-Yan at $Q = m_Z$ and $\ptcut = 20\GeV$ (left)
and $Q = 1\TeV$ and $\ptcut= 25\GeV$ (right) as a function of $\etacut$.
The same observable ($\sigma_0$) is calculated in three different ways, shown by the different bands, as described in the text.
For better readability, all results are normalized to the resummed central value at $\etacut = \infty$.}
\label{fig:compare_approaches_DY}
\end{figure*}

In \fig{compare_approaches_DY} we show the analogous results for Drell-Yan production
at $Q = m_Z$ using $\ptcut = 20 \GeV$ (left panel) and $Q = 1\TeV$ using $\ptcut = 25\GeV$ (right panel).
For better readability, these results are normalized to the resummed 0-jet cross section at $\etacut = \infty$.
While all predictions agree in the slope of the cross section with respect to $\etacut$,
the NLO$(\etacut)$ result has a constant offset and an unrealistically small uncertainty estimate.
At the lower $Q \sim 100 \GeV$, we find practically no difference between the NLL$'(\etacut)$ and NLL$'(\infty)$ calculations, so here
the effects of the jet rapidity cut can safely be included via the fixed-order matching
corrections to the regime~1 resummation.
At higher production energies, the intrinsic NLL$'(\etacut)$ ingredients
become more relevant, similar to gluon-fusion, as shown by the increasing
uncertainty estimates as $\etacut$ decreases. Note that below $\etacut = 2.5$,
$Q e^{-\etacut} \gtrsim 80\GeV$ becomes large compared to this choice of $\ptcut = 25\GeV$,
so resumming logarithms of $\ptcut/(Q e^{-\etacut})$ using the regime~3
factorization given in \sec{no_step_reg3} might help reduce the uncertainties.

%===============================================================================
\subsection{Resummed predictions with a sharp rapidity cut}
\label{sec:resummed_predictions}
%===============================================================================

Here, we compare predictions for different values of $\etacut$ as a function
of $\ptcut$.
Our working order is NLL$'(\etacut)+$NLO$(\etacut)$ in the notation of the previous section,
which from now on we simply refer to as NLL$'+$NLO,
i.e., the $\etacut$ dependence is always included in the resummation.
We stress that the differences we observe between predictions in this subsection
are physical differences due to the different jet rapidity cuts,
and \emph{not} due to different theoretical treatments as in the previous
subsection.

\begin{figure*}
\centering
\includegraphics[width=\WidthTwoSubfigs]{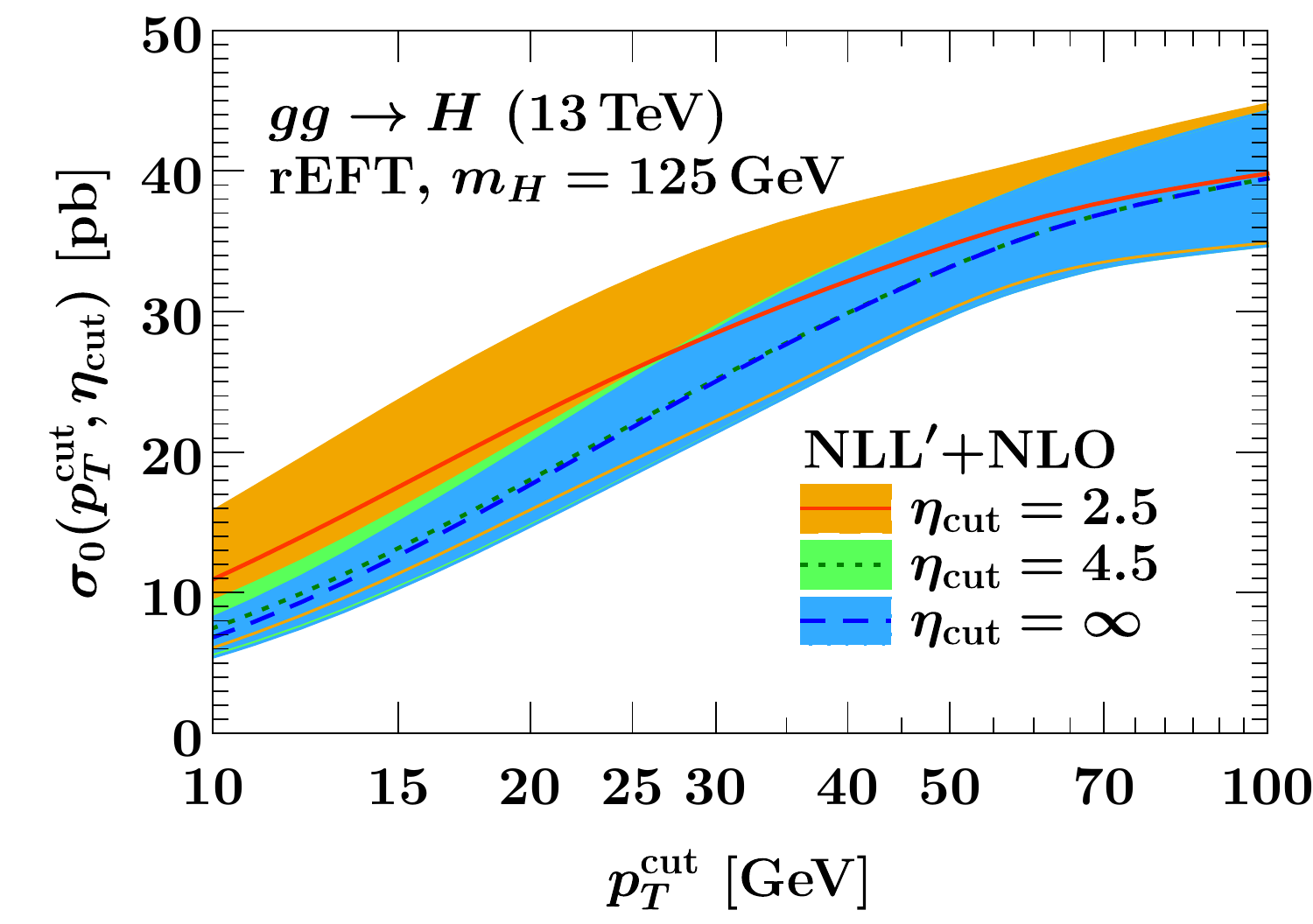}%
\hfill%
\includegraphics[width=\WidthTwoSubfigs]{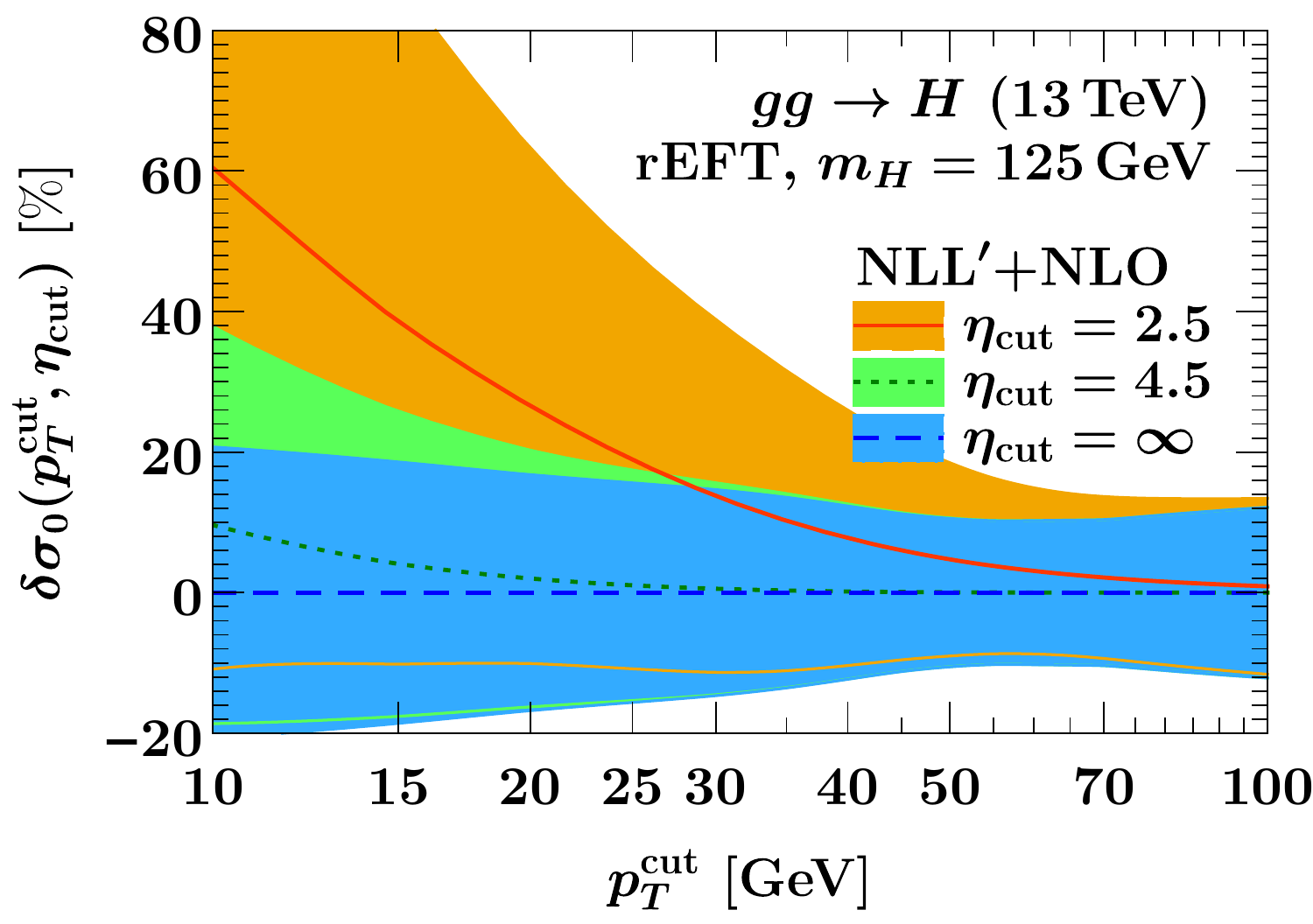}%
\caption{0-jet cross section $\sigma_0(\ptcut, \etacut)$ for $gg\to H$
for $m_H = 125\GeV$ at NLL$'$+NLO for different values of $\etacut$.
The bands indicate the total uncertainty $\Delta_{\mu 0} \oplus \Delta_\varphi \oplus \Delta_\res$.
The absolute cross section is shown on the left. On the right, the same results are shown
as the percent difference relative to the 0-jet cross section at $\etacut = \infty$.}
\label{fig:resummed_ggH}
\end{figure*}

\begin{table}
\centering
\begin{tabular}{l|ll}
\hline\hline
\multicolumn{3}{c}{$\sigma_0(\ptcut, \etacut)~[\!\pb],\, gg \to H\,(13 \TeV),\, \text{rEFT},\, m_H = 125 \GeV$}
\\
\hline\hline
$\etacut$
& \multicolumn{1}{c}{$\ptcut = 25 \GeV$}
& \multicolumn{1}{c}{$\ptcut = 30 \GeV$}
\\
\hline
$2.5$
& $25.9\!\pm\! 3.8_{\mu 0}\!\pm\! 1.5_\varphi\!\pm\! 5.0_\mathrm{res}\, (25.0\%)$
& $28.5\!\pm\! 4.0_{\mu 0}\!\pm\! 1.6_\varphi\!\pm\! 4.6_\mathrm{res}\, (22.0\%)$
\\
$4.5$
& $22.0\!\pm\! 2.0_{\mu 0}\!\pm\! 1.0_\varphi\!\pm\! 2.8_\mathrm{res}\, (16.2\%)$
& $25.2\!\pm\! 2.2_{\mu 0}\!\pm\! 1.2_\varphi\!\pm\! 2.8_\mathrm{res}\, (15.0\%)$
\\
$\infty$
& $21.8\!\pm\! 1.9_{\mu 0}\!\pm\! 1.0_\varphi\!\pm\! 2.7_\mathrm{res}\, (15.6\%)$
& $25.0\!\pm\! 2.2_{\mu 0}\!\pm\! 1.2_\varphi\!\pm\! 2.7_\mathrm{res}\, (14.7\%)$
\\
\hline\hline
\end{tabular}%
\caption{
0-jet cross section for $gg\to H$ for $m_H = 125\GeV$ at NLL$'$+NLO for different
values of $\ptcut$ and $\etacut$ with a breakdown of the uncertainties.}
\label{tab:resummed_ggH}
\end{table}

In \fig{resummed_ggH} and \tab{resummed_ggH} we present results for $gg\to H$.
Going from $\etacut = \infty$ to $\etacut = 4.5$ we find a $1\%$ increase of the cross section
for the typical values of $\ptcut = 25 \GeV$ and $30 \GeV$.
At $\etacut = 2.5$ the increase becomes more sizable, $14 \%$ ($19 \%$)
for $\ptcut = 30 \GeV$ ($25 \GeV$).
The differences vanish as the cross section saturates around $\ptcut \sim 100 \GeV$.

\begin{figure*}
\centering
\includegraphics[width=\WidthTwoSubfigs]{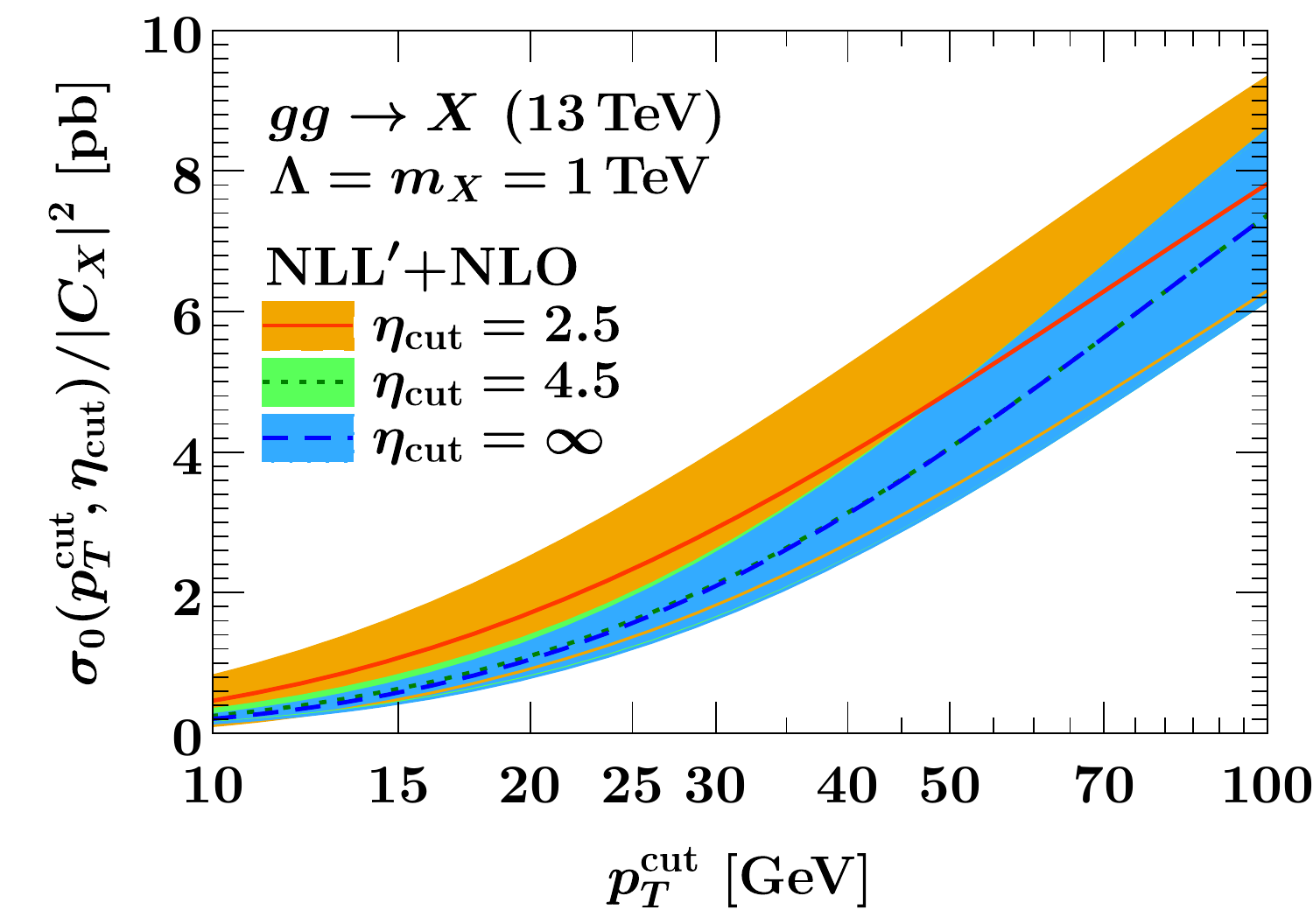}%
\hfill%
\includegraphics[width=\WidthTwoSubfigs]{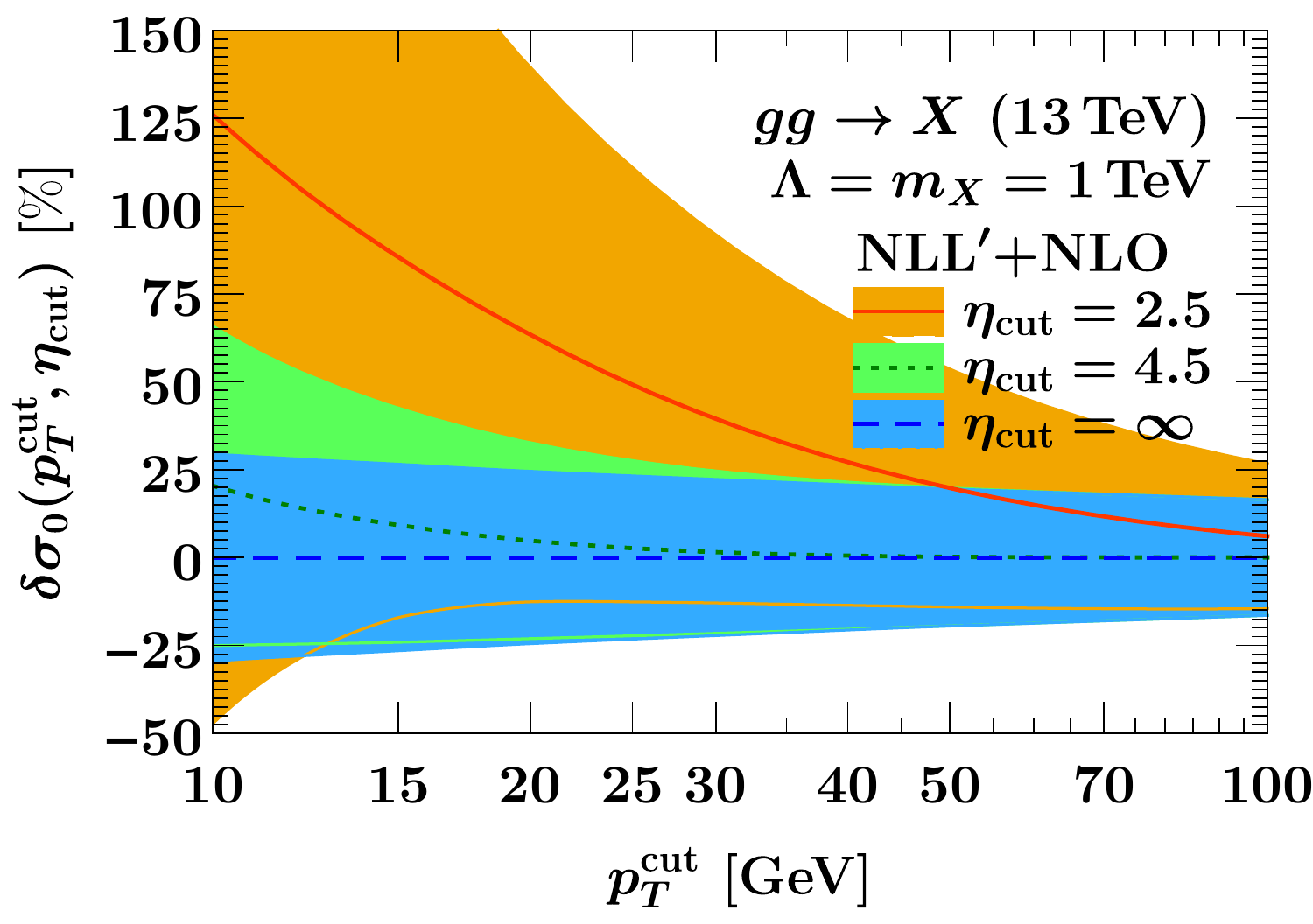}%
\caption{0-jet cross section $\sigma_0(\ptcut, \etacut)$ for $gg\to X$ for $m_X = 1\TeV$
at NLL$'$+NLO for different values of $\etacut$.
The bands indicate the total uncertainty $\Delta_{\mu 0} \oplus \Delta_\varphi \oplus \Delta_\res$.
The absolute cross section is shown on the left. On the right, the same results are shown
as the percent difference relative to the 0-jet cross section at $\etacut = \infty$.}
\label{fig:resummed_ggX}
\end{figure*}

\begin{table}
\centering
\begin{tabular}{l|ll}
\hline\hline
\multicolumn{3}{c}{$\sigma_0(\ptcut, \etacut)/\abs{C_X}^2~[\!\pb],\, gg \to X\,(13 \TeV),\, \Lambda = m_X = 1 \TeV$}
\\
\hline\hline
$\etacut$
& \multicolumn{1}{c}{$\ptcut = 50 \GeV$}
& \multicolumn{1}{c}{$\ptcut = 100 \GeV$}
\\
\hline
$2.5$
& $4.9\!\pm\! 0.7_{\mu 0}\!\pm\! 0.1_\varphi\!\pm\! 1.2_\mathrm{res}\, (28.3\%)$
& $7.8\!\pm\! 0.8_{\mu 0}\!\pm\! 0.1_\varphi\!\pm\! 1.3_\mathrm{res}\, (19.4\%)$
\\
$4.5$
& $4.1\!\pm\! 0.3_{\mu 0}\!\pm\! 0.1_\varphi\!\pm\! 0.7_\mathrm{res}\, (19.6\%)$
& $7.4\!\pm\! 0.6_{\mu 0}\!\pm\! 0.1_\varphi\!\pm\! 1.1_\mathrm{res}\, (16.4\%)$
\\
$\infty$
& $4.1\!\pm\! 0.3_{\mu 0}\!\pm\! 0.1_\varphi\!\pm\! 0.7_\mathrm{res}\, (19.5\%)$
& $7.4\!\pm\! 0.6_{\mu 0}\!\pm\! 0.1_\varphi\!\pm\! 1.1_\mathrm{res}\, (16.4\%)$
\\
\hline\hline
\end{tabular}%
\caption{
0-jet cross section for $gg\to X$ for $m_X = 1\TeV$ at NLL$'$+NLO for different
values of $\ptcut$ and $\etacut$ with a breakdown of the uncertainties.}
\label{tab:resummed_ggX}
\end{table}

The analogous results for $gg\to X$ for $m_X = 1\TeV$
are shown in \fig{resummed_ggX} and \tab{resummed_ggX}.
At such a high hard scale, the uncertainties for $\etacut = 2.5$ become
essentially beyond control for very tight vetoes $\ptcut \lesssim 25 \GeV$,
which would make an additional resummation of $\ln \ptcut/(Q e^{-\etacut})$
as outlined in \sec{no_step_reg3} necessary. As we will see in the next subsection,
this effect can be tamed by replacing the sharp rapidity cut by a step in the
jet veto. However, for any choice of $\etacut$ the cross section is very strongly
Sudakov suppressed for such small values of $\ptcut$. At more realistic values
of the veto, the jet rapidity cut for $\etacut = 2.5$
compared to $\etacut = \infty$ still leads to a sizable increase of
$20 \%$ ($5\%$) for $\ptcut = 50 \GeV$ ($\ptcut = 100\GeV$). In contrast,
the effect for $\etacut = 4.5$ is very small.

\begin{figure*}
\centering
\includegraphics[width=\WidthTwoSubfigs]{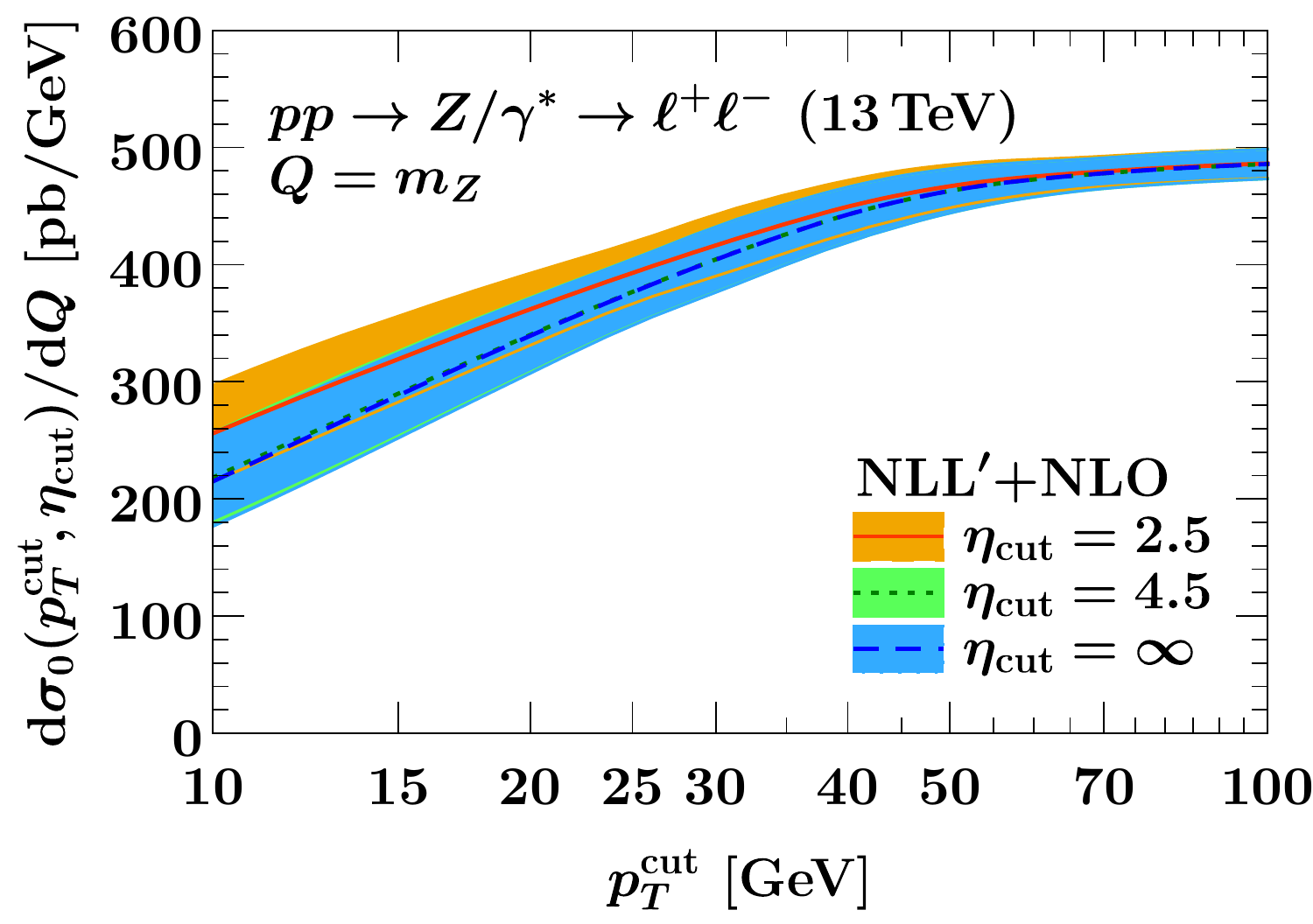}%
\hfill%
\includegraphics[width=\WidthTwoSubfigs]{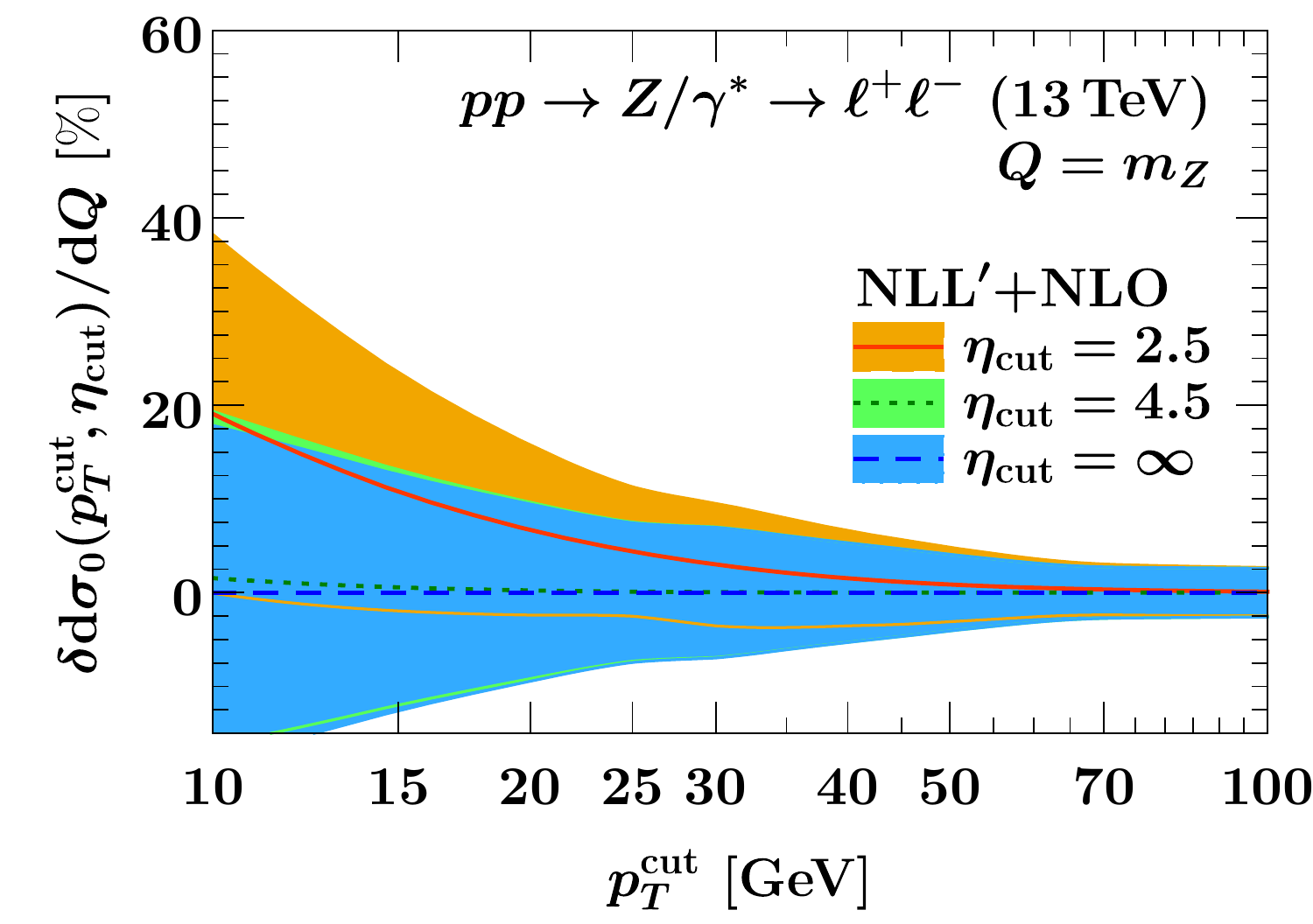}%
\\
\includegraphics[width=\WidthTwoSubfigs]{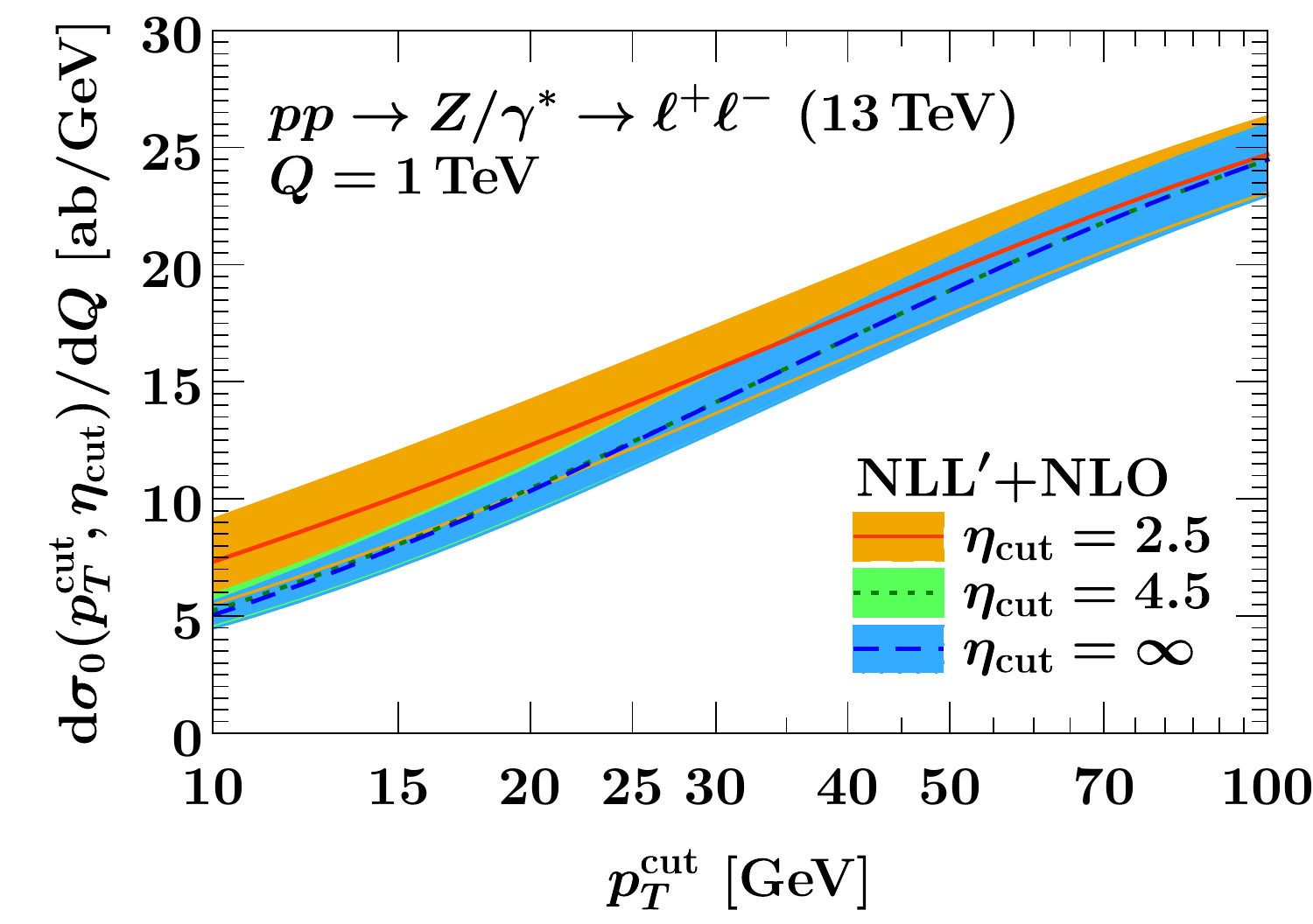}%
\hfill%
\includegraphics[width=\WidthTwoSubfigs]{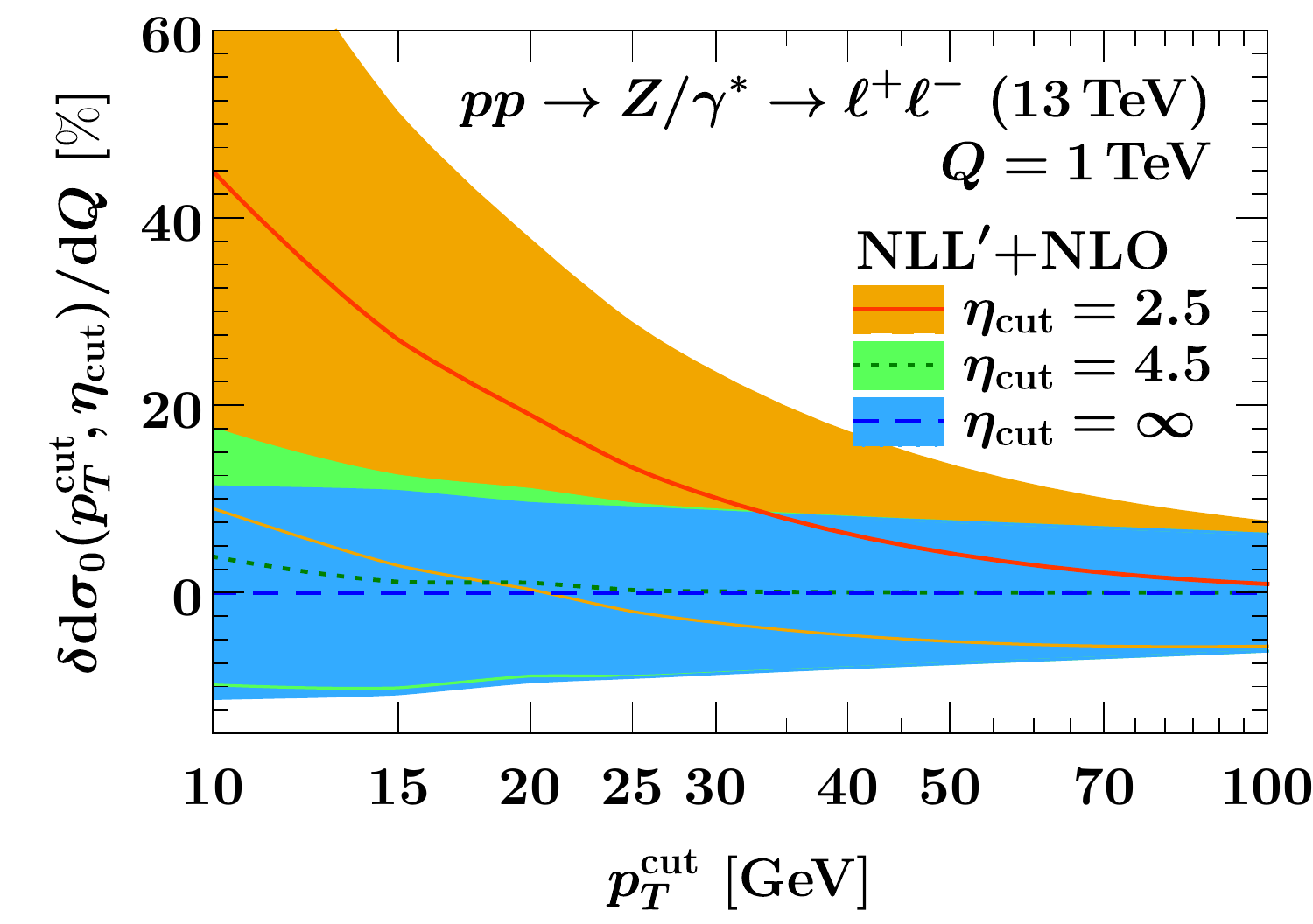}%
\caption{The 0-jet cross section $\df \sigma_0(\ptcut,\etacut) / \df Q$ for
Drell-Yan production at the $Z$ pole $Q = m_Z$ (top row) and at $Q = 1 \TeV$
(bottom row) at NLL$'$+NLO for different values of $\etacut$.
The bands indicate the total uncertainty $\Delta_{\mu 0} \oplus \Delta_\res$.
The absolute cross section is shown on the left. On the right, the same results are shown
as the percent difference relative to the 0-jet cross section at $\etacut = \infty$.}
\label{fig:resummed_DY}
\end{figure*}

\begin{table}
\centering
\begin{tabular}{c|cc}
\hline\hline
\multicolumn{3}{c}{$\df \sigma_0(\ptcut, \etacut)/\df Q~[\!\pb/\!\GeV],\, pp \to Z / \gamma^\ast \to \ell^+ \ell^- \,(13 \TeV),\, Q = m_Z$}
\\
\hline\hline
$\etacut$
& \multicolumn{1}{c}{$\ptcut = 20 \GeV$}
& \multicolumn{1}{c}{$\ptcut = 25 \GeV$}
\\
\hline
$2.5$
& $362\!\pm\! 22_{\mu 0}\!\pm\! 21_\mathrm{res}\, (8.5\%)$
& $393\!\pm\! 22_{\mu 0}\!\pm\! 14_\mathrm{res}\, (6.6\%)$
\\
$4.5$
& $340\!\pm\! 24_{\mu 0}\!\pm\! 22_\mathrm{res}\, (9.4\%)$
& $377\!\pm\! 24_{\mu 0}\!\pm\! 15_\mathrm{res}\, (7.4\%)$
\\
$\infty$
& $339\!\pm\! 24_{\mu 0}\!\pm\! 22_\mathrm{res}\, (9.5\%)$
& $376\!\pm\! 24_{\mu 0}\!\pm\! 15_\mathrm{res}\, (7.4\%)$
\\
\hline\hline
\end{tabular}%
\\[1em]
\begin{tabular}{c|cc}
\hline\hline
\multicolumn{3}{c}{$\df \sigma_0(\ptcut, \etacut)/\df Q~[\!\ab/\!\GeV],\, pp \to Z / \gamma^\ast \to \ell^+ \ell^- \,(13 \TeV),\, Q = 1 \TeV$}
\\
\hline\hline
$\etacut$
& \multicolumn{1}{c}{$\ptcut = 25 \GeV$}
& \multicolumn{1}{c}{$\ptcut = 50 \GeV$}
\\
\hline
$2.5$
& $14.1\!\pm\! 0.8_\mu\!\pm\! 1.7_\mathrm{res}\, (13.6\%)$
& $19.7\!\pm\! 0.6_\mu\!\pm\! 1.7_\mathrm{res}\, (9.0\%)$
\\
$4.5$
& $12.4\!\pm\! 0.4_\mu\!\pm\! 1.1_\mathrm{res}\, (9.2\%)$
& $18.9\!\pm\! 0.4_\mu\!\pm\! 1.4_\mathrm{res}\, (7.6\%)$
\\
$\infty$
& $12.4\!\pm\! 0.4_\mu\!\pm\! 1.1_\mathrm{res}\, (9.1\%)$
& $18.9\!\pm\! 0.4_\mu\!\pm\! 1.4_\mathrm{res}\, (7.6\%)$
\\
\hline\hline
\end{tabular}%
\caption{
The 0-jet cross section for Drell-Yan production at the $Z$ pole $Q = m_Z$ (top)
and at $Q = 1 \TeV$ (bottom) at NLL$'$+NLO for different values of $\ptcut$ and $\etacut$
with a breakdown of the uncertainties.}
\label{tab:resummed_DY}
\end{table}

The results for Drell-Yan production are given in \fig{resummed_DY} and \tab{resummed_DY}.
For $Q = m_Z$ (top rows), we find a $5-7\%$ increase in the cross section at $\etacut = 2.5$
for $\ptcut = 20-25 \GeV$.
Here the uncertainty for $\etacut = 2.5$ is under good control even down to $\ptcut \sim 10\GeV$.
For $Q = 1\TeV$ (bottom rows), the cross section for $\etacut = 2.5$ increases by $14\%$ ($4\%$)
for $\ptcut = 25\GeV$ ($50\GeV$) compared to $\etacut = \infty$.
The Sudakov suppression and the accompanying increase in relative uncertainty
at small $\ptcut$ are weaker than for $gg\to X$ due to the smaller color factor ($C_F$ vs.\ $C_A$)
in the Sudakov exponent, but are still substantial for a quark-induced
process. The effect of the rapidity cut at $\etacut = 4.5$ is negligible.

%===============================================================================
\subsection{Resummed predictions with a step in the jet veto}
\label{sec:resummed_predictions_step}
%===============================================================================

In the previous subsection we have seen that a sharp rapidity
cut at $\etacut = 2.5$ can lead to a substantial loss of precision in the theory
predictions, especially for gluon-induced processes and at high
production energies.

\begin{figure*}
\centering
\includegraphics[width=\WidthTwoSubfigs]{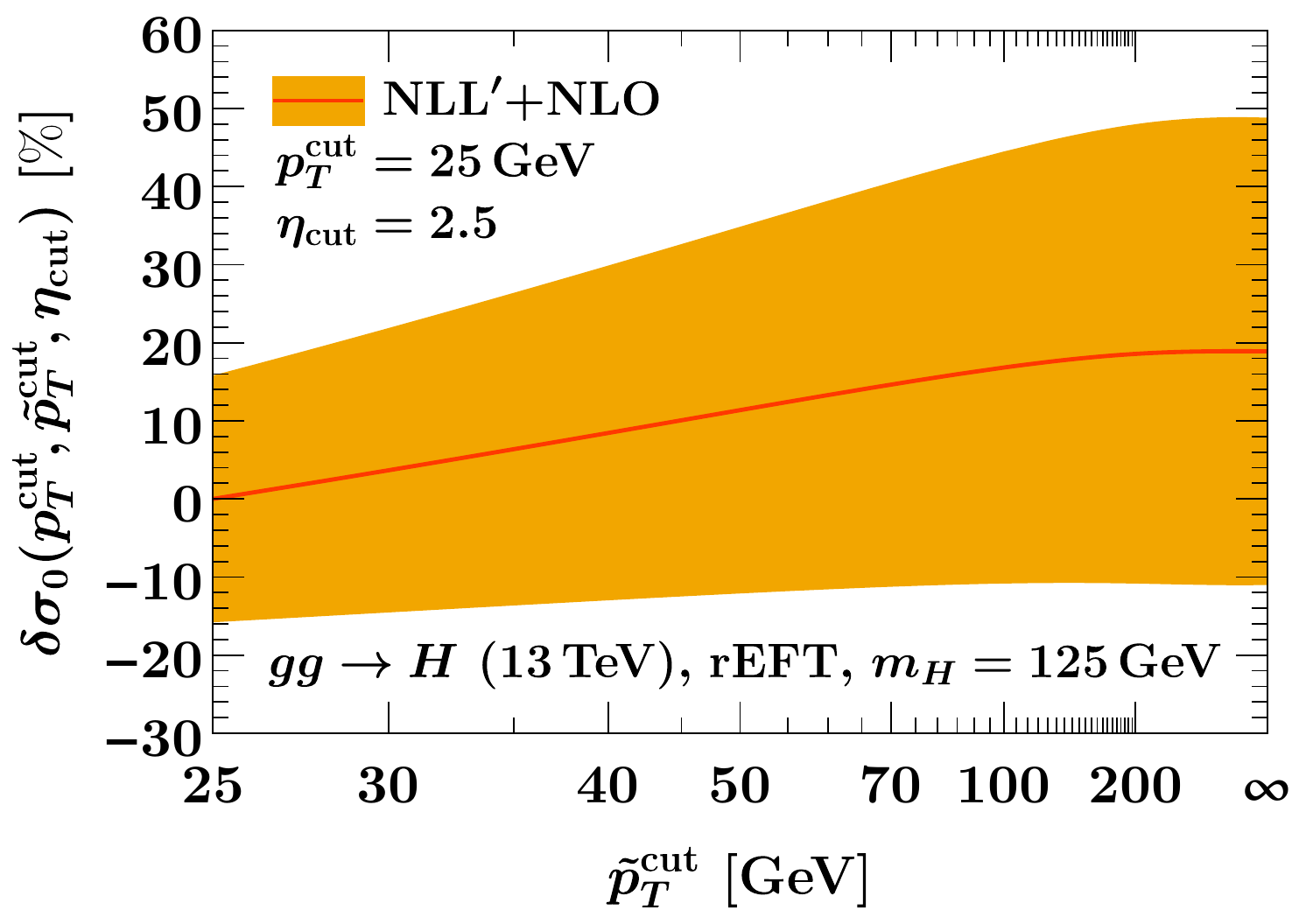}%
\hfill%
\includegraphics[width=\WidthTwoSubfigs]{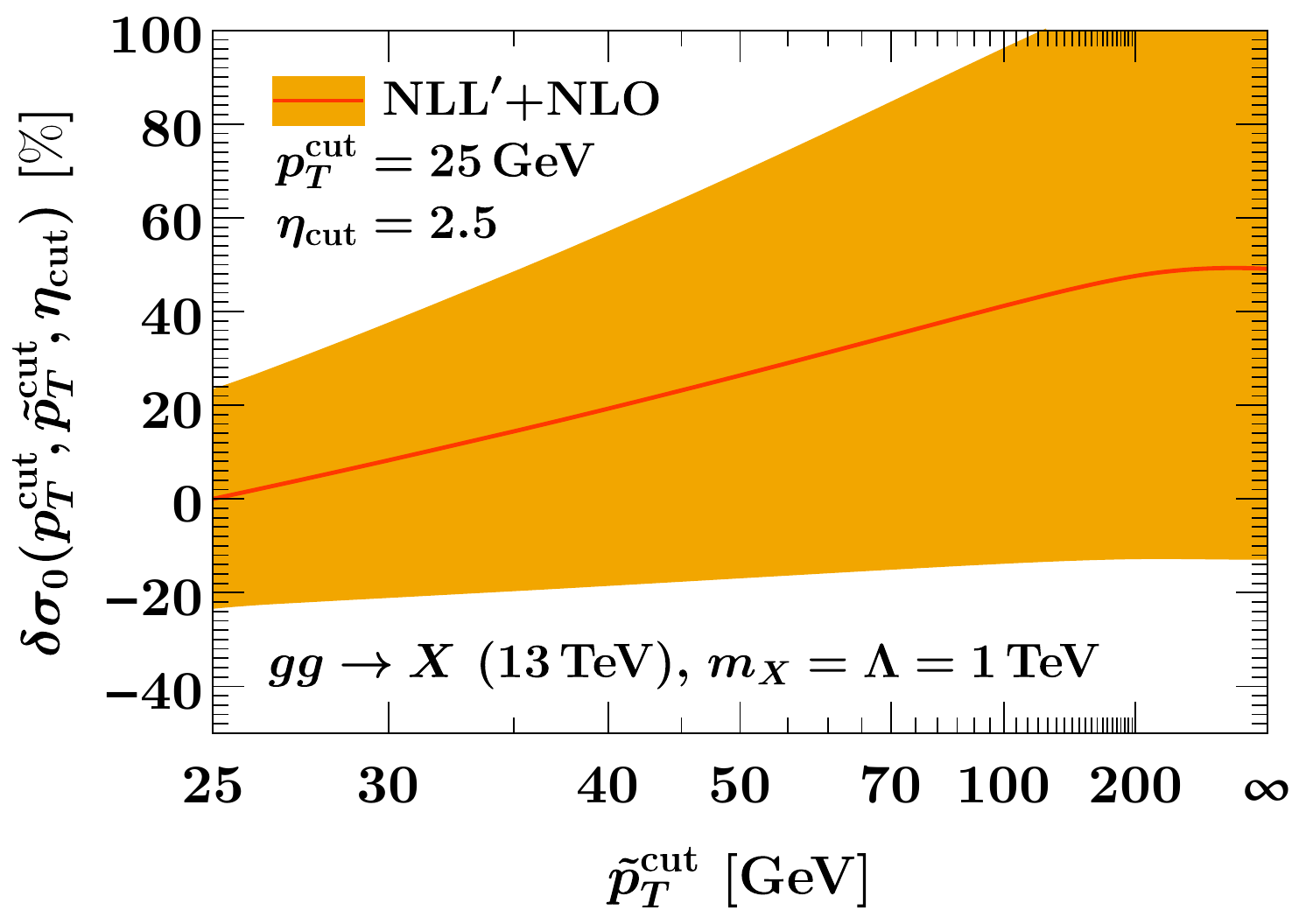}%
\caption{0-jet cross section $\sigma_0(\ptcut, \ptcuttwo, \etacut)$ with a step at $\etacut = 2.5$
for $gg\to H$ (left panel) and $gg\to X$ (right panel) at NLL$'$+NLO. The results
are shown for a fixed central veto at $\ptcut = 25\GeV$ as a function of the
jet veto $\ptcuttwo$ that is applied beyond $\etacut$.
We show the percent differences relative to the result for a uniform veto
$\ptcuttwo = \ptcut$.
The bands indicate the total uncertainty $\Delta_{\mu 0} \oplus \Delta_\varphi \oplus \Delta_\res$.}
\label{fig:resummed_step_ggfus}
\end{figure*}

In \fig{resummed_step_ggfus} we show the resummed 0-jet cross section for
$gg\to H$ and $gg\to X$ with a step in the jet veto at $\etacut = 2.5$ as a function
of the second jet veto parameter $\ptcuttwo$ that is applied beyond $\etacut$.
The central jet veto below $\etacut$ is fixed to $\ptcut = 25\GeV$.
On the left of the plot $\ptcuttwo = \ptcut$, which is equivalent to having
no rapidity cut, in which case the uncertainties are well under control.
In the limit $\ptcuttwo \to \infty$ (towards the right) the step becomes a
sharp cut, corresponding to the results of the previous subsection.
While the step in the jet veto still leads to an increase in the uncertainties,
this can now be controlled by the choice of $\ptcuttwo$. At this order,
a small step from $\ptcut = 25\GeV$ to $\ptcuttwo = 30\GeV$ only leads to a small
increase in uncertainty. For a larger step to $\ptcuttwo = 50\GeV = 2\ptcut$,
the uncertainties already increase substantially but are still much smaller
than for a sharp cut.

%%%%%%%%%%%%%%%%%%%%%%%%%%%%%%%%%%%%%%%%%%%%%%%%%%%%%%%%%%%%%%%%%%%%%%%%%%%%%%%%
\section{Conclusion}
\label{sec:conclusion}
%%%%%%%%%%%%%%%%%%%%%%%%%%%%%%%%%%%%%%%%%%%%%%%%%%%%%%%%%%%%%%%%%%%%%%%%%%%%%%%%

We have developed a systematic framework to
seamlessly incorporate a cut on the rapidity of reconstructed jets,
$\abs{\etajet} < \etacut$, into the theoretical description of jet-vetoed processes at the LHC.
We have shown that the standard jet-veto resummation, which neglects the rapidity cut,
is correct up to power corrections of $\ord{Q e^{-\etacut} / \ptcut}$,
with $Q$ the hard-interaction scale and $\ptcut$ the jet veto cut.

We calculated the necessary $\etacut$-dependent corrections at one loop as well as
all logarithmic contributions to them at two loops (including both small-$R$ clustering
logarithms and all jet-veto logarithms predicted by the RGE; see~\sec{no_step_reg2}).
The remaining ingredients required for a full NNLL$'$ analysis with $\etacut$
effects are finite nonlogarithmic pieces that could be either calculated explicitly
or extracted numerically from the full-QCD results, which we leave to future work.
In addition, we considered for the first time the case of a step in the jet veto,
i.e., an increase in the veto parameter to $\ptcuttwo > \ptcut$ beyond $\etacut$,
and showed how to similarly incorporate it into the jet-veto resummation (see~\sec{collinear_step}).

We also considered the jet veto cross section in the limit $\ptcut \ll Q e^{-\etacut}$,
corresponding to either very tight vetoes or very central rapidity cuts (see~\sec{no_step_reg3}).
In this regime, the jet-veto resummation becomes impaired by the presence of nonglobal logarithms,
requiring a refactorization of the cross section.
However, we have argued that this parametric
region will most likely not play a role for typical  jet binning analyses
at the LHC. If experimentally necessary, it can be avoided by replacing the sharp rapidity cut by
a moderate step in the jet veto, which is free of nonglobal logarithms (see~\sec{soft_collinear_step}).

There are several important outcomes of our analysis. First, a jet rapidity cut at very forward
rapidities due to the finite detector acceptance, $\etacut \simeq 4.5$,
is theoretically safe and unproblematic. In contrast, restricting the
jet veto to the more central region, with a sharp rapidity cut at the end of
the tracking detectors, $\etacut \simeq 2.5$, leads to an increase in the perturbative
uncertainties (which may not be captured if the jet rapidity cut
is not included in the resummation). This loss in theoretical precision can become
particularly severe for gluon-induced processes and for processes at high
scales. It can however be mitigated  by replacing the sharp
rapidity cut by a moderate step in the jet veto. We expect this to be a generic feature
that also holds at higher orders. It will be interesting to extend our resummed predictions
to the next order (NNLL$'$) to confirm this as well as to reduce the overall size of the
theoretical uncertainties.
We encourage our experimental colleagues to take full advantage of
such step-like jet vetoes in order to benefit from suitably tight jet vetoes at
central rapidities, while avoiding the increased pile-up contamination in the forward region.

%%%%%%%%%%%%%%%%%%%%%%%%%%%%%%%%%%%%%%%%%%%%%%%%%%%%%%%%%%%%%%%%%%%%%%%%%%%%%%%%
\acknowledgments
We thank Daekyoung Kang, Yiannis Makris, Thomas Mehen, and Iain Stewart for discussions.
This work was partially supported by the German Science Foundation (DFG) through
the Emmy-Noether Grant No.~TA 867/1-1 and the Collaborative Research Center (SFB)
676 Particles, Strings and the Early Universe.

%%%%%%%%%%%%%%%%%%%%%%%%%%%%%%%%%%%%%%%%%%%%%%%%%%%%%%%%%%%%%%%%%%%%%%%%%%%%%%%%
\appendix
\section{Perturbative ingredients}
\label{app:ingredients}
%%%%%%%%%%%%%%%%%%%%%%%%%%%%%%%%%%%%%%%%%%%%%%%%%%%%%%%%%%%%%%%%%%%%%%%%%%%%%%%%

We collect known results for required anomalous dimensions in \app{anom_dims}
and for the standard $\ptcut$ beam function without a jet rapidity cut in \app{standard_beam_function}.
In \app{etacut_beam_functions_one_loop}
we provide some details on the computation of the one-loop beam function matching coefficients in \eqs{DeltaI}{I_cut}.
In \app{soft_coll} we compute the soft-collinear functions given in \eqs{S_cut}{S_step}.
In \app{compare_Hornig_et_al} we compare to the one-loop results of \refcite{Hornig:2017pud}.
In \app{two_loop_convolutions} we discuss the Mellin convolutions
required in the two-loop $\etacut$ dependent beam function in \eq{delta_I_master_formula}.

%===============================================================================
\subsection{Anomalous dimensions}
\label{app:anom_dims}
%===============================================================================

We expand the $\beta$ function of QCD as
%%%
\begin{equation} \label{eq:beta_expand}
\mu \frac{\df \as(\mu)}{\df \mu}
= \beta[\as(\mu)]
\,, \qquad
\beta(\as) = -2\as \sum_{n = 0}^\infty \beta_n \left( \frac{\as}{4\pi} \right)^{n+1}
\,,\end{equation}
%%%
with the one-loop and two-loop coefficients in the \MSbar scheme given by
%%%
\begin{align} \label{eq:beta_coeff}
\beta_0 = \frac{11}{3}\,C_A -\frac{4}{3}\,T_F\,n_f
\,,\qquad
\beta_1 = \frac{34}{3}\,C_A^2  - \Bigl(\frac{20}{3}\,C_A\, + 4 C_F\Bigr)\, T_F\,n_f
\,.\end{align}
%%%
The cusp and all noncusp anomalous dimensions $\gamma(\as)$ are expanded as
%%%
\begin{equation} \label{eq:anom_dim_expand}
\Gamma^i_\mathrm{cusp}(\as) = \sum_{n = 0}^\infty \Gamma^i_n \Bigl( \frac{\as}{4\pi} \Bigr)^{n+1}
\,,\qquad
\gamma(\as) = \sum_{n = 0}^\infty \gamma_n \Bigl( \frac{\as}{4\pi} \Bigr)^{n+1}
\,.\end{equation}
%%%
The coefficients of the \MSbar cusp anomalous dimension through two loops are
%%%
\begin{align} \label{eq:Gcusp_coeff}
\Gamma^q_n &= C_F \Gamma_n
\,,\qquad
\Gamma^g_n = C_A \Gamma_n
\,,\qquad \text{(for $n = 0,1,2$)}
\,,\nn\\[1ex]
\Gamma_0 &= 4
\,,\nn\\
\Gamma_1
&= 4 \Bigl[ C_A \Bigl( \frac{67}{9} - \frac{\pi^2}{3} \Bigr)  - \frac{20}{9}\,T_F\, n_f \Bigr]
= \frac{4}{3} \bigl[ (4 - \pi^2) C_A + 5 \beta_0 \bigr]
\,.\end{align}
%%%
The PDF anomalous dimension in \eq{DGLAP} is expanded as
%%%
\begin{equation} \label{eq:DGLAP_expand}
P_{ij}(\as, z)
= \sum_{n = 0}^\infty P^{(n)}_{ij}(z) \left( \frac{\as}{4\pi} \right)^{n + 1}
\,.\end{equation}
%%%
Note that we expand the PDF anomalous dimension in $\as/(4\pi)$ and not $\as/(2\pi)$
as is often done.
The one-loop coefficients of the PDF anomalous dimension read
%%%
\begin{align} \label{eq:DGLAP_coeff_lo}
P^{(0)}_{q_iq_j}(z) &= P^{(0)}_{\bar q_i \bar q_j}(z) = 2C_F \,\delta_{ij} \, \theta(z) P_{qq}(z)
\,,\quad
&P^{(0)}_{gg}(z) &= 2 C_A \,\theta(z) P_{gg}(z) + \beta_0 \,\delta(1-z)
\,, \nn \\
P^{(0)}_{q_ig}(z) &= P^{(0)}_{\bar q_i g}(z) = 2T_F \,\theta(z) P_{qg}(z)
\,, \quad
&P^{(0)}_{gq_i}(z) &= P^{(0)}_{g\bar q_i}(z) = 2C_F \,\theta(z) P_{gq}(z)
\,,\end{align}
%%%
in terms of the standard color-stripped one-loop QCD splitting functions
%%%
\begin{alignat}{3} \label{eq:p_ij}
P_{qq}(z) &= 2 \mathcal{L}_0(1-z)-\theta(1-z)(1+z) + \frac{3}{2}\delta(1-z)
&&= \Bigl[ \theta(1-z) \frac{1+z^2}{1-z} \Bigr]_+
\,, \nn \\
P_{gg}(z) &= 2 \mathcal{L}_0(1-z)+\theta(1-z) \Bigl[ 2z(1-z)+\frac{2(1-z)}{z}-2 \Bigr] \,
&&= \, 2\mathcal{L}_0(1-z) \frac{(1-z + z^2)^2}{z}
\,, \nn \\
P_{qg}(z) &= \theta(1-z) \bigl[ 1-2z(1-z) \bigr]
\,, \nn \\
P_{gq}(z) &= \theta(1-z) \frac{1+(1-z)^2}{z}
\,.\end{alignat}
%%%
The two-loop coefficients were calculated in
\refscite{Curci:1980uw, Furmanski:1980cm, Ellis:1996nn}. They can be decomposed as
%%%
\begin{align} \label{eq:DGLAP_coeff_nlo}
P^{(1)}_{q_iq_j}(z) = P^{(1)}_{\bar q_i \bar q_j}(z) &= 4C_F \,\theta(z) \bigl[ \delta_{ij} P^1_{qqV}(z) + P^1_{qqS}(z) \bigr]
\,, \nn \\
P^{(1)}_{q_ig}(z) = P^{(1)}_{\bar q_i g}(z) &= 4 T_F \, \theta(z) \, P^1_{qg}
\,, \nn \\
P^{(1)}_{q_i\bar{q}_j}(z) = P^{(1)}_{\bar{q}_i q_j}(z) &= 4C_F \, \theta(z) \bigl[ \delta_{ij} P^1_{q\bar{q}V}(z) + P^1_{qqS}(z) \bigr]
\,, \nn \\
P^{(1)}_{gg}(z) &= 4 \theta(z) \bigl[ C_A P^1_{ggA} + T_F n_f \, P^1_{ggF} \bigr]
\,, \nn \\
P^{(1)}_{gq_i}(z) = P^{(1)}_{g\bar q_i}(z) &= 4 C_F \theta(z) \, P^1_{gq}
\,,\end{align}
%%%
where explicit expressions for the $P^1$ functions on the right-hand side can be found
in appendices A of \refscite{Gaunt:2014xga, Gaunt:2014cfa}.
[Note that in \refscite{Gaunt:2014xga, Gaunt:2014cfa} the superscript ``$1$''
here is written as ``$(1)$'' there,
and the PDF anomalous dimension is expanded there in $\as/(2\pi)$, which is already accounted for by
the overall factors of 4 on the right-hand side of \eq{DGLAP_coeff_nlo}.]
Explicit results for the Mellin convolutions of two color-stripped leading-order splitting functions
can also be found there.

The coefficients of the noncusp beam anomalous dimension are~\cite{Stewart:2013faa, Tackmann:2016jyb}
%%%
\begin{align}
\gamma^q_{B\,0}
&= 6 C_F
\,,\nn\\
\gamma^q_{B\,1}
&= C_F  \Bigl[  ( 3 - 4 \pi^2 + 48 \zeta_3 ) C_F + \bigl(-14 + 16 (1+\pi^2) \ln 2 - 96 \zeta_3 \bigr) C_A
\nn\\
&\phantom{= C_F  a}
+ \Bigl( \frac{19}{3} - \frac{4}{3} \pi^2 + \frac{80}{3} \ln 2 \Bigr) \beta_0  \Bigr]
\,, \nn \\[1ex]
\gamma_{B\,0}^g
&= 2\beta_0
\,,\nn\\
\gamma_{B\,1}^g
&= 2 \beta_1
+ 8C_A \biggl[ \Bigl(-\frac{5}{4} + 2(1+\pi^2) \ln2 - 6 \zeta_3 \Bigr) C_A
+ \Bigl(\frac{5}{24} - \frac{\pi^2}{3} + \frac{10}{3}\ln2 \Bigr) \beta_0 \biggr]
\end{align}
%%%
The coefficients of the rapidity noncusp anomalous dimension depend on the jet radius $R$.
They read~\cite{Stewart:2013faa}
%%%
\begin{align}
\gamma_{\nu\,0}^i(R)
&= 0
\,, \\
\gamma_{\nu\, 1}^i(R)
&= -16 C_i \biggl[\Bigl(\frac{17}{9} - (1 + \pi^2)\ln2 + \zeta_3 \Bigl) C_A
+ \Bigl(\frac{4}{9} +\frac{\pi^2}{12} - \frac{5}{3}\ln2 \Bigl) \beta_0 \biggr] + C_2^i(R)
\nn \,.\end{align}
%%%
Here $C_i = C_F\,(C_A)$ for $i = q\,(g)$ and $C_2^i(R)$ is the clustering correction due to the jet algorithm
relative to a global $E_T$ veto, as computed in \refscite{Tackmann:2012bt, Stewart:2013faa},
%%%
\begin{align} \label{eq:C2value}
C_2^i(R) &=
16 C_i c^R_{ii} \ln R + 15.62 \, C_i C_A - 9.17 \, C_i \beta_0 + \ord{R^2}
\,.\end{align}
%%%
The small-$R$ clustering coefficient $c_{ii} = c_{gg} = c_{qq}$ is given in \eq{clustering_coefficients_no_etacut}.

%===============================================================================
\subsection[Beam function master formula for \texorpdfstring{$\etacut \to \infty$}{etacut to infinity}]
{\boldmath  Beam function master formula for                 $\etacut \to \infty$}
\label{app:standard_beam_function}
%===============================================================================

In analogy to \eq{rge delta I} the matching coefficient $\cI_{ij}(\ptcut, R, \omega, z, \mu, \nu)$
of the $\etacut \to \infty$ beam functions satisfies (suppressing all other arguments of $\cI_{ij}$)
%%%
\begin{align} \label{eq:rge I}
\mu \frac{\df}{\df \mu} \cI_{ij}(z)
&= \gamma^i_B(\omega, \mu, \nu) \, \cI_{ij}(z) - \sum_k \cI_{ik}(z) \otimes_z 2P_{kj}[\as(\mu), z]
\,, \nn \\
\nu \frac{\df}{\df \nu} \cI_{ij}(z)
&= \gamma^i_{\nu,B}(\ptcut, R, \mu)\, \cI_{ij}(z)
\,.\end{align}
%%%
Solving this order by order in $\as$ yields the beam function master formula,
%%%
\begin{align} \label{eq:I_standard}
\cI_{ij}(z) &=
\delta_{ij}\delta(1-z)
+ \frac{\as(\mu)}{4\pi} \, \cI_{ij}^{(1)}(z)
+ \frac{\as^2(\mu)}{(4\pi)^2} \, \cI_{ij}^{(2)}(z)
+ \ord{\as^3}
\,,\nn \\
%%%
\cI_{ij}^{(1)}(z) &=
\delta_{ij} \delta(1-z) \, L^\mu_B (2 \Gamma^i_0 L^\nu_B+\gamma^i_{B \,0})
-2 L^\mu_B P_{ij}^{(0)}(z)
+I_{ij}^{(1)}(z)
\,,\nn \\
%%%
\cI_{ij}^{(2)}(z) &=
\delta_{ij} \delta(1-z) \biggl\{
   (L^\mu_B)^2 \Bigl[ 2 (\Gamma^i_0)^2 (L^\nu_B)^2 + L^\nu_B (2 \beta_0 \Gamma^i_0+2 \Gamma^i_0 \gamma^i_{B \,0})+\beta_0 \gamma^i_{B \,0}+\frac{(\gamma^i_{B \,0})^2}{2} \Bigr]
\nn\\[-0.25em] &\quad\quad\quad\quad\quad\quad
   +L^\mu_B \Bigl[ 2 \Gamma^i_1 L^\nu_B+\gamma^i_{B \,1} \Bigr]
   -\frac{1}{2} \gamma^i_{\nu\,1}(R) L^\nu_B
   \biggr\}
\nn \\[0.25em] &\quad
+ P_{ij}^{(0)}(z) \, (L^\mu_B)^2 \Bigl[ -4 \Gamma^i_0 L^\nu_B -2 \beta_0-2 \gamma^i_{B \,0} \Bigr]
+ I_{ij}^{(1)}(z) \, L^\mu_B \Bigl[ 2 \Gamma^i_0 L^\nu_B+ 2 \beta_0+\gamma^i_{B \,0} \Bigr]
\nn \\[0.25em] &\quad
-2 L^\mu_B \sum_k I_{ik}^{(1)}(z) \otimes_z P_{kj}^{(0)}(z)
-2 L^\mu_B P_{ij}^{(1)}(z)
+2 (L^\mu_B)^2 \sum_k P_{ik}^{(0)}(z) \otimes_z P_{kj}^{(0)}(z)
\nn \\ &\quad
+I_{ij}^{(2)}(R, z)
\,.\end{align}
%%%
where we abbreviated
%%%
\begin{equation}
L_B^\mu = \ln \frac{\mu}{\ptcut}
\,, \quad
L_B^\nu = \ln \frac{\nu}{\omega}
\,.\end{equation}
%%%
The one-loop finite terms $I_{ij}^{(1)}$ using the $\eta$ regulator~\cite{Chiu:2011qc, Chiu:2012ir}
are given by (see e.g.\ \refscite{Stewart:2013faa, Li:2014ria, Tackmann:2016jyb})
%%%
\begin{align}
I^{(1)}_{q_iq_j}(z) = I^{(1)}_{\bar{q}_i\bar{q}_j}(z) &= C_F \, \delta_{ij} \, \theta(z)\theta(1-z) \, 2(1-z)
\,, \nn \\
I^{(1)}_{q_ig}(z) = I^{(1)}_{\bar{q}_ig}(z) &= T_F \, \theta(z)\theta(1-z) \, 4z(1-z)
\,, \nn \\
I^{(1)}_{gg}(z) &= 0
\,, \nn \\
I^{(1)}_{gq_i}(z) = I^{(1)}_{g\bar{q}_i}(z) &= C_F \, \theta(z)\theta(1-z) \, 2z
\,.\end{align}
%%%
Their convolutions with leading-order splitting functions always appear in the form
%%%
\begin{align}
\bigl[ I^{(1)} \otimes P^{(0)} \bigr]_{ij}(z)
&\equiv \sum_k I_{ik}^{(1)}(z) \otimes_z P_{kj}^{(0)}(z)
\,.\end{align}
%%%
For quark-to-(anti)quark transitions we decompose the above flavor structure as
%%%
\begin{alignat}{3}
\bigl[ I^{(1)} \otimes P^{(0)} \bigr]_{q_i q_j}
&= \bigl[ I^{(1)} \otimes P^{(0)} \bigr]_{\bar{q}_i \bar{q}_j}
&&\equiv \delta_{ij} \, \bigl[ I^{(1)} \otimes P^{(0)} \bigr]_{qqV}
+\bigl[ I^{(1)} \otimes P^{(0)} \bigr]_{qqS}
\,, \nn \\
\bigl[ I^{(1)} \otimes P^{(0)} \bigr]_{q_i \bar{q}_j}
&= \bigl[ I^{(1)} \otimes P^{(0)} \bigr]_{q_i \bar{q}_j}
&&= \bigl[ I^{(1)} \otimes P^{(0)} \bigr]_{qqS}
\,.\end{alignat}
%%%
The building blocks on the right, together with the gluon-to-quark case, are given by
%%%
\begin{align}
\bigl[ I^{(1)} \otimes P^{(0)} \bigr]_{qqV}
&= 4C_F^2 \, \theta(z)\theta(1-z) \, (1-z) \Bigl[ 2 \ln (1-z) - \ln z - \frac{1}{2} \Bigr]
\,, \nn \\
\bigl[ I^{(1)} \otimes P^{(0)} \bigr]_{qqS}
&= 4T_F C_F \, \theta(z)\theta(1-z) \, \Bigl( \frac{4}{3} z^2  + \frac{2}{3z} -2z\ln z - 2\Bigr)
\,, \nn \\[0.5em]
\bigl[ I^{(1)} \otimes P^{(0)} \bigr]_{q_i g}
&= \bigl[ I^{(1)} \otimes P^{(0)} \bigr]_{\bar{q}_i g}
= \theta(z)\theta(1-z) \biggl\{
4C_F T_F\, \bigl[ z^2+z-(2 z+1) \ln z-2 \bigr]
\nn \\ &\quad
+ 4T_F C_A \, \biggl[
\frac{34}{3} z^2 - 10 z + \frac{2}{3z} - 8 z \ln z - 2 + 4 z(1-z) \ln(1-z)
\biggr]
\nn \\
&\quad
+ 4 T_F \beta_0 \, z (1-z) \biggr\}
\,.\end{align}
%%%
The convolutions required for the gluon beam function read
%%%
\begin{align}
&\bigl[ I^{(1)} \otimes P^{(0)} \bigr]_{g g}
= 4 C_F (2n_f) T_F \, \theta(z)\theta(1-z) \bigl( 1 + z - 2z^2 + 2z \ln z \bigr)
\,, \\
&\bigl[ I^{(1)} \otimes P^{(0)} \bigr]_{g q_i}
= \bigl[ I^{(1)} \otimes P^{(0)} \bigr]_{g \bar{q}_i }
= 4 C_F^2 \, \theta(z)\theta(1-z) \Bigl[ 1 + \frac{z}{2} - z \ln z + 2z \ln(1-z) \Bigr]
\,. \nn \end{align}
%%%
These expressions agree with the color-stripped convolutions
given in \refscite{Stewart:2013faa, Li:2014ria},
accounting for different conventions for splitting functions.
The two-loop finite terms in \eq{I_standard} depend on $R$. Expanding them as
%%%
\begin{equation}
I^{(2)}_{ij}(R, z) = \ln R \, I^{(2,\ln R)}_{ij}(z) + I^{(2,c)}_{ij}(z) + \ord{R^2}
\,,\end{equation}
%%%
the coefficient of $\ln R$ can be written as
%%%
\begin{equation} \label{eq:I2_lnR}
I^{(2,\ln R)}_{ij}(z) = c^R_{ij} \Bigl[ 2P_{ij}^{(0)}(z) - \gamma^i_{B\,0} \, \delta_{ij} \delta(1-z)  \Bigr]
\,.\end{equation}
%%%
We explicitly recomputed the coefficients $c^R_{ij}$,
for which we found some discrepancies in the literature.
[See \eq{clustering_coefficients_no_etacut} in the main text.]
Note that the terms proportional to $\delta(1-z)$ cancel in \eq{I2_lnR} when
the distributional structure of the splitting function is written purely in terms
of $\delta(1-z)$, $\cL_n(1-z)$, and regular terms in $1-z$.

%===============================================================================
\subsection{Rapidity cut dependent beam functions}
\label{app:etacut_beam_functions_one_loop}
%===============================================================================

\begin{figure}
\centering
\includegraphics[width=14cm]{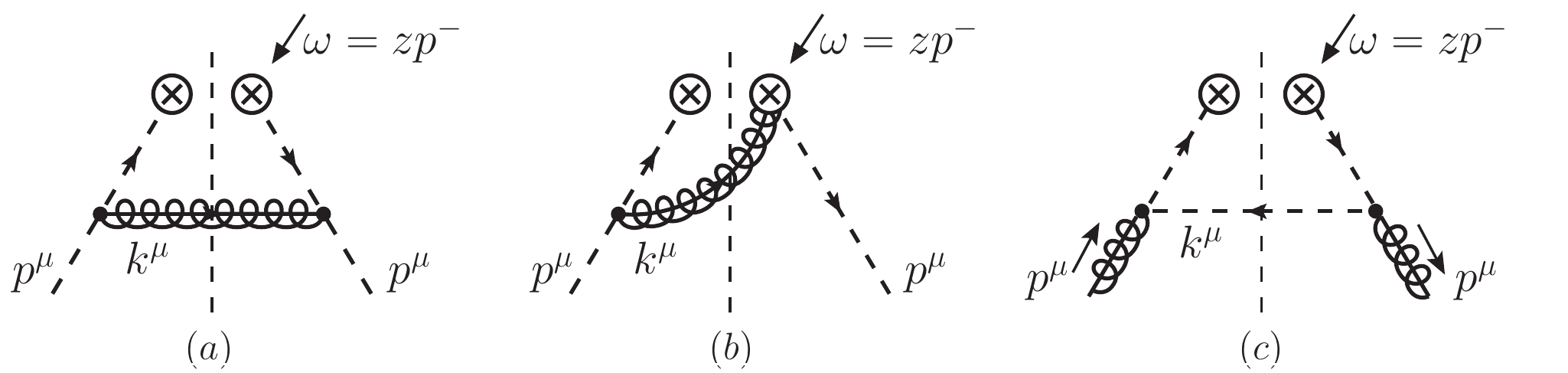}%
\\
\includegraphics[width=14cm]{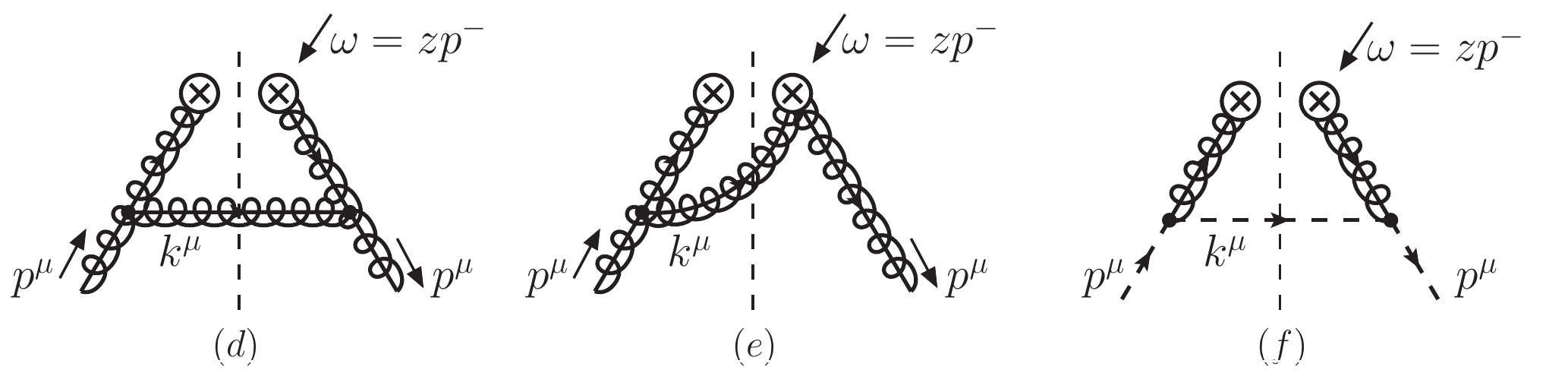}%
\caption{Nonvanishing diagrams for the computation
of the one-loop beam function in pure dimensional regularization and Feynman gauge.
Symmetric configurations are implicit.
The measurement acts on particles crossing the on-shell cut indicated by the vertical
dashed line.}
\label{fig:beamfct_diagrams}
\end{figure}

Here we provide some details on the computation of the one-loop beam function matching coefficients in \eqs{DeltaI}{I_cut}.
We use dimensional regularization for both UV and IR divergences
and the $\eta$ regulator~\cite{Chiu:2011qc,Chiu:2012ir} for rapidity divergences.
This ensures that all virtual diagrams, PDF diagrams, and zero-bin subtractions are scaleless.
We work in Feynman gauge.

The relevant real-radiation diagrams are displayed in \fig{beamfct_diagrams},
and the associated expressions for the spin-contracted amplitudes can be read off
e.g.\ from \refscite{Stewart:2010qs,Berger:2010xi} with a proper replacement of the measurement function.
For the beam function in \eq{fact_pTjet2}, the measurement on a single $n$-collinear emission
with momentum $k^\mu$ and rapidity
%%%
\begin{equation}
\eta = \frac{1}{2} \ln \frac{k^-}{k^+}
\end{equation}
%%%
reads, including label momentum conservation for $\omega = z p^-$, $k^- = (1-z) p^-$,
%%%
\begin{align} \label{eq:measurement_decomp_above_below}
&\mathcal{M}_{B}(k^\mu,\ptcut,\etacut,\omega,z)
\nn \\ &\qquad
=  \biggl[\theta\Bigl(e^{2\etacut}-\frac{k^-}{k^+}\Bigr)\,\theta(\ptcut-|\vec{k}_T|) + \theta\Bigl(\frac{k^-}{k^+}-e^{2\etacut}\Bigr)\biggr] \delta\Bigl(k^- -\frac{\omega(1-z)}{z}\Bigr) \nn \\
&\qquad\equiv
\mathcal{M}^{(\eta < \etacut)}_{B}(k^\mu,\ptcut,\etacut,\omega,z) + \mathcal{M}^{(\eta > \etacut)}_{B}(k^\mu,\etacut,\omega,z)
\,.\end{align}
%%%
Here we will separately display the result for each diagram
with $\mathcal{M}^{(\eta < \etacut)}_{B}$ and $\mathcal{M}^{(\eta > \etacut)}_{B}$ inserted, respectively.
This also allows one to read off the one-loop result for the $B^{(\rm cut)}_{i}$
beam function in \eq{I_cut},
for which the measurement on a single emission is just $\mathcal{M}^{(\eta > \etacut)}_{B}$.
On the other hand, for a direct computation of the finite correction due to the rapidity cut in \eq{DeltaI}
it is more convenient to decompose the measurement function as
%%%
\begin{align} \label{eq:measurement_decomp_etacut_correction}
&\mathcal{M}_{B}(k^\mu,\ptcut,\etacut,\omega,z)
\nn \\ &\qquad
= \biggl[\theta(\ptcut-|\vec{k}_T|) + \theta(|\vec{k}_T| - \ptcut)\, \theta\Bigl(\frac{k^-}{k^+}- e^{2\etacut}\Bigr) \biggr]\delta\Bigl(k^- -\frac{\omega(1-z)}{z}\Bigr)
\nn \\ &\qquad
= \mathcal{M}_{B}(k^\mu,\ptcut,\omega,z) + \Delta \mathcal{M}_{B}(k^\mu,\ptcut,\etacut,\omega,z)
\,.\end{align}
%%%
Inserting the first term into matrix elements yields the known results
for the matching coefficients without any rapidity cut,
while the second term yields the correction.

The relevant diagrams for the computation of the matching coefficient $\mathcal{I}_{qq}$ are (a) and (b).
The on-shell condition and label momentum constraint lead to a trivial $k^+$ integral,
which gives for diagram~(a), after expanding in $\epsilon$,
%%%
\begin{align}
&\langle q_n \vert \theta(\omega) \,\mathcal{O}_q^{\rm bare}(\ptcut,\omega) \vert q_n \rangle^{(a,\eta<\etacut)}
\nn \\ & \qquad
= \frac{\alpha_s C_F}{\pi} \,\theta\Bigl(z- \frac{\omega e^{-\etacut}}{\ptcut +\omega e^{-\etacut}}\Bigr) \,\theta(1-z) \,(1-z) \ln\frac{\ptcut z}{\omega e^{-\etacut}(1-z)}
+ \mathcal{O}(\eps) \,, \nn \\
&\langle q_n \vert \theta(\omega) \,\mathcal{O}_q^{\rm bare}(\ptcut,\omega) \vert q_n \rangle^{(a,\eta>\etacut)}
\nn \\ &\qquad
= \frac{\alpha_s C_F}{\pi} \,\theta(z) \,\theta(1-z) \, (1-z) \biggl[-\frac{1}{2\epsilon}
+ \ln\frac{\omega e^{-\etacut}(1-z)}{\mu\, z} + \frac{1}{2} + \mathcal{O}(\eps)\biggr]
\,.\end{align}
%%%
Diagram~(b) together with its mirror diagram gives, after expanding in $\eta$ and $\eps$,%
\footnote{For the renormalization one needs to account for the full $d$ dimensional
coefficient of the $1/\eta$ divergence, which we do not display here for simplicity.}
%%%
\begin{align}
&\langle q_n \vert \theta(\omega) \,\mathcal{O}_q^{\rm bare}(\ptcut,\omega) \vert q_n \rangle^{(b,\eta<\etacut)}
\\ & \qquad
= \frac{\alpha_s C_F}{\pi} \,\theta\Bigl(z- \frac{\omega e^{-\etacut}}{\ptcut +\omega e^{-\etacut}}\Bigr)  \,\theta(1-z) \biggl\{\delta(1-z)\biggl[\frac{1}{\eta}\biggl(\frac{1}{\eps}-2 \ln\frac{\ptcut}{\mu} +\mathcal{O}(\eps)\biggr)-\frac{1}{2\eps^2}
\nn \\ & \qquad\quad
+ \frac{1}{\eps}\ln\frac{\nu e^{-\etacut}}{\mu} - \ln^2\frac{\omega e^{-\etacut}}{\mu} +2\ln\frac{\ptcut}{\mu} \ln\frac{\omega}{\nu} + \frac{\pi^2}{24}\biggr] +2 \mathcal{L}_0(1-z)\ln\frac{\ptcut z}{\omega e^{-\etacut}}
\nn \\ &\qquad\quad
-2\mathcal{L}_1(1-z)-2 \ln\frac{\ptcut z}{\omega e^{-\etacut}(1-z)} + \mathcal{O}(\eta,\eps) \biggr\}
\,, \nn \\
&\langle q_n \vert \theta(\omega) \,\mathcal{O}_q^{\rm bare}(\ptcut,\omega) \vert q_n \rangle^{(b,\eta>\etacut)}
\nn \\ & \qquad
= \frac{\alpha_s C_F}{\pi} \,\theta(z) \,\theta(1-z) \biggl\{\delta(1-z)\biggl[\frac{1}{2\eps^2}-\frac{1}{\eps} \ln\frac{\omega e^{-\etacut}}{\mu} + \ln^2\frac{\omega e^{-\etacut}}{\mu} - \frac{\pi^2}{24}\biggr]
\nn \\ &\qquad\quad
+ \mathcal{L}_0(1-z)\biggl[-\frac{1}{\eps}+2\ln\frac{\omega e^{-\etacut}}{\mu\,z} \biggr] + 2\mathcal{L}_1(1-z) + \frac{1}{\eps} - 2\ln\frac{\omega e^{-\etacut}(1-z)}{\mu\,z}
+\mathcal{O}(\eps)\biggr\}
\,. \nn\end{align}
%%%
The matching coefficient $\mathcal{I}_{qg}$ is computed from diagram~(c) giving
%%%
\begin{align}
&\langle g_n \vert \theta(\omega) \,\mathcal{O}_q^{\rm bare}(\ptcut,\omega) \vert g_n \rangle^{(c,\eta<\etacut)}
\\ & \qquad
= \frac{\alpha_s T_F}{\pi} \,\theta\Bigl(z- \frac{\omega e^{-\etacut}}{\ptcut +\omega e^{-\etacut}}\Bigr)\,\theta(1-z) \, (1-2z+2z^2)\ln \frac{\ptcut z}{\omega e^{-\etacut}(1-z)}
+ \mathcal{O}(\eps)
\, ,\nn \\
&\langle g_n \vert \theta(\omega) \,\mathcal{O}_q^{\rm bare}(\ptcut,\omega) \vert g_n \rangle^{(c,\eta>\etacut)}
\nn \\ & \qquad
= \frac{\alpha_s T_F}{\pi} \,\theta(z) \,\theta(1-z)  \biggl\{(1-2z+2z^2)\biggl[-\frac{1}{2\eps}
+ \ln\frac{\omega e^{-\etacut}(1-z)}{\mu\, z} \biggr]+z(1-z) + \mathcal{O}(\eps)\biggr\}
\,. \nn\end{align}
%%%
The relevant diagrams for the computation of the matching coefficient $\mathcal{I}_{gg}$ are (d) and (e), which yield
%%%
\begin{align}
&\langle g_n \vert \theta(\omega) \,\mathcal{O}_g^{\rm bare}(\ptcut,\omega) \vert g_n \rangle^{(d,\eta<\etacut)}
\nn \\ & \qquad
= \frac{\alpha_s C_A}{\pi} \,\theta\Bigl(z- \frac{\omega e^{-\etacut}}{\ptcut +\omega e^{-\etacut}}\Bigr) \,\theta(1-z) \,\frac{2-2z+3z^2-2z^3}{z}  \ln\frac{\ptcut z}{\omega e^{-\etacut}(1-z)}
+ \mathcal{O}(\eps)
\, ,\nn \\
&\langle g_n \vert \theta(\omega) \,\mathcal{O}_g^{\rm bare}(\ptcut,\omega) \vert g_n \rangle^{(d,\eta>\etacut)}
\nn \\ &\qquad
= \frac{\alpha_s C_A}{\pi} \,\theta(z) \,\theta(1-z) \, \frac{2-2z+3z^2-2z^3}{z} \biggl[-\frac{1}{2\epsilon} + \ln \frac{\omega e^{-\etacut}(1-z)}{\mu\, z} + \mathcal{O}(\eps)\biggr]
\,,\end{align}
%%%
and, including the symmetric contribution of (e),
%%%
\begin{align}
&\langle g_n \vert \theta(\omega) \,\mathcal{O}_g^{\rm bare}(\ptcut,\omega) \vert g_n \rangle^{(e,\eta<\etacut)}
\nn \\ & \qquad
= \frac{\alpha_s C_A}{\pi} \,\theta\Bigl(z- \frac{\omega e^{-\etacut}}{\ptcut +\omega e^{-\etacut}}\Bigr)  \,\theta(1-z) \biggl\{\delta(1-z)\biggl[\frac{1}{\eta}\biggl(\frac{1}{\eps}-2 \ln\frac{\ptcut}{\mu}
+ \mathcal{O}(\eps)\biggr)-\frac{1}{2\eps^2}
\nn \\ & \qquad\quad
+\frac{1}{\eps}\ln\frac{\nu e^{-\etacut}}{\mu} - \ln^2\frac{\omega e^{-\etacut}}{\mu} + 2\ln\frac{\ptcut}{\mu} \ln\frac{\omega}{\nu} + \frac{\pi^2}{24}\biggr]
+ 2 \mathcal{L}_0(1-z)\ln\frac{\ptcut z}{\omega e^{-\etacut}}
\nn \\* &\qquad\quad
-2\mathcal{L}_1(1-z)- (2+z) \ln\frac{\ptcut z}{\omega e^{-\etacut}(1-z)}
+ \mathcal{O}(\eta,\eps)\biggr\}
\, ,\nn \\
&\langle g_n \vert \theta(\omega) \,\mathcal{O}_g^{\rm bare}(\ptcut,\omega) \vert g_n \rangle^{(e,\eta>\etacut)}
\nn \\ &\qquad
= \frac{\alpha_s C_A}{\pi} \,\theta(z) \,\theta(1-z) \biggl\{\delta(1-z)\biggl[\frac{1}{2\eps^2}
- \frac{1}{\eps} \ln \frac{\omega e^{-\etacut}}{\mu} + \ln^2\frac{\omega e^{-\etacut}}{\mu} -
\frac{\pi^2}{24}\biggr]
\nn \\ &\qquad\quad
+ \mathcal{L}_0(1-z)\biggl[-\frac{1}{\eps} + 2\ln\frac{\omega e^{-\etacut}}{\mu\,z} \biggr] + 2\mathcal{L}_1(1-z) +(2+z)\biggl[\frac{1}{2\eps} - \ln\frac{\omega e^{-\etacut}(1-z)}{\mu\,z} \biggr]
\nn \\ &\qquad\quad
+\mathcal{O}(\eps)\biggr\}
\,.\end{align}
%%%
%%%
The matching coefficient $\mathcal{I}_{gq}$ is computed from diagram~(f), giving
%%%
\begin{align}
&\langle q_n \vert \theta(\omega) \,\mathcal{O}_g^{\rm bare}(\ptcut,\omega) \vert q_n \rangle^{(f,\eta<\etacut)}
\nn \\ &\qquad
= \frac{\alpha_s C_F}{\pi} \,\theta\Bigl(z- \frac{\omega e^{-\etacut}}{\ptcut +\omega e^{-\etacut}}\Bigr)\,\theta(1-z) \, \frac{2-2z+z^2}{z} \ln\frac{\ptcut z}{\omega e^{-\etacut}(1-z)}
+ \mathcal{O}(\eps)
\, ,\nn \\
&\langle q_n \vert \theta(\omega) \,\mathcal{O}_g^{\rm bare}(\ptcut,\omega) \vert q_n \rangle^{(f,\eta>\etacut)}
\nn \\ &\qquad
= \frac{\alpha_s C_F}{\pi} \,\theta(z) \,\theta(1-z)  \biggl\{\frac{2-2z+z^2}{z}\biggl[-\frac{1}{2\eps}
+ \ln\frac{\omega e^{-\etacut}(1-z)}{\mu\, z} \biggr] + \frac{z}{2}+ \mathcal{O}(\eps)\biggr\}
\,.\end{align}
Since PDF diagrams are scaleless in pure dimensional regularization,
the renormalized beam function matching coefficients are given by the $\ord{\eps^0 \eta^0}$ terms in these expressions.
From the results for $\mathcal{M}^{(\eta > \etacut)}_{B}$ we get $\mathcal{I}_{ij}^{(\rm cut,1)}$ in \eq{I_cut},
while adding $\mathcal{M}^{(\eta < \etacut)}_{B}$ gives the sum of \eq{DeltaI} and the second line of \eq{I_standard}.

%===============================================================================
\subsection{Soft-collinear functions}
\label{app:soft_coll}
%===============================================================================

We again use pure dimensional regularization and the $\eta$ regulator,
so virtual diagrams and soft zero-bin subtractions are scaleless.
Note that we expand the $\eta$ regulator to leading power using the soft-collinear scaling,
i.e., for a single emission we insert $\abs{k^-/\nu}^{-\eta}$ rather than $\abs{2k^3/\nu}^{-\eta}$.
This choice leads to a scaleless soft zero bin.
In Feynman gauge the bare one-loop real contribution to the $n$-soft-collinear
function $\mathcal{S}^{(\rm cut)}_i$ in \eq{refactB} is given by
%%%
\begin{align} \label{eq:generic_one_loop_csoft}
\mathcal{S}_{i\,{\rm bare}}^{(\rm cut,1)}(\ptcut, \etacut) &=4g^2 C_i \Bigl(\frac{e^{\gamma_E}\mu^2}{4\pi}\Bigr)^\eps
\int\! \frac{\df^d k}{(2\pi)^d}\,\Bigl|\frac{\nu}{k^-}\Bigr|^\eta \,\frac{2\pi \delta^+(k^\mu)}{k^-k^+}\, \mathcal{M}_{\mathcal{S}}^{(\rm cut)}(k ^\mu,\ptcut,\etacut)
\,,
\end{align}
%%%
where $\delta^+(k^\mu) = \delta(k^2)\,\theta(k^0)$, and the measurement function reads
%%%
\begin{equation}\label{eq:measurement_csoft_cut}
\mathcal{M}_{\mathcal{S}}^{(\rm cut)}(k^\mu,\ptcut,\etacut)
= \theta(\ptcut-|\vec{k}_T|)\, \theta\Bigl(e^{2\etacut}-\frac{k^-}{k^+}\Bigr)
+ \theta\Bigl(\frac{k^-}{k^+} - e^{2\etacut}\Bigr)
\,.\end{equation}
%%%
The second term yields a scaleless contribution, while the first term corresponds to a boosted hemisphere and leads to the result
%%%
\begin{align} \label{eq:soft_collinear_bare}
\mathcal{S}_{i\,{\rm bare}}^{(\rm cut,1)}(\ptcut, \etacut)
&= \frac{\alpha_s C_i}{\pi}\biggl\{\frac{1}{\eta}\biggl[\frac{1}{\eps}-2 \ln\frac{\ptcut}{\mu}
+\mathcal{O}(\eps)\biggr]
-\frac{1}{2\eps^2} + \frac{1}{\eps} \ln\frac{\nu e^{-\etacut}}{\mu}
\nn \\ & \quad
+\ln^2\frac{\ptcut}{\mu} - 2\ln\frac{\ptcut}{\mu} \ln\frac{\nu e^{-\etacut}}{\mu}
+\frac{\pi^2}{24}+ \mathcal{O}(\eta,\eps)\biggr\}
\,.\end{align}
%%%
Absorbing the divergent terms (including contributions of the form $\eps^n/\eta$, which are not shown)
into counterterms yields the renormalized one-loop result in \eq{S_cut}.

The bare one-loop contribution to the soft-collinear function resolving the step
in \eq{fact_soft_collinear_step} is again given by \eq{generic_one_loop_csoft},
but this time the measurement reads
%%%
\begin{equation}\label{eq:measurement_soft_collinear_step}
\mathcal{M}_{\mathcal{S}}^{(\rm step)}(k^\mu,\ptcut,\etacut)
= \theta(\ptcut-|\vec{k}_T|)\,    \theta\Bigl(e^{2\etacut}-\frac{k^-}{k^+}\Bigr)
+ \theta(\ptcuttwo-|\vec{k}_T|)\, \theta\Bigl(\frac{k^-}{k^+} - e^{2\etacut}\Bigr)
\,.\end{equation}
%%%
Successively dropping terms that yield scaleless integrals we can replace ($\mapsto$)
%%%
\begin{align} \label{eq:dim_reg_magic}
\mathcal{M}_{\mathcal{S}}^{(\rm step)}(k^\mu,\ptcut,\etacut)
&\mapsto  \theta\Bigl(\frac{k^-}{k^+} - e^{2\etacut}\Bigr) \Bigl[ \theta(\ptcuttwo-|\vec{k}_T|) - \theta(\ptcut-|\vec{k}_T|) \Bigr]
\nn \\
&\mapsto \theta\Bigl(e^{2\etacut} - \frac{k^-}{k^+}\Bigr) \Bigl[ \theta(\ptcut-|\vec{k}_T|) - \theta(\ptcuttwo-|\vec{k}_T|) \Bigr]
\nn \\[1ex]
&= \mathcal{M}_{\mathcal{S}}^{(\rm cut)}(k^\mu,\ptcut,\etacut) - \mathcal{M}_{\mathcal{S}}^{(\rm cut)}(k^\mu,\ptcuttwo,\etacut)
\,,\end{align}
%%%
so at one loop we find a simple relation between bare results,
%%%
\begin{equation}
\mathcal{S}_{i\,{\rm bare}}^{(1)}(\ptcut, \ptcuttwo, \etacut)
= \mathcal{S}_{i\,{\rm bare}}^{(\rm cut,1)}(\ptcut, \etacut) - \mathcal{S}_{i\,{\rm bare}}^{(\rm cut,1)}(\ptcuttwo, \etacut)
\,.\end{equation}
Remapping the measurement on the primary emission as in \eq{dim_reg_magic}
yields the analogous relation for the small-$R$ clustering contributions.

%===============================================================================
\subsection{Comparison to quark beam function results in the literature}
\label{app:compare_Hornig_et_al}
%===============================================================================

In \refcite{Hornig:2017pud} the regime $\ptcut \sim Q e^{-\etacut}$ was accounted for
by adding a finite contribution $\Delta B_{i/j}^{(1)}$ from so-called out-of-jet radiation
to the unmeasured beam function in \eq{refactB} as
%%%
\begin{equation} \label{eq:beam_decomposition_Hornig_et_al}
\mathcal{I}_{ij}^{(\rm cut,1)}(\etacut, \omega, z, \mu)
\mapsto \mathcal{I}_{ij}^{(\rm cut,1)}(\etacut, \omega, z, \mu) + \Delta B_{i/j}^{(1)}(\ptcut, z, \omega, e^{-\etacut})
\,.\end{equation}
%%%
One-loop consistency with our \eq{fact_pTjet2} reads, at the level of bare ingredients,
%%%
\begin{align} \label{eq:consistency_Hornig_et_al}
&\mathcal{I}_{ij\,{\rm bare}}^{(1)}(\ptcut, \etacut, \omega, z)
\\\nn & \qquad
=\mathcal{I}_{ij\,{\rm bare}}^{(\rm cut,1)}(\etacut, \omega, z)
+ \Delta B_{i/j}^{(1)}(\ptcut, z, \omega, e^{-\etacut})
+ \delta_{ij}\delta(1-z) \,\mathcal{S}_{i\,{\rm bare}}^{(\rm cut,1)}(\ptcut, \etacut)
\,,\end{align}
where $\mathcal{S}_i^{(\rm cut, 1)}$ is the bare soft-collinear function at one loop, see \eq{soft_collinear_bare}.
By \eq{measurement_decomp_above_below} we have, in terms of bare collinear matrix elements up to scaleless PDF diagrams,
%%%
\begin{align}
\mathcal{I}_{qq\,{\rm bare}}^{(1)}(\ptcut, \etacut, \omega, z)
= \mathcal{I}_{qq\,{\rm bare}}^{(\rm cut,1)}(\etacut, \omega, z)
+ \langle q_n \vert \theta(\omega) \,\mathcal{O}_q^{\rm bare}(\ptcut,\omega) \vert q_n \rangle^{(\eta<\etacut)}
\,,\end{align}
%%%
and similarly for $\mathcal{I}_{qg}$.
With this, \eq{consistency_Hornig_et_al} simplifies to
%%%
\begin{align}
\langle q_n \vert \theta(\omega) \,\mathcal{O}_q^{\rm bare}(\ptcut,\omega) \vert q_n \rangle^{(\eta<\etacut)}
&= \Delta B_{q/q}^{(1)}(\ptcut, z, \omega, e^{-\etacut})
+ \delta(1-z) \,\mathcal{S}_{q\,{\rm bare}}^{(\rm cut,1)}(\ptcut, \etacut)
\,,\nn\\
\langle g_n \vert \theta(\omega) \,\mathcal{O}_q^{\rm bare}(\ptcut,\omega) \vert g_n \rangle^{(\eta<\etacut)}
&= \Delta B_{q/g}^{(1)}(\ptcut, z, \omega, e^{-\etacut})
\,.\end{align}
%%%
Both relations are readily checked after summing over all contributing diagrams.

%===============================================================================
\subsection{Mellin convolutions in the two-loop rapidity dependent beam function}
\label{app:two_loop_convolutions}
%===============================================================================

The PDF and beam function RGEs together predict Mellin convolutions of the following form
in the two-loop matching kernels \eq{delta_I_master_formula} for the rapidity dependent beam function:
%%%
\begin{equation}
\sum_k \Delta I_{ik}^{(1)} ( \zetacut, z )
   \otimes_z P^{(0)}_{kj}(z)
\equiv [\Delta I^{(1)} \otimes P^{(0)}]_{ij} ( \zetacut, z )
\,.\end{equation}
%%%
The relevant partonic channels read, leaving all arguments implicit,
%%%
\begin{alignat}{3}
&[\Delta I^{(1)} \otimes P^{(0)}]_{q_i q_j}
&&= \delta_{i j} \,8C_F^2 \, P^{w}_{qq} \otimes_z P_{qq}
+ 8T_F C_F \, P^{w}_{qg} \otimes_z P_{gq}
= [\Delta I^{(1)} \otimes P^{(0)}]_{\bar{q}_i \bar{q}_j}
\,, \nn \\[0.5em]
&[\Delta I^{(1)} \otimes P^{(0)}]_{q_i \bar{q}_j}
&&= 8T_F C_F \, P^{w}_{qg} \otimes_z P_{gq}
= [\Delta I^{(1)} \otimes P^{(0)}]_{\bar{q}_i q_j}
\,, \nn \\[0.5em]
&[\Delta I^{(1)} \otimes P^{(0)}]_{q_i g}
&&= 8C_F T_F \, P^{w}_{qq} \otimes_z P_{qg}
+ 8T_F \Bigl[ C_A \, P^{w}_{qg} \otimes_z P_{gg} + \frac{\beta_0}{2} P^{w}_{qg} \Bigr]
= [\Delta I^{(1)} \otimes P^{(0)}]_{\bar{q}_i g}
\,, \nn \\[0.5em]
&[\Delta I^{(1)} \otimes P^{(0)}]_{gg}
&&= 8C_A \Bigl[ C_A \, P^{w}_{gg} \otimes_z P_{gg} + \frac{\beta_0}{2} P^{w}_{gg} \Bigr]
+ 8 C_F T_F (2 n_f) \, P^{w}_{gq} \otimes_z P_{qg}
\,, \nn \\[0.5em]
&[\Delta I^{(1)} \otimes P^{(0)}]_{g q_i}
&&= 8C_A C_F P^{w}_{gg} \otimes_z P_{gq}
+ 8C_F^2 P^{w}_{gq} \otimes_z P_{qq}
= [\Delta I^{(1)} \otimes P^{(0)}]_{g \bar{q}_i}
\,,\end{alignat}
%%%
where $n_f$ is the number of light quark flavors.
Here we introduced a shorthand for weighted color-stripped splitting functions that depend on $\zetacut$ in addition to $z$,
%%%
\begin{equation}
P^w_{ij}(\zetacut, z) = \theta\Bigl(\frac{\zetacut}{1+\zetacut} - z\Bigr) \ln \frac{\zetacut(1-z)}{z} P_{ij}(z)
\,.\end{equation}
%%%
The Mellin convolutions $P^{w}_{ik} \otimes_z P_{kj}$ are straightforward to evaluate
analytically, but the resulting expressions are lengthy.
They are available from the authors upon request.

%%%%%%%%%%%%%%%%%%%%%%%%%%%%%%%%%%%%%%%%%%%%%%%%%%%%%%%%%%%%%%%%%%%%%%%%%%%%%%%%
\section{\boldmath Jet rapidity cuts in \texorpdfstring{$\Tau_B$}{TauB} and \texorpdfstring{$\Tau_C$}{TauC} vetoes }
\label{app:TauCut}
%%%%%%%%%%%%%%%%%%%%%%%%%%%%%%%%%%%%%%%%%%%%%%%%%%%%%%%%%%%%%%%%%%%%%%%%%%%%%%%%

Here we comment on how the factorization setup for the smoothly rapidity dependent
jet vetoes introduced in \refcite{Gangal:2014qda}
is modified when an additional sharp jet rapidity cut is introduced.
The restriction on reconstructed jets reads in this case
%%%
\begin{align} \label{eq:TauBjetdef}
\max_{k \in \text{jets}:\,\abs{\eta_k} < \etacut} \bigl\{ \abs{\vec{p}_{T,k}} \, f(\eta_k) \bigr\} < \Taucut
\,,\end{align}
%%%
where $f(\eta) e^{|\eta|} \to 1 $ for $\eta \to \pm \infty$.
Examples are the beam thrust veto with $f(\eta) = e^{-|\eta|}$ and the
C-parameter veto with $f(\eta) = 1/(2 \cosh \eta)$. The discussion of an
additional sharp rapidity cut largely parallels the case of the $\ptcut$ veto in
\sec{no_step}. We again distinguish three hierarchies between $\sqrt{\Taucut/Q}$
and $e^{-\etacut}$, where now $\sqrt{\Taucut/Q}$ replaces $\ptcut/Q$ as the
characteristic angular size of collinear radiation constrained by the jet veto.
The hierarchy $\sqrt{\Taucut/Q} \gg e^{-\etacut}$ (regime~1) reduces to the
factorization for $\etacut \to \infty$~\cite{Tackmann:2012bt, Gangal:2014qda,
Gangal:2016kuo}, up to power corrections of
$\mathcal{O}(e^{-\etacut}\sqrt{Q/\Taucut})$.

For $\sqrt{\Taucut/Q} \sim e^{-\etacut}$ (regime~2) the relevant EFT modes scale as
%%%
\begin{align}
\text{soft:}
&\quad
p^\mu\sim (\Taucut, \Taucut, \Taucut)
\,, \nn \\
n_a \text{-collinear:}
&\quad
p^\mu \sim \Bigl( \Taucut, Q, \sqrt{\Taucut Q} \Bigr) \sim \Bigl( Qe^{-2\etacut}, Q, Qe^{-\etacut} \Bigr)
\, ,\nn \\
n_b \text{-collinear:}
&\quad
p^\mu\sim \Bigl( Q, \Taucut, \sqrt{\Taucut Q} \Bigr) \sim \Bigl( Q, Qe^{-2\etacut}, Qe^{-\etacut} \Bigr)
\,.\end{align}
%%%
The factorized 0-jet cross section reads
%%%
\begin{align} \label{eq:fact_Tau2}
\sigma_0 (\Taucut, \etacut, R, \Phi)
&=
H_\kappa(\Phi, \mu) \,
B_{a}(\Taucut, \etacut, R, \omega_a, \mu) \,
B_{b}(\Taucut, \etacut, R, \omega_b, \mu) \,
S_{\kappa} (\Taucut, R, \mu) \nn \\
&\quad \times \biggl[1+\mathcal{O}\Bigl(\frac{\Taucut}{Q}, e^{-\etacut},R^2\Bigr)\biggr]
\,.\end{align}
%%%
The beam and soft function are different from the $\ptcut$ veto.
The rapidity cut again affects only the beam functions without changing their RG
structure or anomalous dimension.
In analogy to \eq{I_tot} we can write the matching coefficients as
%%%
\begin{align} \label{eq:I_tot_Tau}
\mathcal{I}_{ij}(\Taucut, \etacut, R, \omega, z, \mu)
= \mathcal{I}_{ij}(\omega\Taucut, R, z, \mu)
+ \Delta \mathcal{I}_{ij}(\Taucut, \etacut, R, \omega, z, \mu)
\,,\end{align}
%%%
where the first term on the right-hand side is the $\etacut \to \infty$ matching coefficient
as calculated to two loops in \refcite{Gangal:2016kuo},
which only depends on the boost-invariant product $\omega \Taucut$.
The correction $\Delta \mathcal{I}_{ij}$ vanishes for $\omega e^{-2\etacut} \ll\Taucut$
and at one loop is given by
%%%
\begin{align}\label{eq:DeltaI_Tau}
\Delta \mathcal{I}_{ij} (\Taucut, \etacut, R, \omega, z, \mu)
&= \frac{\as(\mu)}{4\pi} \, \theta\Bigl(\frac{\omega e^{-2\etacut}}{\omega e^{-2\etacut} + \Taucut} - z\Bigr)  \, P^{(0)}_{ij}(z) \, \ln \frac{\omega e^{-2\etacut}(1-z)}{z\Taucut}
\nn \\
&\quad + \ord{\as^2}
\,.\end{align}
%%%

For $\sqrt{\Taucut/Q} \ll e^{-\etacut}$ (regime~3) we again distinguish two types of collinear modes,
%%%
\begin{align}
n_a\text{-collinear:}
&\quad
p^\mu \sim \Bigl(Qe^{-2\etacut},Q,Qe^{-\etacut}\Bigr)
\, ,\nn \\
n_a\text{-soft-collinear:}
&\quad
p^\mu \sim \Bigl(\Taucut , \Taucut e^{2\etacut}, \Taucut e^{\etacut}\Bigr)
\,.\end{align}
%%%
The contributions from these modes can be encoded in a function $\mathcal{B}_i$
which can be refactorized in analogy to \eq{refactB} to resum Sudakov logarithms of $\Taucut e^{2\etacut}/Q$,
%%%
\begin{align}\label{eq:refactB_Tau}
\mathcal{B}_{i} (\Taucut, \etacut, R, \omega, z, \mu)
&= B^{(\rm cut)}_{i}(\etacut, \omega, \mu) \,\mathcal{S}^{(\rm cut)}_{i}\bigl(\Taucut e^{\etacut}, R, \mu\bigr) \nn \\
& \quad \times  \biggl[1+ \mathcal{B}^{\rm (NG)}_{i}\Bigl(\frac{\Taucut e^{2\etacut}}{\omega}, \omega, R\Bigr)\biggr]
\,.\end{align}
%%%
Here, the $\etacut$ dependent and $\Taucut$ independent piece $B^{(\rm cut)}_{i}$ is identical to the one in \eq{refactB},
while the soft-collinear function $\mathcal{S}^{(\rm cut)}_{i}$ is different and reads
%%%
\begin{align}\label{eq:S_cut_Tau}
\mathcal{S}^{(\rm cut)}_{i}\bigl(\Taucut e^{\etacut}, \mu\bigr)
= 1 + \frac{\alpha_s C_i}{4\pi}  \biggl(4 \ln^2 \frac{\Taucut e^{\etacut}}{\mu} - \frac{\pi^2}{6}\biggr)
 + \ord{\as^2}
\,.\end{align}
%%%
The $\mathcal{B}^{\rm (NG)}_{i}$ piece, which contains nonglobal logarithms starting at $\ord{\as^2}$,
is again different from the one in \eq{refactB}.
We verified that, up to power corrections, the explicit one-loop expressions in \eqs{I_cut}{S_cut_Tau}
reproduce the sum of \eq{DeltaI_Tau} and the matching coefficients
without a rapidity cut given in app.~B of \refcite{Gangal:2014qda}.

\addcontentsline{toc}{section}{References}
\bibliographystyle{jhep}
\bibliography{RapidityCut}

\providecommand{\href}[2]{#2}\begingroup\raggedright\begin{thebibliography}{10}

\bibitem{Ebert:2016idf}
M.~A. Ebert, S.~Liebler, I.~Moult, I.~W. Stewart, F.~J. Tackmann, K.~Tackmann
  et~al., \emph{{Exploiting jet binning to identify the initial state of
  high-mass resonances}},
  \href{https://doi.org/10.1103/PhysRevD.94.051901}{\emph{Phys. Rev.}
  {\bfseries D94} (2016) 051901}
  [\href{https://arxiv.org/abs/1605.06114}{{\ttfamily 1605.06114}}].

\bibitem{Berger:2010xi}
C.~F. Berger, C.~Marcantonini, I.~W. Stewart, F.~J. Tackmann and W.~J.
  Waalewijn, \emph{{Higgs Production with a Central Jet Veto at NNLL+NNLO}},
  \href{https://doi.org/10.1007/JHEP04(2011)092}{\emph{JHEP} {\bfseries 04}
  (2011) 092} [\href{https://arxiv.org/abs/1012.4480}{{\ttfamily 1012.4480}}].

\bibitem{Stewart:2011cf}
I.~W. Stewart and F.~J. Tackmann, \emph{{Theory Uncertainties for Higgs and
  Other Searches Using Jet Bins}},
  \href{https://doi.org/10.1103/PhysRevD.85.034011}{\emph{Phys. Rev.}
  {\bfseries D85} (2012) 034011}
  [\href{https://arxiv.org/abs/1107.2117}{{\ttfamily 1107.2117}}].

\bibitem{Stewart:2009yx}
I.~W. Stewart, F.~J. Tackmann and W.~J. Waalewijn, \emph{{Factorization at the
  LHC: From PDFs to Initial State Jets}},
  \href{https://doi.org/10.1103/PhysRevD.81.094035}{\emph{Phys. Rev.}
  {\bfseries D81} (2010) 094035}
  [\href{https://arxiv.org/abs/0910.0467}{{\ttfamily 0910.0467}}].

\bibitem{Stewart:2010pd}
I.~W. Stewart, F.~J. Tackmann and W.~J. Waalewijn, \emph{{The Beam Thrust Cross
  Section for Drell-Yan at NNLL Order}},
  \href{https://doi.org/10.1103/PhysRevLett.106.032001}{\emph{Phys. Rev. Lett.}
  {\bfseries 106} (2011) 032001}
  [\href{https://arxiv.org/abs/1005.4060}{{\ttfamily 1005.4060}}].

\bibitem{Banfi:2012yh}
A.~Banfi, G.~P. Salam and G.~Zanderighi, \emph{{NLL+NNLO predictions for
  jet-veto efficiencies in Higgs-boson and Drell-Yan production}},
  \href{https://doi.org/10.1007/JHEP06(2012)159}{\emph{JHEP} {\bfseries 06}
  (2012) 159} [\href{https://arxiv.org/abs/1203.5773}{{\ttfamily 1203.5773}}].

\bibitem{Becher:2012qa}
T.~Becher and M.~Neubert, \emph{{Factorization and NNLL Resummation for Higgs
  Production with a Jet Veto}},
  \href{https://doi.org/10.1007/JHEP07(2012)108}{\emph{JHEP} {\bfseries 07}
  (2012) 108} [\href{https://arxiv.org/abs/1205.3806}{{\ttfamily 1205.3806}}].

\bibitem{Tackmann:2012bt}
F.~J. Tackmann, J.~R. Walsh and S.~Zuberi, \emph{{Resummation Properties of Jet
  Vetoes at the LHC}},
  \href{https://doi.org/10.1103/PhysRevD.86.053011}{\emph{Phys. Rev.}
  {\bfseries D86} (2012) 053011}
  [\href{https://arxiv.org/abs/1206.4312}{{\ttfamily 1206.4312}}].

\bibitem{Banfi:2012jm}
A.~Banfi, P.~F. Monni, G.~P. Salam and G.~Zanderighi, \emph{{Higgs and Z-boson
  production with a jet veto}},
  \href{https://doi.org/10.1103/PhysRevLett.109.202001}{\emph{Phys. Rev. Lett.}
  {\bfseries 109} (2012) 202001}
  [\href{https://arxiv.org/abs/1206.4998}{{\ttfamily 1206.4998}}].

\bibitem{Liu:2012sz}
X.~Liu and F.~Petriello, \emph{{Resummation of jet-veto logarithms in hadronic
  processes containing jets}},
  \href{https://doi.org/10.1103/PhysRevD.87.014018}{\emph{Phys. Rev.}
  {\bfseries D87} (2013) 014018}
  [\href{https://arxiv.org/abs/1210.1906}{{\ttfamily 1210.1906}}].

\bibitem{Liu:2013hba}
X.~Liu and F.~Petriello, \emph{{Reducing theoretical uncertainties for
  exclusive Higgs-boson plus one-jet production at the LHC}},
  \href{https://doi.org/10.1103/PhysRevD.87.094027}{\emph{Phys. Rev.}
  {\bfseries D87} (2013) 094027}
  [\href{https://arxiv.org/abs/1303.4405}{{\ttfamily 1303.4405}}].

\bibitem{Becher:2013xia}
T.~Becher, M.~Neubert and L.~Rothen, \emph{{Factorization and
  $N^{3}LL_{p}$+NNLO predictions for the Higgs cross section with a jet veto}},
  \href{https://doi.org/10.1007/JHEP10(2013)125}{\emph{JHEP} {\bfseries 10}
  (2013) 125} [\href{https://arxiv.org/abs/1307.0025}{{\ttfamily 1307.0025}}].

\bibitem{Stewart:2013faa}
I.~W. Stewart, F.~J. Tackmann, J.~R. Walsh and S.~Zuberi, \emph{{Jet $p_T$
  resummation in Higgs production at NNLL$'+$NNLO}},
  \href{https://doi.org/10.1103/PhysRevD.89.054001}{\emph{Phys. Rev.}
  {\bfseries D89} (2014) 054001}
  [\href{https://arxiv.org/abs/1307.1808}{{\ttfamily 1307.1808}}].

\bibitem{Banfi:2013eda}
A.~Banfi, P.~F. Monni and G.~Zanderighi, \emph{{Quark masses in Higgs
  production with a jet veto}},
  \href{https://doi.org/10.1007/JHEP01(2014)097}{\emph{JHEP} {\bfseries 01}
  (2014) 097} [\href{https://arxiv.org/abs/1308.4634}{{\ttfamily 1308.4634}}].

\bibitem{Boughezal:2013oha}
R.~Boughezal, X.~Liu, F.~Petriello, F.~J. Tackmann and J.~R. Walsh,
  \emph{{Combining Resummed Higgs Predictions Across Jet Bins}},
  \href{https://doi.org/10.1103/PhysRevD.89.074044}{\emph{Phys. Rev.}
  {\bfseries D89} (2014) 074044}
  [\href{https://arxiv.org/abs/1312.4535}{{\ttfamily 1312.4535}}].

\bibitem{Gangal:2014qda}
S.~Gangal, M.~Stahlhofen and F.~J. Tackmann, \emph{{Rapidity-Dependent Jet
  Vetoes}}, \href{https://doi.org/10.1103/PhysRevD.91.054023}{\emph{Phys. Rev.}
  {\bfseries D91} (2015) 054023}
  [\href{https://arxiv.org/abs/1412.4792}{{\ttfamily 1412.4792}}].

\bibitem{Banfi:2015pju}
A.~Banfi, F.~Caola, F.~A. Dreyer, P.~F. Monni, G.~P. Salam, G.~Zanderighi
  et~al., \emph{{Jet-vetoed Higgs cross section in gluon fusion at
  N$^{3}$LO+NNLL with small-$R$ resummation}},
  \href{https://doi.org/10.1007/JHEP04(2016)049}{\emph{JHEP} {\bfseries 04}
  (2016) 049} [\href{https://arxiv.org/abs/1511.02886}{{\ttfamily
  1511.02886}}].

\bibitem{Shao:2013uba}
D.~Y. Shao, C.~S. Li and H.~T. Li, \emph{{Resummation Prediction on Higgs and
  Vector Boson Associated Production with a Jet Veto at the LHC}},
  \href{https://doi.org/10.1007/JHEP02(2014)117}{\emph{JHEP} {\bfseries 02}
  (2014) 117} [\href{https://arxiv.org/abs/1309.5015}{{\ttfamily 1309.5015}}].

\bibitem{Li:2014ria}
Y.~Li and X.~Liu, \emph{{High precision predictions for exclusive $VH$
  production at the LHC}},
  \href{https://doi.org/10.1007/JHEP06(2014)028}{\emph{JHEP} {\bfseries 06}
  (2014) 028} [\href{https://arxiv.org/abs/1401.2149}{{\ttfamily 1401.2149}}].

\bibitem{Moult:2014pja}
I.~Moult and I.~W. Stewart, \emph{{Jet Vetoes interfering with $H \to WW$}},
  \href{https://doi.org/10.1007/JHEP09(2014)129}{\emph{JHEP} {\bfseries 09}
  (2014) 129} [\href{https://arxiv.org/abs/1405.5534}{{\ttfamily 1405.5534}}].

\bibitem{Jaiswal:2014yba}
P.~Jaiswal and T.~Okui, \emph{{Explanation of the $WW$ excess at the LHC by
  jet-veto resummation}},
  \href{https://doi.org/10.1103/PhysRevD.90.073009}{\emph{Phys. Rev.}
  {\bfseries D90} (2014) 073009}
  [\href{https://arxiv.org/abs/1407.4537}{{\ttfamily 1407.4537}}].

\bibitem{Becher:2014aya}
T.~Becher, R.~Frederix, M.~Neubert and L.~Rothen, \emph{{Automated NNLL+NLO
  resummation for jet-veto cross sections}},
  \href{https://doi.org/10.1140/epjc/s10052-015-3368-y}{\emph{Eur. Phys. J.}
  {\bfseries C75} (2015) 154}
  [\href{https://arxiv.org/abs/1412.8408}{{\ttfamily 1412.8408}}].

\bibitem{Wang:2015mvz}
Y.~Wang, C.~S. Li and Z.~L. Liu, \emph{{Resummation prediction on gauge boson
  pair production with a jet veto}},
  \href{https://doi.org/10.1103/PhysRevD.93.094020}{\emph{Phys. Rev.}
  {\bfseries D93} (2016) 094020}
  [\href{https://arxiv.org/abs/1504.00509}{{\ttfamily 1504.00509}}].

\bibitem{Tackmann:2016jyb}
F.~J. Tackmann, W.~J. Waalewijn and L.~Zeune, \emph{{Impact of Jet Veto
  Resummation on Slepton Searches}},
  \href{https://doi.org/10.1007/JHEP07(2016)119}{\emph{JHEP} {\bfseries 07}
  (2016) 119} [\href{https://arxiv.org/abs/1603.03052}{{\ttfamily
  1603.03052}}].

\bibitem{Fuks:2017vtl}
B.~Fuks and R.~Ruiz, \emph{{A comprehensive framework for studying $W'$ and
  $Z'$ bosons at hadron colliders with automated jet veto resummation}},
  \href{https://doi.org/10.1007/JHEP05(2017)032}{\emph{JHEP} {\bfseries 05}
  (2017) 032} [\href{https://arxiv.org/abs/1701.05263}{{\ttfamily
  1701.05263}}].

\bibitem{Aad:2015ina}
{\scshape ATLAS} collaboration, \emph{{Performance of pile-up mitigation
  techniques for jets in $pp$ collisions at $\sqrt{s}=8$ TeV using the ATLAS
  detector}}, \href{https://doi.org/10.1140/epjc/s10052-016-4395-z}{\emph{Eur.
  Phys. J.} {\bfseries C76} (2016) 581}
  [\href{https://arxiv.org/abs/1510.03823}{{\ttfamily 1510.03823}}].

\bibitem{Aaboud:2018xdt}
{\scshape ATLAS} collaboration, \emph{{Measurements of Higgs boson properties
  in the diphoton decay channel with 36 fb$^{-1}$ of $pp$ collision data at
  $\sqrt{s} = 13$ TeV with the ATLAS detector}},
  \href{https://doi.org/10.1103/PhysRevD.98.052005}{\emph{Phys. Rev.}
  {\bfseries D98} (2018) 052005}
  [\href{https://arxiv.org/abs/1802.04146}{{\ttfamily 1802.04146}}].

\bibitem{Gangal:2016kuo}
S.~Gangal, J.~R. Gaunt, M.~Stahlhofen and F.~J. Tackmann, \emph{{Two-Loop Beam
  and Soft Functions for Rapidity-Dependent Jet Vetoes}},
  \href{https://doi.org/10.1007/JHEP02(2017)026}{\emph{JHEP} {\bfseries 02}
  (2017) 026} [\href{https://arxiv.org/abs/1608.01999}{{\ttfamily
  1608.01999}}].

\bibitem{Bauer:2000ew}
C.~W. Bauer, S.~Fleming and M.~E. Luke, \emph{{Summing Sudakov logarithms in $B
  \to X_s\gamma$ in effective field theory}},
  \href{https://doi.org/10.1103/PhysRevD.63.014006}{\emph{Phys. Rev.}
  {\bfseries D63} (2000) 014006}
  [\href{https://arxiv.org/abs/hep-ph/0005275}{{\ttfamily hep-ph/0005275}}].

\bibitem{Bauer:2000yr}
C.~W. Bauer, S.~Fleming, D.~Pirjol and I.~W. Stewart, \emph{{An Effective field
  theory for collinear and soft gluons: Heavy to light decays}},
  \href{https://doi.org/10.1103/PhysRevD.63.114020}{\emph{Phys. Rev.}
  {\bfseries D63} (2001) 114020}
  [\href{https://arxiv.org/abs/hep-ph/0011336}{{\ttfamily hep-ph/0011336}}].

\bibitem{Bauer:2001ct}
C.~W. Bauer and I.~W. Stewart, \emph{{Invariant operators in collinear
  effective theory}},
  \href{https://doi.org/10.1016/S0370-2693(01)00902-9}{\emph{Phys. Lett.}
  {\bfseries B516} (2001) 134}
  [\href{https://arxiv.org/abs/hep-ph/0107001}{{\ttfamily hep-ph/0107001}}].

\bibitem{Bauer:2001yt}
C.~W. Bauer, D.~Pirjol and I.~W. Stewart, \emph{{Soft collinear factorization
  in effective field theory}},
  \href{https://doi.org/10.1103/PhysRevD.65.054022}{\emph{Phys. Rev.}
  {\bfseries D65} (2002) 054022}
  [\href{https://arxiv.org/abs/hep-ph/0109045}{{\ttfamily hep-ph/0109045}}].

\bibitem{Hornig:2016ahz}
A.~Hornig, Y.~Makris and T.~Mehen, \emph{{Jet Shapes in Dijet Events at the LHC
  in SCET}}, \href{https://doi.org/10.1007/JHEP04(2016)097}{\emph{JHEP}
  {\bfseries 04} (2016) 097}
  [\href{https://arxiv.org/abs/1601.01319}{{\ttfamily 1601.01319}}].

\bibitem{Hornig:2017pud}
A.~Hornig, D.~Kang, Y.~Makris and T.~Mehen, \emph{{Transverse Vetoes with
  Rapidity Cutoff in SCET}},
  \href{https://doi.org/10.1007/JHEP12(2017)043}{\emph{JHEP} {\bfseries 12}
  (2017) 043} [\href{https://arxiv.org/abs/1708.08467}{{\ttfamily
  1708.08467}}].

\bibitem{Kang:2018agv}
D.~Kang, Y.~Makris and T.~Mehen, \emph{{From Underlying Event Sensitive To
  Insensitive: Factorization and Resummation}},
  \href{https://doi.org/10.1007/JHEP09(2018)055}{\emph{JHEP} {\bfseries 09}
  (2018) 055} [\href{https://arxiv.org/abs/1803.04413}{{\ttfamily
  1803.04413}}].

\bibitem{Kang:2016mcy}
Z.-B. Kang, F.~Ringer and I.~Vitev, \emph{{The semi-inclusive jet function in
  SCET and small radius resummation for inclusive jet production}},
  \href{https://doi.org/10.1007/JHEP10(2016)125}{\emph{JHEP} {\bfseries 10}
  (2016) 125} [\href{https://arxiv.org/abs/1606.06732}{{\ttfamily
  1606.06732}}].

\bibitem{scetlib}
M.~A. Ebert, J.~K.~L. Michel, F.~J. Tackmann et~al., \emph{{SCETlib: A C++
  Package for Numerical Calculations in QCD and Soft-Collinear Effective
  Theory}}, {\emph{DESY-17-099} (2018) }.

\bibitem{Campbell:1999ah}
J.~M. Campbell and R.~K. Ellis, \emph{{An Update on vector boson pair
  production at hadron colliders}},
  \href{https://doi.org/10.1103/PhysRevD.60.113006}{\emph{Phys. Rev.}
  {\bfseries D60} (1999) 113006}
  [\href{https://arxiv.org/abs/hep-ph/9905386}{{\ttfamily hep-ph/9905386}}].

\bibitem{Campbell:2011bn}
J.~M. Campbell, R.~K. Ellis and C.~Williams, \emph{{Vector boson pair
  production at the LHC}},
  \href{https://doi.org/10.1007/JHEP07(2011)018}{\emph{JHEP} {\bfseries 07}
  (2011) 018} [\href{https://arxiv.org/abs/1105.0020}{{\ttfamily 1105.0020}}].

\bibitem{Campbell:2015qma}
J.~M. Campbell, R.~K. Ellis and W.~T. Giele, \emph{{A Multi-Threaded Version of
  MCFM}}, \href{https://doi.org/10.1140/epjc/s10052-015-3461-2}{\emph{Eur.
  Phys. J.} {\bfseries C75} (2015) 246}
  [\href{https://arxiv.org/abs/1503.06182}{{\ttfamily 1503.06182}}].

\bibitem{Harlander:2012pb}
R.~V. Harlander, S.~Liebler and H.~Mantler, \emph{{SusHi: A program for the
  calculation of Higgs production in gluon fusion and bottom-quark annihilation
  in the Standard Model and the MSSM}},
  \href{https://doi.org/10.1016/j.cpc.2013.02.006}{\emph{Comput. Phys. Commun.}
  {\bfseries 184} (2013) 1605}
  [\href{https://arxiv.org/abs/1212.3249}{{\ttfamily 1212.3249}}].

\bibitem{Harlander:2016hcx}
R.~V. Harlander, S.~Liebler and H.~Mantler, \emph{{SusHi Bento: Beyond NNLO and
  the heavy-top limit}},
  \href{https://doi.org/10.1016/j.cpc.2016.10.015}{\emph{Comput. Phys. Commun.}
  {\bfseries 212} (2017) 239}
  [\href{https://arxiv.org/abs/1605.03190}{{\ttfamily 1605.03190}}].

\bibitem{Harlander:2002wh}
R.~V. Harlander and W.~B. Kilgore, \emph{{Next-to-next-to-leading order Higgs
  production at hadron colliders}},
  \href{https://doi.org/10.1103/PhysRevLett.88.201801}{\emph{Phys. Rev. Lett.}
  {\bfseries 88} (2002) 201801}
  [\href{https://arxiv.org/abs/hep-ph/0201206}{{\ttfamily hep-ph/0201206}}].

\bibitem{Harlander:2005rq}
R.~Harlander and P.~Kant, \emph{{Higgs production and decay: Analytic results
  at next-to-leading order QCD}},
  \href{https://doi.org/10.1088/1126-6708/2005/12/015}{\emph{JHEP} {\bfseries
  12} (2005) 015} [\href{https://arxiv.org/abs/hep-ph/0509189}{{\ttfamily
  hep-ph/0509189}}].

\bibitem{Butterworth:2015oua}
J.~Butterworth et~al., \emph{{PDF4LHC recommendations for LHC Run II}},
  \href{https://doi.org/10.1088/0954-3899/43/2/023001}{\emph{J. Phys.}
  {\bfseries G43} (2016) 023001}
  [\href{https://arxiv.org/abs/1510.03865}{{\ttfamily 1510.03865}}].

\bibitem{Dulat:2015mca}
S.~Dulat, T.-J. Hou, J.~Gao, M.~Guzzi, J.~Huston, P.~Nadolsky et~al.,
  \emph{{New parton distribution functions from a global analysis of quantum
  chromodynamics}},
  \href{https://doi.org/10.1103/PhysRevD.93.033006}{\emph{Phys. Rev.}
  {\bfseries D93} (2016) 033006}
  [\href{https://arxiv.org/abs/1506.07443}{{\ttfamily 1506.07443}}].

\bibitem{Harland-Lang:2014zoa}
L.~A. Harland-Lang, A.~D. Martin, P.~Motylinski and R.~S. Thorne, \emph{{Parton
  distributions in the LHC era: MMHT 2014 PDFs}},
  \href{https://doi.org/10.1140/epjc/s10052-015-3397-6}{\emph{Eur. Phys. J.}
  {\bfseries C75} (2015) 204}
  [\href{https://arxiv.org/abs/1412.3989}{{\ttfamily 1412.3989}}].

\bibitem{Ball:2014uwa}
{\scshape NNPDF} collaboration, R.~D. Ball et~al., \emph{{Parton distributions
  for the LHC Run II}},
  \href{https://doi.org/10.1007/JHEP04(2015)040}{\emph{JHEP} {\bfseries 04}
  (2015) 040} [\href{https://arxiv.org/abs/1410.8849}{{\ttfamily 1410.8849}}].

\bibitem{Gao:2013bia}
J.~Gao and P.~Nadolsky, \emph{{A meta-analysis of parton distribution
  functions}}, \href{https://doi.org/10.1007/JHEP07(2014)035}{\emph{JHEP}
  {\bfseries 07} (2014) 035} [\href{https://arxiv.org/abs/1401.0013}{{\ttfamily
  1401.0013}}].

\bibitem{Carrazza:2015aoa}
S.~Carrazza, S.~Forte, Z.~Kassabov, J.~I. Latorre and J.~Rojo, \emph{{An
  Unbiased Hessian Representation for Monte Carlo PDFs}},
  \href{https://doi.org/10.1140/epjc/s10052-015-3590-7}{\emph{Eur. Phys. J.}
  {\bfseries C75} (2015) 369}
  [\href{https://arxiv.org/abs/1505.06736}{{\ttfamily 1505.06736}}].

\bibitem{Hatta:2013iba}
Y.~Hatta and T.~Ueda, \emph{{Resummation of non-global logarithms at finite
  $N_c$}}, \href{https://doi.org/10.1016/j.nuclphysb.2013.06.021}{\emph{Nucl.
  Phys.} {\bfseries B874} (2013) 808}
  [\href{https://arxiv.org/abs/1304.6930}{{\ttfamily 1304.6930}}].

\bibitem{Caron-Huot:2015bja}
S.~Caron-Huot, \emph{{Resummation of non-global logarithms and the BFKL
  equation}}, \href{https://doi.org/10.1007/JHEP03(2018)036}{\emph{JHEP}
  {\bfseries 03} (2018) 036}
  [\href{https://arxiv.org/abs/1501.03754}{{\ttfamily 1501.03754}}].

\bibitem{Larkoski:2015zka}
A.~J. Larkoski, I.~Moult and D.~Neill, \emph{{Non-Global Logarithms,
  Factorization, and the Soft Substructure of Jets}},
  \href{https://doi.org/10.1007/JHEP09(2015)143}{\emph{JHEP} {\bfseries 09}
  (2015) 143} [\href{https://arxiv.org/abs/1501.04596}{{\ttfamily
  1501.04596}}].

\bibitem{Becher:2016mmh}
T.~Becher, M.~Neubert, L.~Rothen and D.~Y. Shao, \emph{{Factorization and
  Resummation for Jet Processes}},
  \href{https://doi.org/10.1007/JHEP11(2016)019,
  10.1007/JHEP05(2017)154}{\emph{JHEP} {\bfseries 11} (2016) 019}
  [\href{https://arxiv.org/abs/1605.02737}{{\ttfamily 1605.02737}}].

\bibitem{Kolodrubetz:2016dzb}
D.~W. Kolodrubetz, P.~Pietrulewicz, I.~W. Stewart, F.~J. Tackmann and W.~J.
  Waalewijn, \emph{{Factorization for Jet Radius Logarithms in Jet Mass Spectra
  at the LHC}}, \href{https://doi.org/10.1007/JHEP12(2016)054}{\emph{JHEP}
  {\bfseries 12} (2016) 054}
  [\href{https://arxiv.org/abs/1605.08038}{{\ttfamily 1605.08038}}].

\bibitem{Ellis:2010rwa}
S.~D. Ellis, C.~K. Vermilion, J.~R. Walsh, A.~Hornig and C.~Lee, \emph{{Jet
  Shapes and Jet Algorithms in SCET}},
  \href{https://doi.org/10.1007/JHEP11(2010)101}{\emph{JHEP} {\bfseries 11}
  (2010) 101} [\href{https://arxiv.org/abs/1001.0014}{{\ttfamily 1001.0014}}].

\bibitem{Procura:2011aq}
M.~Procura and W.~J. Waalewijn, \emph{{Fragmentation in Jets: Cone and
  Threshold Effects}},
  \href{https://doi.org/10.1103/PhysRevD.85.114041}{\emph{Phys. Rev.}
  {\bfseries D85} (2012) 114041}
  [\href{https://arxiv.org/abs/1111.6605}{{\ttfamily 1111.6605}}].

\bibitem{Chiu:2011qc}
J.-y. Chiu, A.~Jain, D.~Neill and I.~Z. Rothstein, \emph{{The Rapidity
  Renormalization Group}},
  \href{https://doi.org/10.1103/PhysRevLett.108.151601}{\emph{Phys. Rev. Lett.}
  {\bfseries 108} (2012) 151601}
  [\href{https://arxiv.org/abs/1104.0881}{{\ttfamily 1104.0881}}].

\bibitem{Chiu:2012ir}
J.-y. Chiu, A.~Jain, D.~Neill and I.~Z. Rothstein, \emph{{A Formalism for the
  Systematic Treatment of Rapidity Logarithms in Quantum Field Theory}},
  \href{https://doi.org/10.1007/JHEP05(2012)084}{\emph{JHEP} {\bfseries 05}
  (2012) 084} [\href{https://arxiv.org/abs/1202.0814}{{\ttfamily 1202.0814}}].

\bibitem{Hornig:2011iu}
A.~Hornig, C.~Lee, I.~W. Stewart, J.~R. Walsh and S.~Zuberi, \emph{{Non-global
  Structure of the $O(\alpha_s^2)$ Dijet Soft Function}},
  \href{https://doi.org/10.1007/JHEP08(2011)054,
  10.1007/JHEP10(2017)101}{\emph{JHEP} {\bfseries 08} (2011) 054}
  [\href{https://arxiv.org/abs/1105.4628}{{\ttfamily 1105.4628}}].

\bibitem{Balsiger:2018ezi}
M.~Balsiger, T.~Becher and D.~Y. Shao, \emph{{Non-global logarithms in jet and
  isolation cone cross sections}},
  \href{https://doi.org/10.1007/JHEP08(2018)104}{\emph{JHEP} {\bfseries 08}
  (2018) 104} [\href{https://arxiv.org/abs/1803.07045}{{\ttfamily
  1803.07045}}].

\bibitem{Bell:2015lsf}
G.~Bell, R.~Rahn and J.~Talbert, \emph{{Automated Calculation of Dijet Soft
  Functions in Soft-Collinear Effective Theory}},
  \href{https://doi.org/10.22323/1.235.0052}{\emph{PoS} {\bfseries RADCOR2015}
  (2016) 052} [\href{https://arxiv.org/abs/1512.06100}{{\ttfamily
  1512.06100}}].

\bibitem{Bell:2018jvf}
G.~Bell, R.~Rahn and J.~Talbert, \emph{{Automated Calculation of Dijet Soft
  Functions in the Presence of Jet Clustering Effects}},
  \href{https://doi.org/10.22323/1.290.0047}{\emph{PoS} {\bfseries RADCOR2017}
  (2018) 047} [\href{https://arxiv.org/abs/1801.04877}{{\ttfamily
  1801.04877}}].

\bibitem{Ligeti:2008ac}
Z.~Ligeti, I.~W. Stewart and F.~J. Tackmann, \emph{{Treating the b quark
  distribution function with reliable uncertainties}},
  \href{https://doi.org/10.1103/PhysRevD.78.114014}{\emph{Phys. Rev.}
  {\bfseries D78} (2008) 114014}
  [\href{https://arxiv.org/abs/0807.1926}{{\ttfamily 0807.1926}}].

\bibitem{Abbate:2010xh}
R.~Abbate, M.~Fickinger, A.~H. Hoang, V.~Mateu and I.~W. Stewart, \emph{{Thrust
  at N$^3$LL with Power Corrections and a Precision Global Fit for
  $\alpha_s(m_Z)$}},
  \href{https://doi.org/10.1103/PhysRevD.83.074021}{\emph{Phys. Rev.}
  {\bfseries D83} (2011) 074021}
  [\href{https://arxiv.org/abs/1006.3080}{{\ttfamily 1006.3080}}].

\bibitem{Ebert:2017uel}
M.~A. Ebert, J.~K.~L. Michel and F.~J. Tackmann, \emph{{Resummation Improved
  Rapidity Spectrum for Gluon Fusion Higgs Production}},
  \href{https://doi.org/10.1007/JHEP05(2017)088}{\emph{JHEP} {\bfseries 05}
  (2017) 088} [\href{https://arxiv.org/abs/1702.00794}{{\ttfamily
  1702.00794}}].

\bibitem{Curci:1980uw}
G.~Curci, W.~Furmanski and R.~Petronzio, \emph{{Evolution of Parton Densities
  Beyond Leading Order: The Nonsinglet Case}},
  \href{https://doi.org/10.1016/0550-3213(80)90003-6}{\emph{Nucl. Phys.}
  {\bfseries B175} (1980) 27}.

\bibitem{Furmanski:1980cm}
W.~Furmanski and R.~Petronzio, \emph{{Singlet Parton Densities Beyond Leading
  Order}}, \href{https://doi.org/10.1016/0370-2693(80)90636-X}{\emph{Phys.
  Lett.} {\bfseries 97B} (1980) 437}.

\bibitem{Ellis:1996nn}
R.~K. Ellis and W.~Vogelsang, \emph{{The Evolution of parton distributions
  beyond leading order: The Singlet case}},
  \href{https://arxiv.org/abs/hep-ph/9602356}{{\ttfamily hep-ph/9602356}}.

\bibitem{Gaunt:2014xga}
J.~R. Gaunt, M.~Stahlhofen and F.~J. Tackmann, \emph{{The Quark Beam Function
  at Two Loops}}, \href{https://doi.org/10.1007/JHEP04(2014)113}{\emph{JHEP}
  {\bfseries 04} (2014) 113} [\href{https://arxiv.org/abs/1401.5478}{{\ttfamily
  1401.5478}}].

\bibitem{Gaunt:2014cfa}
J.~Gaunt, M.~Stahlhofen and F.~J. Tackmann, \emph{{The Gluon Beam Function at
  Two Loops}}, \href{https://doi.org/10.1007/JHEP08(2014)020}{\emph{JHEP}
  {\bfseries 08} (2014) 020} [\href{https://arxiv.org/abs/1405.1044}{{\ttfamily
  1405.1044}}].

\bibitem{Stewart:2010qs}
I.~W. Stewart, F.~J. Tackmann and W.~J. Waalewijn, \emph{{The Quark Beam
  Function at NNLL}},
  \href{https://doi.org/10.1007/JHEP09(2010)005}{\emph{JHEP} {\bfseries 09}
  (2010) 005} [\href{https://arxiv.org/abs/1002.2213}{{\ttfamily 1002.2213}}].

\end{thebibliography}\endgroup

\end{document}